\documentclass[%
 aip,
 amsmath,amssymb,
preprint,%
]{revtex4-1}

\usepackage{graphicx}
\usepackage{dcolumn}
\usepackage{bm}

\usepackage[utf8]{inputenc}
\usepackage[T1]{fontenc}
\usepackage{mathptmx}
\usepackage{etoolbox}
\usepackage{physics}
\usepackage{mathdesign}
\usepackage{xcolor}

\renewcommand{\r}{\bm r}
\newcommand{\R}{\bm R}
\newcommand{\vi}{\bm v}
\newcommand{\C}{ \mathcal{C}}
\newcommand{\tens}[1]{\mathbf{#1}}

\newcommand{\g}{\textsl{g}}

\renewcommand{\eqref}[1]{Eq.~(\ref{#1})}
\newcommand{\figref}[1]{Fig.~\ref{#1}}
\newcommand{\secref}[1]{Sec.~\ref{#1}}

\newcommand{\rev}[1]{{\color{black} #1}}
\newcommand{\revv}[1]{{\color{black} #1}}
\newcommand{\revvv}[1]{{\color{black} #1}}

\makeatletter
\def\@email#1#2{%
 \endgroup
 \patchcmd{\titleblock@produce}
  {\frontmatter@RRAPformat}
  {\frontmatter@RRAPformat{\produce@RRAP{*#1\href{mailto:#2}{#2}}}\frontmatter@RRAPformat}
  {}{}
}%
\makeatother
\begin{document}

\preprint{AIP/123-QED}

\title[Implementation of the Improved Sugama Collision Operator]{Numerical Implementation of the Improved Sugama Collision Operator Using a Moment Approach}

\author{B. J. Frei}
 \email{baptiste.frei@epfl.ch}
 \affiliation{ Ecole Polytechnique F\'ed\'erale de Lausanne (EPFL), Swiss
Plasma Center, CH-1015 Lausanne, Switzerland
}%

\author{S. Ernst}%
\affiliation{ Ecole Polytechnique F\'ed\'erale de Lausanne (EPFL), Swiss
Plasma Center, CH-1015 Lausanne, Switzerland
}%

\author{P. Ricci}
 \affiliation{ Ecole Polytechnique F\'ed\'erale de Lausanne (EPFL), Swiss
Plasma Center, CH-1015 Lausanne, Switzerland
}%

\date{\today}

\begin{abstract}
The numerical implementation of the linearized gyrokinetic (GK) and drift-kinetic (DK) improved Sugama (IS) collision operators, recently introduced by Sugama \textit{et al.} [Phys. Plasmas \textbf{26}, 102108 (2019)], is reported. The  IS collision operator extends the validity of the widely-used original Sugama (OS) operator [Sugama \textit{et al.}, Phys. Plasmas \textbf{16}, 112503 (2009)] to the Pfirsch-Schlüter collisionality regime. Using a Hermite-Laguerre velocity-space decomposition of the perturbed gyrocenter distribution function that we refer to as the gyro-moment approach, the IS collision operator is written in a form of algebraic coefficients that depend on the mass and temperature ratios of the colliding species and perpendicular wavenumber. A comparison between the IS, OS, and Coulomb collision operators is performed, showing that the IS collision operator is able to approximate the Coulomb collision operator in the case of trapped electron mode (TEM) in H-mode pedestal conditions better than the OS operator. In addition, the IS operator leads to a level of zonal flow (ZF) residual which has an intermediate value between the Coulomb and the OS collision operators. \rev{The IS operator is also shown to predict a parallel electrical conductivity that approaches the one of the Coulomb operator within less than $1 \%$, while the OS operator can underestimate the parallel electron current by at least $10 \%$. Finally, closed analytical formulae of the lowest-order gyro-moments of the IS, OS and Coulomb operators are given that are ready to use to describe collisional effects in reduced gyro-moment fluid models. }
\end{abstract}

\maketitle

%

\section{Introduction}

While the core of fusion devices such as tokamaks and stellarators is sufficiently hot that plasma collisional processes are less important, the lower temperature in the boundary region enhances the role of collisions, calling for their accurate description. In fact, while the turbulent particle and heat transport in fusion devices is primarily anomalous, that is driven by small scale electromagnetic instabilities, collisions between particles can still play a crucial role since they may significantly affect the linear properties of these instabilities \cite{Belli2017,Barnes2009,manas2015enhanced,Pan2020,pan2021importance}, their saturation mechanisms via zonal flow generation \cite{Pan2020,pan2021importance,frei2021} and the velocity-space structures of the particle distribution functions. In addition, collisions drive the neoclassical processes when turbulent transport is suppressed \cite{belli2011full}. An accurate description of collisions might be also important in the transport of heavy impurities, even in the core region because of their large atomic number. 

The relatively large value of the Coulomb logarithm ($\ln \Lambda \gtrsim 10$) in fusion devices, where small-angle deflections dominate, allows for the use of the Fokker-Planck collision operator \cite{Rosenbluth1957} to describe binary Coulomb collisions between particles. In this work, we refer to the Fokker-Planck collision operator as the Coulomb collision operator. The integro-differential nature of the Coulomb collision operator makes its analytical and numerical treatment challenging. Hence, for practical applications, it is usually assumed that the distribution function is close to a Maxwellian allowing for the linearization of the Coulomb collision operator up to first order in the perturbed quantities. However, the numerical implementation and analytical treatment of the linearized Coulomb collision operator remains also challenging. For this reason, approximated linearized collision operator models have been proposed in the literature \cite{Dougherty1964,Hirshman1976b,abel2008linearized,francisquez2021improved}. 

One of the most widely used collision operator models, which we refer to as the linearized original Sugama (OS) collision operator in this work, is originally derived in Ref. \onlinecite{sugama2009}. The OS collision operator is constructed from the linearized Coulomb collision operator to fulfill the conservation laws (particle, momentum and energy), the entropy production criterion and the self-adjoint relations for arbitrary mass and temperatures ratios of the colliding species. This operator extends previous models \cite{abel2008linearized} to the case of colliding species with different temperatures. While the OS operator is implemented in numerous GK codes and tested in neoclassical and turbulent studies \cite{nunami2015development,nakata2015improved}, its deviation with respect to the Coulomb operator is expected to be enhanced when applied in the boundary plasma conditions. For instance, the OS produces a stronger collisional zonal flow damping \cite{pan2021,frei2021} and can therefore yield turbulent transport levels that can significantly differ from the ones obtained by the Coulomb collision operator \cite{pan2021}. A recent study of the collisional effects of ITG reports that the OS Sugama predicts a smaller ITG growth rate at perpendicular wavelength of the order of the ion gyroscale than the Coulomb collision operator \cite{frei2022}. Additionally, previous collisional simulations of ZF demonstrate that the OS operator yields a stronger ZF damping than the Coulomb operator in both banana and Pfirsch-Schlüter regime \cite{pan2021,frei2021}. 

To simulate highly collisional plasmas while still avoiding the use the Coulomb collision operator, Ref. \onlinecite{sugama2019improved} recently reported on the development of an operator that we refer to as the improved Sugama (IS) collision operator. The IS operator is designed to reproduce the same friction-flow relations (that we define below) of the linearized Coulomb collision operator by adding a correction term to the OS \cite{hirshman1981neoclassical,honda2014impact}. While the IS operator has been successfully tested and implemented recently in neoclassical simulations using the GT5D code \cite{matsuoka2021neoclassical} where like-species collisions are considered, no direct comparison between the IS, OS and Coulomb collision operators have been reported yet on, e.g., microinstabilities and collisional zonal flow damping.

In this paper, we take advantage of recent analytical and numerical progress made in the development of GK collision operators based on a Hermite-Laguerre expansion \cite{frei2021} , which we refer to as the gyro-moment approach, of the perturbed distribution function \cite{frei2020gyrokinetic}, and present the derivation and the expansion of the GK and DK IS collision operators on the same basis. \rev{In particular, we leverage the gyro-moment expansion of the OS operator reported in Ref. \onlinecite{frei2021}, which is benchmarked with GENE \cite{jenko2000}. Despite that the gyro-moment expansion of the GK Coulomb is available in Ref. \onlinecite{frei2021}, simpler collision operator models are still important to develop accurate collisional descriptions of the plasma dynamics in the boundary region, yet simpler than the Coulomb operator.} Using the Hermite-Laguerre approach, the integro-differential nature of collision operator model is reduced to the evaluation of closed analytical expressions involving numerical coefficients that depend on the mass and temperature ratios of the colliding species and, when finite Larmor radius (FLR) terms are included, on the perpendicular wavenumber. The numerical implementation of the IS collision operator allows us to perform its comparison with the OS as well as the Coulomb collision operators on the study of instabilities and ZF damping for the first time. In particular, the linear properties of the trapped electron modes (TEMs) at steep pressure gradients, similar to H-mode conditions, are investigated and reveal that, indeed, the IS can approach better the Coulomb operator in the Pfirsch-Schlüter regime than the OS operator. Also, all operators yield results that agree within $10 \%$ at least for the parameters explored in this work. Nevertheless, larger deviations are expected in the case of heavy impurities \cite{casson2014theoretical} with temperatures of the colliding species that can be significantly different. Additionally, we show that the IS yields a ZF damping intermediate between the OS and Coulomb collision operators in the Pfirsch-Schlüter regime. \rev{Finally, we evaluate the electrical Spitzer conductivity using the IS operator showing a good agreement with the Coulomb collision operator, while it is found that the OS operator underestimates the parallel electric current compared to the Coulomb operator by at least $10 \%$.} Taking advantage of the gyro-moment expansion, we explicitly evaluate the lowest-order gyro-moments of the IS, OS and Coulomb collision operators that can be used to model collisional effects in reduced gyro-moment models valid under the high-collisionality assumption.

The remainder of the present paper is organised as follows. In \secref{sec:IS}, the IS collision operator is introduced. Then, in \secref{sec:GKandDKIS}, we derive the spherical harmonic expansion of the IS collision operator that allows us to evaluate the GK and DK limits of the same operator. In \secref{sec:braginksii}, we derive closed analytical expressions of the Braginksii matrices of the Coulomb and OS collision operators necessary for the evaluation of the correction terms added to the OS operator. Then, the gyro-moment method is detailed in \secref{sec:gyromoment} where the Hermite-Laguerre expansion of the GK and DK IS collision operators are obtained analytically. In \secref{sec:numericaltest}, numerical tests and comparisons are performed focusing on the TEM at steep pressure gradients, on the study of the collisional ZF damping in the Pfirsch-Schlüter regime\rev{, and on the Spitzer electrical conductivity.} We conclude by discussing the results and future applications in \secref{sec:conclusion}. \rev{Appendix \ref{appendix:A} and \ref{appendix:Abis} details the Coulomb and OS collision operators, and Appendix \ref{appendix:B} reports on the analytical expressions of the lowest-order gyro-moments of the Coulomb, OS and IS collision operators, which are useful to derive reduced gyro-moment models for high collisional plasmas.}

\section{Improved Sugama Collision Operator}
\label{sec:IS}

 We start by presenting the linearized IS collision operator following the notation and definitions of Ref.  \onlinecite{sugama2019improved}. We assume that the particle distribution function is perturbed with respect to a Maxwellian distribution of the particle of species $a$, $f_{Ma} = f_{Ma}(\bm r, \bm v) = n_a(\bm  r) / \pi^{3/2} v_{Ta}(\bm r)^{3/2} e^{- v^2 /v_{Ta}(\bm r)^2 }$, with $n_a(\bm r)$ the particle density, $v_{Ta}(\bm r)^2 = 2 T_a(\bm r) / m_a$ the thermal particle velocity and $\bm z = (\bm r, \bm v)$ the particle phase-space coordinates, being $\bm r$ the particle position and $\bm v$ the particle velocity such that $v^2 = \vi \cdot \vi$. The small-amplitude perturbation, $f_a = f _a(\bm r,\bm v)$, i.e. $f_a / f_{Ma} \ll 1$, allows us to describe the collisions between species $a$ and $b$ by linearizing a nonlinear collision operator model $C_{ab}^{NL} (f_a,f_b)$. The linearized operator is denoted by $C_{ab} = C_{ab}(\bm r, \bm v) = C_{ab} (f_a,f_b)$ and can be written as 

\begin{align} \label{eq:linCab}
   C_{ab} (f_a,f_b) =  C_{ab}^T(f_a) + C_{ab}^F (f_b),
\end{align}
\\
where $ C_{ab}^T(f_a)$ and $ C_{ab}^F (f_b)$ are the test and field components of $C_{ab} (f_a,f_b)$, and are defined from the nonlinear collision operator model, as $C_{ab}^T(f_a) = C_{ab}^{NL }(f_a, f_{Mb})$ and  $C_{ab}^F(f_b) = C_{ab}^{NL }(f_{Ma}, f_{b})$.

Because the test and field components of the Coulomb collision operator, denoted by $C_{ab}^L$ and defined in Appendix \ref{appendix:A}, involve complex velocity-space derivatives of $f_a$ (e.g. in $C_{ab}^{LT}$ given in \eqref{eq:CLT}) and integrals $f_b$ (e.g. in $C_{ab}^{LF}$ given in \eqref{eq:CLF}), approximated linearized collision operators have been proposed for implementation in numerical codes and analytical purposes in the past years \cite{Dougherty1964,abel2008linearized,sugama2009}. Among these simplified models, the OS collision operator model \cite{sugama2009}, that we denote as $C_{ab}^S$, is widely used in present GK codes. \rev{The definitions of the test and field components of the OS collision operator, $C_{ab}^S$, are introduced in Appendix \ref{appendix:A}.} The original Sugama operator, $C_{ab}^S$, is derived from the linearized Coulomb collision operator, $C_{ab}^L$, to conserve the three lowest-order velocity moments of $C_{ab}$, i.e.

\begin{subequations} \label{eq:conservationlaws}
\begin{align}
    \int d \vi C_{ab}(f_a, f_b) & = 0,  \label{eq:particlecons} \\
     m_a \int d \vi \vi  C_{ab}^T(f_a) & =  - m_b \int d \vi \vi C_{ba}^F(f_a) ,\label{eq:momentumconservation} \\
      m_a \int d \vi  v^2 C_{ab}^T(f_a) & =  - m_b \int d \vi  v^2 C_{ba}^F(f_a),
\end{align}
\end{subequations}
\\
and satisfy the H-theorem and the adjointess relations even in the case of collisions between particles with temperatures $T_a \neq T_b$, given by
 
 \begin{align} \label{eq:Htheorem}
T_a \int d \bm{v} \frac{f_a}{f_{Ma}} C_{ab} (f_a,f_b) + T_b  \int d \bm{v} \frac{f_b}{f_{Mb}}C_{ba} (f_b ,f_a) \leq 0,
 \end{align}
and 
\begin{subequations} \label{eq:selfadjoint}
\begin{align}
\int d \bm{v} \frac{f_a}{f_{Ma}} C_{ab}^T(\g_a )  & = \int d \bm{v} \frac{\g_a}{f_{Ma}} C_{ab}^T(f_a ), \\
T_a \int d \bm{v} \frac{f_a}{f_{Ma}} C_{ab}^F(\g_b ) &  =T_b \int d \bm{v} \frac{\g_b}{f_{Mb}} C_{ba}^F(f_a ),
\end{align}
\end{subequations}
\\
respectively, where $f_a$ and $g_a$ are two arbitrary phase-space $(\bm r, \bm v)$ functions. The complete analytical derivation of  $C_{ab}^S$ in Ref. \onlinecite{sugama2009}. 

The difference between the OS and the Coulomb collision operators is expected to have a larger impact as the collisionality increases, in particular, in the case of particles with different mass and temperature \cite{Belli2017}. Hence, Ref. \onlinecite{sugama2019improved} proposes to improve the OS collision operator by adding a correction term to $C_{ab}^S$, thus defining a new improved operator, that we refer to as the IS collision operator. The IS operator is designed such that it produces to same friction-flow relations \cite{hirshman1981neoclassical}, given by 

\begin{align} \label{eq:ForcesFai}
    \bm F_{ai} = (-1)^{i-1} \int d \bm{v} m_a \bm v L_{i-1}^{3/2} (s_a^2) \sum_{b} C_{ab}(f_a, f_b),
\end{align}
\\
than the Coulomb collision operator. In \eqref{eq:ForcesFai}, $s_a^2 = v^2 / v_{Ta}^2$ is the energy coordinate and $L_{i}^{3/2 }(x)$ (with $i = 0,1,2, \dots$) is \rev{the associated Laguerre polynomial defined by

\begin{align} \label{eq:Laguerre}
L_i^{\alpha}(x) = \sum_{l=0}^i L_{il}^{\alpha - 1/2} x^l,
\end{align}
\\
where $L_{il}^{\alpha - 1/2}  = (-1)^l (\alpha + i )! / [(i - l)!(l +\alpha )! l! ]$ (with $\alpha > -1$).} The derivation of the correction term added to $C_{ab}^S$ can be found in Ref. \onlinecite{sugama2019improved} and we report the results here.

The linearized IS collision operator, denoted by $C_{ab}^{IS}$, is obtained by adding the correction term $\Delta C_{ab}(f_a,f_b)$ to $C_{ab}^S$ (see Appendix \ref{appendix:A}), i.e.

\begin{align} \label{eq:ISugama}
C_{ab}^{IS}(f_a, f_b) = C_{ab}^{S}(f_a, f_b) + \Delta C_{ab}(f_a, f_b),
\end{align}
\\
where $\Delta C_{ab}$ is defined as

\begin{align} \label{eq:DeltaCabLS}
\Delta C_{ab} =  \Delta C_{ab}^T + \Delta C_{ab}^F,
\end{align}
\\
with the test and field components of the correction term being

\begin{align} \label{eq:DeltaCabT}
     \Delta C_{ab}^T  &=   
 \sum_{\ell =0}^{L}    \sum_{k=0}^K  \frac{m_a}{T_a}\frac{f_{Ma}}{\bar{\tau}_{ab}}   c_\ell \Delta M_{ab}^{\ell k}L_\ell^{3/2}(s_a^2) \bm v \cdot \bm{u}_{ak}(f_a), 
\end{align}
\\
and 

 \begin{align} \label{eq:DeltaCabF}
     \Delta C_{ab}^F & =
\sum_{\ell=0}^{L}    \sum_{k=0}^K   \frac{m_a}{T_a}\frac{f_{Ma}}{\bar{\tau}_{ab}}  c_\ell  \Delta N_{ab}^{\ell k} L_\ell^{3/2}(s_a^2) \bm v \cdot \bm{u}_{bk}(f_b),
\end{align}
\\
respectively. Here, $\bar{\tau}_{ab} = 3 \sqrt{\pi}/( 4 \nu_{ab})$ is the collisional time between the colliding species $a$ and $b$, with $\nu_{ab}$ the associated collision frequency $\nu_{ab} =  4 \sqrt{\pi} n _b q_a^2 q_b^2 \ln \Lambda/[3 m_a^{1/2} T_a^{3/2}]$ (see Ref. \onlinecite{frei2021}). While the IS is derived in the limit $L \to \infty$ and $K \to \infty$, here we consider $(L,K)$ to be two positive integers that we choose equal, i.e. $L = K$. Since previous neoclassical transport studies suggest that accurate friction coefficients in the Pfirsch-Schlüter regime require that $L  = K \gtrsim 2$ (see Ref. \onlinecite{honda2014impact}), we consider the cases of $L = K = 2$, $5$ and $10$. \rev{Despite that neoclassical studies have revealed that the energy ($\sim v^2$) expansion in Laguerre polynomials has slow convergence in the banana regime \cite{landreman2013new}, we show here that small $L$ and $K$ are required at high-collisionality in the test cases considered in \secref{sec:numericaltest}}. We note that the first neoclassical studies reported in Ref. \onlinecite{matsuoka2021neoclassical} are performed with the like-species IS using only $L = K = 1$, $2$.

The quantities $\bm{u}_{ak}(f_a)$ appearing in Eqs. (\ref{eq:DeltaCabT}) and (\ref{eq:DeltaCabF}) are defined as the flow vectors and are expressed by \citep{honda2014impact}

\begin{align} \label{eq:uak}
 \bm{u}_{ak} (f_a) = \frac{c_k}{n_a} \int d \vi f_a L_k^{3/2}(s_a^2) \vi ,
\end{align}
\\
with $c_k = 3 \cdot  2^k k ! /(2k +3 )!!$. Finally, we introduce the correction Braginskii matrix elements, $\Delta M_{ab}^{\ell k}$ and $\Delta N_{ab}^{\ell k}$, defined by

\begin{subequations} \label{eq:correctionmatrix}
\begin{align}
\Delta M_{ab}^{\ell k} &  =  M_{ab}^{L\ell k} -  M_{ab}^{S\ell k},\\
\Delta N_{ab}^{\ell k} &  =  N_{ab}^{L\ell k} -  N_{ab}^{S \ell k},
\end{align}
\end{subequations}
\\
where the Braginskii matrices, $M_{ab}^{A\ell k}$ and $N_{ab}^{A\ell k}$ (being $A = L$ for the Coulomb and $A = S$ for the OS operators) are obtained from the test and field components of the operator collision model $C_{ab}^{AT}$ and $C_{ab}^{AF}$, respectively, and are defined by \cite{helander2005collisional,honda2014impact}

\begin{subequations} \label{eq:matrixelements}
\begin{align} \label{eq:MabAjk}
\frac{n_a}{\bar{\tau}_{ab}}M_{ab}^{A\ell k} & = \int d \vi v_{\parallel} L_\ell^{3/2}(s_a^2) C_{ab}^{AT}\left(f_{Ma} \frac{m_a v_\parallel}{T_a} L_k^{3/2}(s_a^2)  \right), \\
\frac{n_a}{\bar{\tau}_{ab}} N_{ab}^{A\ell k} & = \int d \vi v_{\parallel} L_\ell^{3/2}(s_a^2) C_{ab}^{AF}\left( f_{Mb} \frac{ m_b v_\parallel}{T_b} L_k^{3/2}(s_b^2)\right),\label{eq:NabAjk}
\end{align}
\end{subequations}
\\
with $v_\parallel  = \bm b \cdot \bm v$ the parallel component of the velocity along the magnetic field and $\bm b = \bm B / B$. 

The Braginskii matrices $M_{ab}^{A\ell k}$ and $N_{ab}^{A\ell k}$ satisfy a set of relations stemming from the conservation laws and symmetries of the collision operator. In particular, from the momentum conservation law in \eqref{eq:momentumconservation}, one obtains that

\begin{align} \label{eq:momentumconsbraginskii}
M_{a b}^{A0 k}+\frac{T_{a} v_{T a}}{T_{b} v_{T b}} N_{b a}^{A0 k}=0 \quad(k=0,1,2, \ldots).
\end{align}
\\
In addition, in the case of collisions between particle species with the same temperatures ($T_a = T_b$), the Braginskii matrices admit symmetry properties because of the self-adjoint relations of \rev{$C_{ab}^{AT}$ and  $C_{ab}^{AF}$} given in \eqref{eq:selfadjoint}. From Eqs. (\ref{eq:matrixelements}) and (\ref{eq:selfadjoint}), one obtains that 

\begin{align}
M_{a b}^{A \ell k} &=M_{a b}^{A k \ell}, \\
\frac{N_{a b}^{A \ell k}}{T_{a} v_{T a}} &=\frac{N_{b a}^{A k\ell}}{T_{b} v_{T b}} \quad(\ell, k=0,1,2, \ldots),
\end{align}
\\
which implies, \rev{ from \eqref{eq:correctionmatrix}}, that

\revv{
\begin{align}
\Delta M_{a b}^{\ell k} &=0, \\
\Delta N_{a b}^{\ell k} &=\frac{N_{a b}^{A00} N_{a b}^{A\ell k}-N_{a b}^{A\ell 0} N_{a b}^{A0 k}}{N_{ab}^{A00}} \quad(\ell, k=0,1,2, \ldots),\\ 
\Delta N_{a b}^{00}& =\Delta N_{a b}^{\ell 0}=\Delta N_{ab}^{0 k}=0,
\end{align}}
\\
for $\ell, k=1,2, \ldots$ in the case $T_a = T_b$. While the analytical expressions of the Braginskii matrices $M_{a}^{A\ell k}$ and $N_{a}^{A\ell k}$ up to $(\ell, k) \leq 2$ can be found in Ref. \onlinecite{sugama2019improved}, we provide a new derivation of these coefficients here that allows us to extend them to arbitrary order $(\ell, k)$ in \secref{sec:braginksii}. This allows us to evaluate gyro-moment expansion of the IS for any $(L,K)$. Using these analytical expressions, we demonstrate numerically in \secref{sec:numericaltest} that the relations and symmetry properties of the Braginskii matrices are satisfied.

\section{Spherical harmonic Expansion and Gyro-Average of the Improved Sugama Operator}
\label{sec:GKandDKIS}

We expand the IS collision operator in terms of spherical harmonic particle moments in \secref{sec:sphericalharmonicexpansion} \cite{jorge2019nonlinear,frei2021}. This allow us to evaluate its gyro-average and to derive the GK and DK limits of the IS collision operators in Secs. \ref{sec:GKIS} and \ref{sec:DKIS}, respectively. We notice that, while the GK form of the IS collision operator is presented in Ref. \onlinecite{sugama2019improved}, here we follow a different methodology, which is based in the spherical harmonic technique used to obtain the GK and DK Coulomb collision operators in Ref. \onlinecite{frei2021}. \rev{We note that the GK formulations of the Coulomb and of the OS collision operators used in this work are reported in Appendix \ref{appendix:A}.} Finally, we note that the spherical harmonic expansion is particularly useful in deriving the expressions of the Braginksii matrix elements in \secref{sec:braginksii}.

\subsection{Spherical Harmonic Expansion of the Improved Sugama Colllsion Operator}
\label{sec:sphericalharmonicexpansion}

 The perturbed particle distribution function, $f_a (\bm r, \bm v)$, is expanded in the spherical harmonic basis according to \citep{jorge2019nonlinear,frei2021}

\begin{align} \label{eq:momentexpansion}
f_a  (\bm r, \bm v)= f_{Ma} \sum_{p,j} \frac{1}{\sigma_j^p} \tens{M}_a^{pj} (\bm r) \cdot \tens{Y}^{pj}(\bm{s}_a),
\end{align}
\\
with \revv{$\bm s_a = \bm v / v_{Ta}$} and $\sigma_j^p = p!(p + j +1/2)! / [2^p (p +1/2)! j!]$. In \eqref{eq:momentexpansion}, the spherical harmonic basis is defined by $\tens{Y}^{pj}(\bm s_a) = \tens{Y}^p(\bm s_a) L_j^{p+1/2}(s_a^2)$, where we introduce the spherical harmonic tensors of order $p$, $\tens{Y}^p$, and the associated Laguerre polynomials, $L_j^{p+1/2}(s_a^2)$, defined in \eqref{eq:Laguerre}. The tensor $\tens{Y}^p $ can be explicitly defined by introducing the spherical harmonic basis, $\tens e^{pm}$, that satisfies the orthogonality relation $\tens e^{pm} \cdot \tens e^{pm'} = (-1)^m \delta_{-m}^{m'}$ \cite{Snider2017}, such that

\begin{align} \label{eq:Yp}
 \tens Y^p (\bm s_a)= s_a^p  \sqrt{\frac{2 \pi^{3/2} p!}{2^p (p+1/2)!}}\sum_{m=-p}^{p }  Y_p^m(\xi,\theta) \tens e^{pm},
 \end{align}
 \\
where $Y_p^m(\xi,\theta)$ are the scalar harmonic functions. \rev{The spherical harmonic basis, $\bm e^{1m}$, can be related to the orthogonal cartesian velocity-space basis $(\bm e_x,\bm  e_y,\bm e_z )$ with $\bm e_z = \bm b$, by $\bm e^{1-1} = (\bm e_x - i \bm e_y)/\sqrt{2}$, $\bm e^{10} = \bm b$ and $\bm e^{11} = - (\bm e_x + i \bm e_y)/\sqrt{2} $ for $p =1$, while $\bm e^{pm}$ with $p > 1$ can be expressed in terms of $\bm e^{1m}$ by using the closed formula \cite{Snider2017}

\begin{align}\label{eq:bmepm}
\bm e^{pm} = N_{pm} \sum_{n=0}^{[(p-m)/2]} a_n^{pm} \left\{(\bm e^{11} )^{m+n} (\bm e^{1-1})^n (\bm e^{10})^{p-m-2n} \right\}_{S} ,
\end{align}
\\
where $N_{pm} = \sqrt{(p+m)!(p-m)!2^{p-m}/(2p)!}$, $a_n^{pm} = p!/[2^n n! (m+n)!(p-m-2n)!]$, and $\{\dots \}_{S}$ denotes the symmetric part. A complete introduction to the spherical harmonic tensors can be found in Ref. \onlinecite{Snider2017}.}

The spherical harmonic basis in \eqref{eq:momentexpansion}, satisfies the orthogonality relation \cite{Snider2017}

\begin{align} \label{eq:orthYpLj} & \frac{1}{\pi^{3/2} \sigma^p_j }
\int d \bm v e^{- v^2}  L_j^{p+1/2}(v^2) \tens Y^{p'}(\bm v)  L_{j'}^{p'+1/2}(v^2) \tens{Y}^{p}(\bm v) \cdot \tens{T}^{p} \nonumber \\ 
&  = \delta_{pp'} \delta_{jj'}  \tens{T}^{p},
\end{align}
\\
with $\tens{T}^p$ an arbitrary $p$-th order tensor. From the orthogonality relation in \eqref{eq:orthYpLj}, it follows that the spherical harmonic particle moments $\tens{M}_a^{pj}(\bm r)$ are defined as

\begin{align} \label{eq:defMabpj}
  \tens{M}_a^{pj}(\bm r)  = \frac{1}{n_a}\int d \vi f_a(\bm r, \bm v) \tens Y^{pj}(\bm s_a) . 
\end{align}
\\
Here, we emphasize that the spherical harmonic particle moments depend on the particle position $\bm r$ only. 

We now use the definition of the spherical harmonic particle moments, $\tens{M}_a^{pj}$ given \eqref{eq:defMabpj}, and relate them to the flow vectors expressed in \eqref{eq:uak}, to obtain the spherical harmonic expansion of the IS collision operator. Therefore, from \eqref{eq:uak} and noticing that $\bm s_a = \tens Y^1(\bm s_a)$, we derive 

\begin{align} \label{eq:uaktoM1k}
 \tens{u}_{ak} (f_a) & =  \frac{c_k  v_{Ta}}{n_a} \int d \vi f_a L_k^{3/2}(s_a^2) \tens Y^1(\bm s_a) \nonumber \\ 
 & =  c_k v_{Ta}  \tens{M}_a^{1k}(\bm r).
\end{align}
\\
Inserting \eqref{eq:uaktoM1k} into Eqs. (\ref{eq:DeltaCabT}) and (\ref{eq:DeltaCabF}) yields the spherical harmonic expansion of the IS collision operator, which is useful to evaluate its GK limit.

\subsection{Gyrokinetic Improved Sugama Collision Operator}
\label{sec:GKIS}

We now consider the GK limit of the IS collision operator where the fast particle gyro-motion is analytically averaged out. Contrary to the IS collision operator, defined on the particle phase-space $\bm z = (\r, \bm v)$ (see \eqref{eq:ISugama}), the GK IS collision operator, which we denote by $\mathcal{C}_{ab}^{IS}$, is defined on the gyrocenter phase-space coordinates $\bm Z = (\bm R, \mu, v_\parallel, \theta, t)$ where $\bm R$ is the gyrocenter position, $\mu = m v_\perp^2/[2 B]$ is the magnetic moment and $\theta$ is the gyroangle. More precisely, $\mathcal{C}_{ab}^{IS}$, is obtained by performing the gyro-average of the IS collision operator $C_{ab}^{IS}$, i.e.

\begin{align} \label{eq:GKISdef}
    \mathcal{C}_{ab}^{IS} =  \left< C_{ab}^{IS} \right>_{\bm R } = \int_{0}^{2 \pi} \bigg \rvert_{\bm R} \frac{d \theta}{2 \pi} C_{ab}^{IS}(\bm z(\bm Z)) .
\end{align}
\\
where the integral over the gyroangle appearing in \eqref{eq:GKISdef} is performed holding $\bm R$ constant, while collisions occur at the particle position $\bm r$. We detail the transformation given in \eqref{eq:GKISdef} in Appendix \ref{appendix:Abis} that we use to obtain, for instance, the GK Coulomb operator. In general, the coordinate transformation that relates the gyrocenter and particles coordinates, $\bm Z$ and $\bm z$ respectively, can be written as $\bm Z = \bm z + \delta \bm z $, where $\delta \bm z$ are functions of phase-space coordinates and perturbed fields and contain terms at all orders in the GK expansion parameter $\epsilon \sim e \phi / T_e$ ($\phi$ being the small amplitude and small scale electrostatic fluctuating potential \cite{brizard2007foundations, frei2021}). At the lowest order in the GK expansion, the coordinate transformation reduces to $\delta v_\parallel = \delta \mu = \delta \theta = 0$ and $\delta \r \simeq - \bm \rho_{a}$. Hence, the IS collision operator can be gyro-averaged holding $\bm R$ constant in gyrocenter coordinates, such that $\r= \R(\r, \bm v) + \bm \rho_a(\r, \vi)$ \cite{frei2020gyrokinetic}. 

Focusing first on the test component of $\Delta C_{ab}^T$, we perform the gyro-average in \eqref{eq:GKISdef} in gyrocenter coordinates using $\r= \R(\r, \bm v) + \bm \rho_a(\r, \vi)$ in the spatial argument of $\tens{M}_a^{1 k}(\bm r)$, such that in Fourier space it yields $\tens{M}_a^{1 k}(\bm r)  = \int d \bm k \tens{M}_a^{1 k}(\bm k) e^{i \bm k \cdot \bm R} e^{ i \bm k \cdot \bm \rho_a}$. For a single Fourier component, we derive

\begin{align} \label{eq:DeltacurlyCabT}
\Delta \C_{ab}^{T} & = \sum_{\ell =0}^{L}    \sum_{k=0}^K  \frac{m_a}{T_a}\frac{f_{Ma}}{\bar{\tau}_{ab}}   c_\ell c_k v_{Ta} \Delta M_{ab}^{\ell k}L_\ell^{3/2}(s_a^2)   e^{i \bm k \cdot \bm R}  \nonumber \\
& \times  \left< e^{i \bm k \cdot \bm \rho_a} \bm v \right>_{\bm R}\cdot \tens{M}_a^{1k}(\bm k)    \nonumber \\ 
= &  \sum_{\ell =0}^{L}    \sum_{k=0}^K  \frac{m_a}{T_a}\frac{f_{Ma}}{\bar{\tau}_{ab}}   c_\ell c_k v_{Ta} \Delta M_{ab}^{\ell k}L_\ell^{3/2}(s_a^2)   e^{i \bm k \cdot \bm R} \nonumber \\
& \times \left[ v_{\parallel } J_{0a} \bm b \cdot\tens{M}_a^{1k}(\bm k)  + i v_\perp J_{1a} \bm e_{2}\cdot\tens{M}_a^{1k}(\bm k) \right],
\end{align}
\\
where we introduce the basis vectors $(\bm e_1, \bm e_2)$, such that $\bm v_\perp = v_\perp ( \bm e_2 \cos \theta - \bm e_1 \sin \theta )$. Additionally, $J_{0a} = J_0(b_a \sqrt{x_a})$ and  $J_{1a} = J_1(b_a \sqrt{x_a})$ are the zeroth and first order Bessel functions (with $b_a = k_\perp \rho_a$ the normalized perpendicular wavenumber and $x_a = v_\perp^2 / v_{Ta}^2$), resulting from the presence of FLR effects in the IS collision operator. \rev{We remark that the equilibrium quantities in \eqref{eq:DeltacurlyCabT} (e.g., $f_{Ma}$, $T_a$ and $\bar{\tau}_{ab}$) are assumed to vary on spatial scale lengths much larger that $\bm \rho_a$. Therefore, they are evaluated at $\bm R$ and are not affected by the gyro-average operator, in contrast to $\tens{M}_a^{1 k}(\bm r)$.}

We now relate the spherical harmonic moments $\tens{M}_a^{1k}$ of the perturbed particle to the gyrocenter perturbed distribution functions. More precisely, we express the $\tens{M}_a^{1k}$ in terms of the nonadiabatic part, $h_a$, of the perturbed gyrocenter distribution function $\g_a$. The two gyrocenter distribution functions, $h_a$ and $\g_a$, are related by \cite{frei2021}

\begin{align} \label{eq:hatoga}
     h_a(\R,\mu,v_\parallel) = g_a(\R,\mu,v_\parallel)  + \frac{q_a}{T_{a}}     F_{Ma}(\R, \mu, v_\parallel)   \left< \phi \right>_{\R},
 \end{align}
\\
in the electrostatic limit. \rev{ Here, $F_{Ma} = N_a(\bm R)  / \pi^{3/2} v_{Ta}^3(\bm R) e^{- v_\parallel^2 /v_{Ta}^2(\bm R) - \mu B(\bm R) / T_a(\bm R) }$ is the gyrocenter Maxwellian distribution function}. The perturbed particle distribution function $f_a$ is related to the perturbed gyrocenter distribution function $g_a$ by the scalar invariance of the full particle and gyrocenter distribution functions, i.e.

 \begin{align} \label{eq:fafctofga}
  f_a(\bm r, \bm v)  =g_a(\bm R, \mu, v_\parallel) +  F_{Ma}(\R, \mu, v_\parallel) - f_{Ma}(\r(\bm Z), \bm v(\bm Z)) .
 \end{align}
 \\
Using the pull-back operator $\mathcal{T}$, such that the functional forms of $f_{Ma}$ and $F_{Ma}$ are related by $f_{Ma} = \mathcal{T} F_{Ma} $ \cite{frei2020gyrokinetic}, we derive that

 \begin{align} \label{eq:fafctofga}
  f_a(\bm r, \bm v)  =g_a(\bm R, \mu, v_\parallel) +  ( F_{Ma} - \mathcal{T} F_{Ma}) (\bm Z)  = g_a^{gc}(\bm Z).
 \end{align}
\\
We remark that, while both $g_a$ and $h_a$ are gyrophase independent functions, $g_a^{gc}$ is gyrophase dependent via its arguments $\bm Z$. An expression of the pull-back transformation $\mathcal{T}$ can be obtained at the leading order in the GK expansion parameter $\epsilon \sim e \phi / T_e $, yielding \cite{frei2021}

\begin{align} \label{eq:fatoha}
 f_a(\r, \bm v)   & = g_a^{gc}(\bm Z(\bm z))  \nonumber \\
&   =  g_a(\R(\r, \bm v),\mu,v_\parallel) - \frac{q_a }{T_a}  F_{Ma} \left(  \phi (\r)-   \left< \phi \right>_{\R}\right) + O(\epsilon^2) \nonumber \\
&  =  h_a(\R(\r, \bm v),\mu,v_\parallel) - \frac{q_a }{T_a}\phi (\r)  F_{Ma} + O(\epsilon^2) , 
\end{align}
\\
being $\r = \R + \bm \rho_a(\mu, \theta)$. Using \eqref{eq:fatoha} allows us to finally express $\tens{M}_a^{1k}$ in terms of $h_a$, 

\begin{align} \label{eq:Mawithha}
v_{Ta} \tens{M}_a^{1k}(\bm r) &  =  \frac{v_{Ta}}{n_a} \int d \vi f_a(\bm r, \bm v ) L_k^{3/2}(s_a^2) \tens Y^1(\bm s_a) 
 \nonumber \\
  & = \frac{1}{n_a} \int d \vi \int d \bm r' \delta ( \bm r' - \bm r) f_a(\bm r', \bm v ) L_k^{3/2}(s_a^2) \bm v \nonumber \\
  & = \frac{1}{N_a} \int d \bm R d v_\parallel d \mu d \theta \frac{B}{m_a} \delta ( \bm R + \bm \rho_a - \bm r) \nonumber \\
   & \times h_a(\bm R(\r, \bm v), \mu, v_\parallel ) L_k^{3/2}(s_a^2) \bm v \nonumber \\
&  =  \frac{e^{i \bm k \cdot \bm r}}{N_a} \int d v_\parallel d \mu d \theta \frac{B}{m_a} L_k^{3/2}(s_a^2) \nonumber \\
& \times  h_a(\bm k, \mu, v_\parallel ) 
\left( \bm b v_\parallel J_{0a} - i v_\perp \bm e_2 J_{1a}\right),
\end{align}
\\
\rev{where $N_a = \int d v_\parallel d \mu d \theta F_{Ma}$ is the equilibrium gyrocenter density. \revv{In the third line of \eqref{eq:Mawithha}}, the lowest-order guiding-center contribution to the gyro-center phase-space volume element is neglected such that $B_\parallel^* / m_a = B (1 + v_\parallel \bm b \cdot \grad \times \bm b /  \Omega_a) / m_a \simeq B / m_a$, since the last term in $B_\parallel^*$ is of the order $v_\parallel \bm b \cdot \grad \times \bm b / \Omega_a \sim \rho_a / L_B \ll 1$ (with $L_B$ the typical equilibrium scale length of the magnetic field  $B$).} We remark that the contribution from the terms proportional to $\phi$, appearing in \eqref{eq:fatoha}, are neglected in \eqref{eq:Mawithha}. In fact, while these terms are of the same order as $h_a$ (i.e. they are order $\epsilon$), they yield a small contribution in the collision operator \cite{sugama2009}. We neglect them here, but notice that their contributions to the Coulomb collision operator are included in Ref. \onlinecite{frei2021} and have little effects at the gyroradius scale.
With Eqs. (\ref{eq:Mawithha}) and (\ref{eq:DeltacurlyCabT}), the GK test component of the correction term , $\Delta \C_{ab}^T = \left< \Delta C_{ab}^T \right>_{\bm R}$, can be obtained in terms of the gyrocenter distribution function $h_a$. Focusing on the field component  of $\Delta \C_{ab}$, i.e. on $ \Delta \C_{ab}^F$, we remark that a similar derivation of its expression can be carried out as for $ \Delta \C_{ab}^T$ . In particular, the expression of the spherical harmonic moment $\tens{M}_b^{1k}$, appearing in \eqref{eq:DeltaCabF}, is obtained with \eqref{eq:Mawithha} having replaced $a$ with $b$.

Finally, the GK IS collision operator can be expressed as \cite{sugama2019improved}

\begin{align} \label{eq:GKCabIS}
    \C_{ab}^{IS} = \C_{ab}^S + \Delta \C^T_{ab} + \Delta \C^F_{ab} ,
\end{align}
\\
where $\C^S_{ab}$ is the OS GK collision operator, given in Ref. \onlinecite{sugama2009}, and $\Delta \C^T_{ab}$ and $\Delta \C^F_{ab}$ given by

\begin{subequations} \label{eq:DeltaGKCabTF}
    \begin{align}\label{eq:DeltaGKCabT}
    \Delta \C_{ab}^T &= \sum_{\ell=0}^L \sum_{k=0}^K \frac{c_\ell}{\bar{\tau}_{ab}} \frac{m_a F_{Ma}}{T_a} L_\ell^{3/2}(s_a^2) \Delta M_{ab}^{\ell k} 
    \nonumber \\
     &\times \qty(\bar u_{\parallel a}^k\left(h_a\right) J_{0a} v_\parallel + \bar u_{\perp a}^k\left(h_a\right) J_{1a} v_\perp) ,  \\
    \Delta \C_{ab}^F &= \sum_{\ell =0}^L \sum_{k=0}^K   \frac{c_\ell }{\bar{\tau}_{ab}} \frac{m_a F_{Ma}}{T_a} L_\ell^{3/2}(s_a^2)   \Delta N_{ab}^{\ell k} \nonumber \\
     &\times \qty(\bar u_{\parallel b}^{k} \left(h_b\right)J_{0a} v_\parallel + \bar u_{\perp b}^k\left(h_b\right) J_{1a} v_\perp). \label{eq:DeltaGKCabF}
\end{align}
\end{subequations}
\\
where we introduce the quantities

\begin{subequations} \label{eq:Uparallelperpsk}
    \begin{align}
        \bar u_{\parallel s}^k\left(h_s\right) &= \frac{c_k}{n_s}\int   d v_\parallel d \mu d \theta  \frac{B}{m_s} L_k^{3/2} (s_s^2) h_s J_{0s} v_{\parallel},  \\
        \bar u_{\perp s}^k\left(h_s\right) &= \frac{c_k}{n_s}\int  d v_\parallel d \mu d \theta  \frac{B}{m_s} L_k^{3/2} (s_s^2) h_s J_{1s} v_\perp.
    \end{align}
\end{subequations}
\\
\rev{The GK IS collision operator can be obtained by adding to the GK OS collision operator, $\C_{ab}^S$, the terms in Eqs. (\ref{eq:DeltaGKCabTF}) and (\ref{eq:Uparallelperpsk}). In Appendix \ref{appendix:Abis}, we discuss the GK formulation of the Coulomb and OS operators that we use to perform the numerical tests in Sec. \ref{sec:numericaltest}. In particular, we note that the GK operators considered in this work are derived from the linearized collision operators applied to the perturbed particle distribution function $f_a$, following the derivation of the full-F nonlinear GK Coulomb collision operator in Ref. \onlinecite{jorge2019nonlinear} (see Appendix \ref{appendix:Abis}).}


 
 \subsection{Drift-Kinetic Improved Sugama Collision Operator}
 \label{sec:DKIS}
 
 We now derive the DK IS operator from the GK IS collision operator, given in \eqref{eq:GKCabIS}, by neglecting the difference between the particle and gyrocenter position, such that $\bm r \simeq \bm R$. Hence, the DK IS collision operator is derived in the zero gyroradius limit of the GK IS operator approximating $J_{0s} \simeq 1$, $J_{1s} =0$ and  $f_s \simeq g_s$ (see \eqref{eq:hatoga}). The DK OS is given in Ref. \onlinecite{frei2021} and the DK limits of the GK test and field components of $\Delta \C_{ab}$ are 

\begin{subequations} \label{eq:DeltaCabTFDK}
    \begin{align}
    \Delta \C_{ab}^T &= \sum_{\ell=0}^L \sum_{k=0}^K \frac{m_a}{T_a}\frac{F_{Ma}}{\bar{\tau}_{ab}}  c_\ell c_k v_{Ta}L_\ell^{3/2}(s_a^2) \Delta M_{ab}^{\ell k} v_\parallel \bm b \cdot  \tens{M}_a^{1k}  , \label{eq:DeltaCabTDK}\\
    \Delta \C_{ab}^F &= \sum_{\ell=0}^L \sum_{k=0}^K \frac{m_a}{T_a}\frac{F_{Ma}}{\bar{\tau}_{ab}}  c_\ell c_k v_{Tb}L_\ell^{3/2}(s_a^2)   \Delta N_{ab}^{\ell k} v_\parallel \bm b \cdot  \tens{M}_b^{1k}, \label{eq:DeltaCabFDK}
\end{align}
\end{subequations}
\\
respectively. In \eqref{eq:DeltaCabTFDK}, the spherical particle moments, $\tens{M}_s^{1k}$ ($s = {a,b}$), are expressed in terms of $g_s$, such that

\begin{align} 
  \bm b \cdot \tens{M}_s^{1k}  = \frac{1}{N_s}\int d \mu d v_\parallel d \theta \frac{B}{m_s} g_s L_k^{3/2}(s_s^2) \frac{v_\parallel}{v_{Ts}} . 
\end{align}
\\
The GK and DK expressions of the IS collision operator, given in Eqs. (\ref{eq:DeltaGKCabTF}) and (\ref{eq:DeltaCabTFDK}), can be implemented in continuum GK codes using a discretization scheme in velocity-space (e.g., based on a finite volume approach or a finite difference scheme) to evaluate numerically the velocity-integrals. In this work, we use a gyro-moment approach to carry out these velocity integrals and implement these operators numerically. For this purpose, we derive closed analytical expressions of the Braginksii matrices, required to evaluate the quantities $\Delta M_{ab}^{\ell k}$ and $\Delta N_{ab}^{\ell k}$, in the next section.

\section{Braginksii Matrices}
\label{sec:braginksii}

In order to evaluate the correction term $\Delta C_{ab}$ given in \eqref{eq:DeltaCabLS}, analytical expressions for the Braginskii matrices, \rev{$M_{ab}^{A\ell k}$ and $N_{ab}^{A\ell k}$}, associated with the Coulomb and OS collision operators are derived. This extends the evaluation of the $(\ell,k) \leq 2$ Braginksii matrices reported in Ref. \onlinecite{sugama2019improved} to arbitrary $(\ell,k)$. For this calculation, we leverage the spherical harmonic expansions of the Coulomb and OS collision operators presented in Ref. \onlinecite{frei2021}. More precisely, we use the spherical harmonic expansion of $f_a$ (see \secref{sec:sphericalharmonicexpansion}) to obtain the Braginskii matrices of the Coulomb and OS collision operators in Secs. \ref{sec:braginskiicoulomb} and \ref{sec:braginskiisugama}, respectively. 

\subsection{Braginskii Matrix of the Coulomb Collision Operator}
\label{sec:braginskiicoulomb}

We first derive the Braginskii matrix associated with the Coulomb collision operator, \rev{namely $M_{ab}^{L\ell k}$ and $N_{ab}^{L\ell k}$}, appearing in \eqref{eq:correctionmatrix}. For this purpose, we use the expansion of the perturbed particle distribution function, given in \eqref{eq:momentexpansion}, to obtain the spherical harmonic expansion of the test and field components of the Coulomb collision operator, $C_{ab}^{LT}$ and $C_{ab}^{LF}$, derived in Ref. \onlinecite{frei2021}. \rev{These expressions, reported here, are given by

\begin{subequations}\label{eq:cmomentexpansion}
\begin{align} 
C^{LT}_{ab}(f_a) & = C^{LT}_{ab}(\bm r, \bm v) =  \sum_{p=0}^\infty \sum_{j=0}^\infty   \tens{M}_a^{pj}(\bm r) \cdot  C^{LT}_{ab} \left( \frac{f_{Ma}}{\sigma_j^p}  \tens{Y}^{pj}(\bm{s}_a)\right) ,\label{eq:CabTmoment} \\
  C^{LF}_{ab}(f_b) & =C^{LF}_{ab}(\bm r, \bm v)  =\sum_{p=0}^\infty \sum_{j=0}^\infty   \tens{M}_b^{pj} (\bm r) \cdot  C^{LF}_{ab} \left( \frac{f_{Mb}}{\sigma_j^p} \tens{Y}^{pj}(\bm{s}_b)\right)  , \label{eq:CabFmoment}
\end{align}
\end{subequations}
\\
with 

\begin{subequations} \label{eq:cmomentexpansion_pj}
\begin{align}
  C^{LT}_{ab} \left( \frac{f_{Ma}}{\sigma_j^p}  \tens{Y}^{pj}(\bm{s}_a)\right) &   = \frac{f_{Ma}}{\sigma_j^p} \tens Y^p(\hat{\bm v })\nu_{ab}^{Tpj}(v), \label{eq:CabTmomentpj} \\
 C^{LF}_{ab} \left( \frac{f_{Mb}}{\sigma_j^p} \tens{Y}^{pj}(\bm{s}_b)\right)   &  = \frac{f_{Ma}}{\sigma_j^p} \tens Y^p(\hat{\vi})\nu_{ab}^{Fpj}(v), \label{eq:CabFmomentpj} 
\end{align}
\end{subequations}
\\
where the expressions of the test and field speed functions, $\nu_{ab}^{Tpj} (v)$ and $\nu_{ab}^{Fpj}(v)$, can be found in Appendix A of Ref. \onlinecite{frei2021}.} We first use the expression in \eqref{eq:CabTmomentpj} to evaluate $M_{ab}^{L\ell  k}$ (see \eqref{eq:MabAjk}). We remark that the GK Coulomb collision operator, which we compare with the GK IS operator in \secref{sec:numericaltest}, is derived in Ref. \onlinecite{frei2021} by gyro-averaging \eqref{eq:cmomentexpansion} according to \eqref{eq:GKISdef}.

\rev{ By noticing that $\tens Y^1(\bm s_a) = \vi / v_{Ta}$ and by considering the component parallel to $\bm b$ of \eqref{eq:CabTmomentpj}, we obtain the test part of the Coulomb collision operator evaluated with $ f_{Ma} m_a v_\parallel L_k^{3/2}(s_a^2) / T_a$, i.e.}
 
\begin{align}  \label{eq:CabLTpj}
C^{LT}_{ab}\left( f_{Ma} \frac{m_a v_\parallel}{T_a} L_k^{3/2}(s_a^2)\right)  = \frac{2 f_{Ma}}{v_{Ta}} \frac{v_\parallel}{v}\nu_{ab}^{T1k}(v).
\end{align}
\\
Then, the matrix element $M_{ab}^{L\ell k}$, defined in \eqref{eq:MabAjk}, can be computed by expanding the associated Laguerre polynomial using \eqref{eq:Laguerre} with $p=1$ and by performing the velocity integrals over the speed function $\nu_{ab}^{T1k}(v)$. It yields

\begin{equation}\label{eq:M_ab_jk}
    M_{ab}^{L\ell k} = \sum_{l=0}^\ell  \frac{2}{3}\frac{\bar{\tau}_{ab}}{n_a} L_{\ell l}^1 \bar\nu_{*ab}^{T1kl}.
\end{equation}
\\
A similar derivation can be carried out to evaluate $N_{ab}^{L \ell k}$ (see \eqref{eq:NabAjk}) by using the expression in \eqref{eq:CabFmomentpj}, i.e.

\begin{equation}\label{eq:N_ab_jk}
    N_{ab}^{L\ell k} = \sum_{l=0}^\ell \frac{2}{3}\frac{\bar{\tau}_{ab}}{n_a}\chi_{ab} L_{\ell l}^1 \bar\nu_{*ab}^{F1kl},
\end{equation}
\\
with $\chi_{ab} = v_{Ta} / v_{Tb}$ the ratio between the thermal velocities. The closed analytical expressions of test and field speed integrated functions $\bar\nu_{*ab}^{T1kl}$ and $\bar\nu_{*ab}^{F1kl}$ appearing in Eqs. (\ref{eq:M_ab_jk}) and (\ref{eq:N_ab_jk}), respectively, are given in Ref. \onlinecite{frei2021}. Eqs. (\ref{eq:M_ab_jk}) and (\ref{eq:N_ab_jk}) allows for the evaluation of the terms associated with the Coulomb collision operator appearing in the correction matrix elements, $\Delta M_{ab}^{\ell k}$ and $\Delta N_{ab}^{\ell k}$ defined in \eqref{eq:correctionmatrix}. They are evaluated in terms of mass and temperature ratios of the colliding species.

\subsection{Braginskii Matrix of the Original Sugama Collision Operator}
\label{sec:braginskiisugama}

We now evaluate the Braginskii matrix associated with the OS collision operator, namely $M_{ab}^{S\ell k}$ and $N_{ab}^{S\ell k}$, appearing in \eqref{eq:correctionmatrix}. For $f_a = f_{Ma} m_a v_\parallel L_k^{3/2}(s_a^2)  / T_a$, using the spherical harmonic expansion of the OS operator in Ref. \onlinecite{frei2021}, the test component of the OS collision operator yields

\begin{align} \label{eq:CabTs}
    C_{ab}^{ST}\left(f_{Ma}L_k^{3/2}(s_a^2)\frac{m_a v_\parallel}{T_a} \right) &  = \frac{2}{v_{Ta}} f_{Ma} \frac{v_\parallel}{v}\nu_{ab}^{S1k}(v) + \sum_{i=1}^3 X_{ab}^i,
\end{align}
\\
where the quantities $X_{ab}^i$ are defined by

\begin{subequations}
\begin{align}
X_{ab}^1 &= - \frac{16}{3\sqrt{\pi}}(1+\chi_{ab}^2)(\theta_{ab}-1)  f_{Ma} \frac{m_a v_\parallel}{T_a} \sum_{l=0}^k L_{kl}^1 \bar\nu_{ab}^{\parallel l+3}, \\   
   X_{ab}^2 &=  - 2(1+\chi_{ab}^2) (\theta_{ab}-1)\frac{m_a}{T_a} u_{\parallel a}^k v_\parallel  f_{Ma} \nu_{ab}^\parallel(v) s_a^2, \\ 
    X_{ab}^3 &= -\frac{2}{\bar{\tau}_{ab}} f_{Ma} \frac{\chi_{ab}(\theta_{ab}-1)^2}{\sqrt{1+\chi_{ab}^2}}\frac{m_a}{T_a} v_\parallel u_{\parallel a}^k ,  
    \end{align}
    \end{subequations}
    \\
 with $\nu_{ab}^{S1k}(v)$ the velocity dependent speed function (whose expression is given in Ref. \onlinecite{frei2021}), being $\bar\nu_{ab}^{\parallel k} = \int_0^\infty \dd s_a s_a^{2k} \nu_{ab}^\parallel(v) e^{-s_a^2}$ (with $\nu_{ab}^{\parallel}(v)  = 2 \nu_{ab}  \left[ \erf(s_b) - s_b \erf'(s_b)\right]/(2 s_b^2 s_a^3) $ the velocity dependent energy diffusion frequency, and $u_{\parallel a}^k = 4 /(3\sqrt{\pi}) (k+3/2)! /k!\delta_k^0$. In deriving \eqref{eq:CabTs}, we remark that the terms proportional to $ T_a \int d \vi f_a \left( 2s_a^2/3 - 1 \right) /n_a$ vanish exactly when applied to $f_a = f_{Ma} m_a v_\parallel L_k^{3/2}(s_a^2)  / T_a$ because of the velocity integration over the pitch-angle variable $v_\parallel / v$.

Similarly, the field component of the OS collision for $f_b = f_{Mb} m_b v_\parallel L_k^{3/2}(s_b^2)  / T_b$ yields

\begin{align} \label{eq:CabFS}
    & C_{ab}^{SF}\left( f_{Mb}L_k^{3/2}(s_b^2)\frac{m_b v_\parallel}{T_a}\right) = \dfrac{2 \theta_{ab}}{\bar{\tau}_{ab}}(1+\chi_{ab}^2)f_{Ma} \frac{m_a v_\parallel }{T_a} \nonumber \\ 
    & \times \qty[\frac{3\sqrt{\pi}}{2}\frac{\Phi(s_b)}{s_a}+\frac{\chi_{ab}(\theta_{ab}-1)}{(1+\chi_{ab}^2)^{3/2}}]     V_{ab}\left( f_{Mb}L_k^{3/2}(s_b^2)\frac{m_b v_\parallel}{T_b} \right) ,
\end{align}
\\
where 

\begin{align} \label{eq:Vab}
&V_{ab}\left(f_{Mb}L_k^{3/2}(s_b^2)\frac{m_b v_\parallel}{T_b} \right)    
    = -\frac{\theta_{ba}}{\bar{\tau}_{ba}}(1+\chi_{ba}^2) \frac{m_b}{\gamma_{ab}}\nonumber \\
    & \times \int d \vi L_k^{3/2}(s_b^2) \frac{m_b v_\parallel^2}{T_b} f_{Mb}\qty[\frac{3\sqrt{\pi}}{2}\frac{\Phi(s_a)}{s_b}+\frac{\chi_{ba}(\theta_{ba}-1)}{(1+\chi_{ba}^2)^{3/2}}],
    \end{align}
\\
with $\Phi(x) =  \left[ \erf(x) - x \erf'(x)\right]/(2 x^2)$. The velocity integral in \eqref{eq:Vab} can be performed analytically using \eqref{eq:Laguerre}, and leads to 

\begin{align}\label{eq:Vab_expansion}
    V_{ab}\left(f_{Mb}L_k^{3/2}(s_b^2)\frac{m_b v_\parallel}{T_b} \right) 
    &= \frac{m_b}{\gamma_{ab}} \mathcal V_{ab}^l,
\end{align}
\\
with the following definition

\begin{align}
    \mathcal V_{ab}^l & = -2\theta_{ab}(1+\chi_{ab}^2) \frac{n_b}{\bar{\tau}_{ab}} \left[\frac{1}{2} \frac{\chi_{ab}(\theta_{ab}-1)}{(1+\chi_{ab}^2)^{3/2}}u_{\parallel a}^l \right. \nonumber \\ & \left.  + \sum_{m=0}^l L_{lm}^1 \qty(\frac{1}{\chi_{ab}^2}E_{ab}^m - \frac{1}{\chi_{ab}}e_{ab}^{m+1}) \right],
    \end{align}
\\
where we introduce $e_{ba}^k = \int_0^\infty d s_b s_b^{2k} \erf'(s_a) e^{-s_b^2} $ and $E_{ba}^k = \int_0^\infty d s_b s_b^{2k+1} \erf(s_a)e^{-s_b^2}$.
The test and field components of the OS collision operator, given in Eqs. (\ref{eq:CabTs}) and (\ref{eq:CabFS}), are now in a suitable form to evaluate the analytical expressions  of $M_{ab}^{S\ell k}$ and $N_{ab}^{S\ell k}$ defined by \eqref{eq:matrixelements}.

Starting with the Braginksii matrix element associated with $C_{ab}^{TS}$, the velocity integral in $M_{ab}^{Sjk}$, given in \eqref{eq:MabAjk}, is evaluated using the series expansion of the associated Laguerre polynomials, \eqref{eq:Laguerre}. Thus, we derive 

\begin{equation}\label{eq:M_ab_Sjk}
    M_{ab}^{S\ell k} = \sum_{n=1}^3 M_{abn}^{S\ell k},
\end{equation}
\\
where we introduce the quantities

\begin{subequations}
\begin{align}
    \label{eq:M_ab0_Sjk} M_{ab1}^{S\ell k} &= \frac{\bar{\tau}_{ab}}{n_a} \sum_{l=0}^\ell \frac{2}{3}L_{\ell l}^1 \bar\nu_{*ab}^{S1kl},\\
    \label{eq:M_ab1_Sjk} M_{ab2}^{S\ell k} &= -\frac{16}{3\sqrt{\pi}}\bar{\tau}_{ab}(\theta_{ab}-1)(1+\chi_{ab}^2) \nonumber \\
    & \times \qty[u_{\parallel a}^k\sum_{l=0}^\ell L_{\ell l}^1 \bar\nu_{ab}^{\parallel l+3} + u_{\parallel a}^\ell  \sum_{l=0}^k L_{kl}^1 \bar\nu_{ab}^{\parallel l+3}],\\
    \label{eq:M_ab2_Sjk} M_{ab3}^{S\ell k} &= -\frac{2\chi_{ab}(\theta_{ab}-1)^2}{\sqrt{1+\chi_{ab}^2}} u_{\parallel a}^k u_{\parallel a}^\ell.
\end{align}
\end{subequations}
\\
In \eqref{eq:M_ab0_Sjk}, the analytical expression of the speed integrated function, $\bar{\nu}_{*ab}^{S1kl}$, is reported in Ref. \onlinecite{frei2021}. Similarly for $N_{ab}^{S\ell j}$, using \eqref{eq:CabFS} and employing the expansion of the associated Laguerre polynomials in \eqref{eq:Laguerre} yield

\begin{equation}\label{eq:N_ab_Sjk}
    N_{ab}^{S\ell k} = -\frac{2\bar{\tau}_{ab}}{n_a} \frac{m_b}{\gamma_{ab}}\mathcal V_{ba}^k \mathcal V_{ab}^\ell.
\end{equation}
\\
The Braginskii matrix elements associated with the Coulomb collision operator, given in Eqs. (\ref{eq:M_ab_jk}) and (\ref{eq:N_ab_jk}), and the ones associated with the OS operator, given in Eqs. (\ref{eq:M_ab_Sjk}) and (\ref{eq:N_ab_Sjk}), allow us to obtain the correction Braginskii matrix elements  $\Delta M_{ab}^{\ell k}$ and $\Delta N_{ab}^{\ell k}$ for arbitrary $(\ell,k)$.

\section{Gyro-Moment Expansion of the Improved Sugama Collision Operator}
\label{sec:gyromoment}

We now project the GK and DK IS collision operators onto a Hermite-Laguerre polynomial basis, a technique that we refer to as the gyro-moment approach. Previous works \citep{frei2021,jorge2017drift,frei2022} demonstrate the advantage of the gyro-moment approach in modelling the plasma dynamics in the boundary region, where the time evolution of the gyro-moments is obtained by projecting the GK Boltzmann equation onto the Hermite-Laguerre basis yielding an infinite set of fluid-like equations \cite{frei2020gyrokinetic}. At high-collisionality, high-order gyro-moments are damped such that only the lowest-order ones are sufficient to evolve the dynamics. As a consequence, in these conditions, the gyro-moment hierarchy can be reduced to a fluid model where collisional effects are obtained using the Hermite-Laguerre expansion of advanced collision operators at the lowest-order in the ratio between the particle mean-free-path to the parallel scale length. \rev{In Appendix \ref{appendix:B}, we evaluate the lowest-order gyro-moments of the DK IS, OS and Coulomb collision operators that enter in the evolution equations of the lowest-order gyro-moments associated with fluid quantities. Ultimately, this allows us to compare analytically the fundamental differences between the IS, OS and the Coulomb collision operators. In addition, the closed analytical expressions reported in Appendix \ref{appendix:B} can be used to derive high-collisional closures of the gyro-moment hierarchy \cite{frei2022}.}

 Since the gyro-moment expansion of the OS GK and DK Sugama collision operator is obtained in Ref. \onlinecite{frei2021}, we focus here on the projections of the corrections $\Delta \C^T_{ab}$ and $\Delta \C^F_{ab}$, given in \eqref{eq:DeltaGKCabTF}. The GK IS collision operator $\C_{ab}^{IS}$ is formulated in terms of moments of $h_a$, the non-adiabatic part of the perturbed gyrocenter distribution function $g_a$ (see \eqref{eq:hatoga}). Therefore, we expand the collision operator in terms of gyro-moments of $h_a$. More precisely, $h_a$ is written as

\begin{align} \label{eq:hHL}
  h_a = \sum_{p = 0}^\infty \sum_{j = 0}^\infty n_a^{pj} \frac{H_p(s_{\parallel a}) L_j(x_a)}{\sqrt{2^p p!}} F_{Ma},
 \end{align}
\\
with $s_{\parallel a} = v_\parallel / v_{Ta}$. In \eqref{eq:hHL}, the Hermite and Laguerre polynomials, $H_p$ and $L_j$, are defined via their Rodrigues' formulas $H_p(x)  = (-1)^p e^{x^2} d^p\left( e^{- x^2} \right)/ d x^p$ and $L_j(x)  = e^{x}/ j! d^j\left ( e^{- x } x^j\right)/ d x^j$, and are orthogonal over the intervals,\rev{ $[ -\infty, \infty]$ weighted by $e^{-x^2}$, and $[0,+ \infty]$ weighted by $e^{-x}$}, respectively, such that 
 
 \begin{align} \label{eq:orthogonality}
 \int_{- \infty}^\infty d x H_p(x) H_{p'}(x) e^{- x^2} & = 2^p p! \sqrt{\pi} \delta_{p}^{p'}, \\
 \quad  \int_0^\infty d x L_j(x) L_{j'}(x) e^{-x} & = \delta_j^{j'}.
 \end{align}
\\
Because of the orthogonality relations in \eqref{eq:orthogonality}, the non-adiabatic gyro-moments of $h_a$, $n_a^{pj}$, are defined by

\begin{align} \label{eq:npjdef}
n_a^{pj} = \frac{1}{N_a} \int d \mu  d v_\parallel  d \theta\frac{B}{m_a} h_a \frac{H_p(s_{\parallel a}) L_j(x_a)}{\sqrt{2^p p!}}.
\end{align}
\\
where $N_a $ is the gyrocenter density and $F_{Ma}$ the gyrocenter Maxwellian distribution function. \rev{ We remark that the velocity-dependence contained in the gyrocenter phase-space Jacobian (neglected in \eqref{eq:Mawithha}) can be retained in the gyro-moment expansion, carried out below, by introducing the modified gyro-moments $n_a^{*pj}$, defined by \eqref{eq:npjdef} having replaced $B/ m_a$ by $B_\parallel^* / m_a$ \cite{jorge2017drift,frei2020gyrokinetic}, such that 

\begin{align}
    n_a^{*pj} =  n_a^{pj} + \frac{v_{Ta} }{\sqrt{2}} \frac{\bm b \cdot \grad \times \bm b}{\Omega_a} \left( \sqrt{p +1} n_a^{p+1j} + \sqrt{p} n_a^{p-1 j}\right).
\end{align}
}
\\
We detail the projection of the GK IS collision operator in \secref{subsec:expansionofGKIS}, and obtain the DK limit of the same operator in \secref{subsec:expansionofDKIS} in terms of $n_a^{pj}$.

\subsection{Expansion of the GK IS Collision Operator}
\label{subsec:expansionofGKIS}

 We first derive the gyro-moment expansion of the GK IS collision operator in \eqref{eq:GKCabIS}, that is
 
 \begin{align} \label{eq:CabISpj}
     \C_{ab}^{ISpj} = \C_{ab}^{Spj} + \Delta \C_{ab}^{Tpj} + \Delta \C_{ab}^{Fpj}.
 \end{align}
 \\
 where the Hermite-Laguerre projection of $\C_{ab}^{IS}$ is defined by 
 
 \begin{align}
     \C_{ab}^{ISpj} =  \frac{1}{N_a} \int d \mu  d v_\parallel d \theta\frac{B}{m_a}   \C_{ab}^{IS} \frac{H_p(s_{\parallel a}) L_j(x_a)}{\sqrt{2^p p!}},
 \end{align}
 \\
 and similar definitions are used for the remaining terms in \eqref{eq:CabISpj}, in particular
 
 \begin{align} 
     \Delta \C_{ab}^{Tpj} & =  \frac{1}{N_a} \int d \mu  d v_\parallel d \theta\frac{B}{m_a}   \Delta \C_{ab}^{T} \frac{H_p(s_{\parallel a}) L_j(x_a)}{\sqrt{2^p p!}}  \label{eq:DeltaCabTpjdef}
     \end{align}
\\
and
     \begin{align}
      \Delta \C_{ab}^{Fpj} & =  \frac{1}{N_a} \int d \mu  d v_\parallel d \theta\frac{B}{m_a}   \Delta \C_{ab}^{F} \frac{H_p(s_{\parallel a}) L_j(x_a)}{\sqrt{2^p p!}}.\label{eq:DeltaCabFpjdef}
 \end{align}
 \\
Since the expression of $\C_{ab}^{Spj}$ is reported in Ref. \onlinecite{frei2021}, we focus here on the gyro-moment expansion of $\Delta \C_{ab}^T$ and $\Delta \C_{ab}^F$, defined in Eqs. (\ref{eq:DeltaCabTpjdef}) and (\ref{eq:DeltaCabFpjdef}), respectively. First, we derive the Hermite-Laguerre projection of the test component of the correction term $\Delta \C_{ab}^{Tpj}$. As an initial step, we express $\bar u_{\parallel s}^k $ and $\bar u_{\perp s}^k $, defined in \eqref{eq:Uparallelperpsk} and appearing in \eqref{eq:DeltaGKCabTF}, in terms of the non-adiabatic gyro-moments $n_a^{pj}$. Injecting the expansion of $h_a$ into $\bar u_{\parallel s}^k $ and $\bar u_{\perp s}^k $ yields 
 
 \begin{subequations} \label{eq:uparauaperp}
 \begin{align}
\bar u_{\parallel s}^k &= v_{Ts}c_k \sum_{p=0}^\infty \sum_{j=0}^\infty n_a^{pj} I_{\parallel s}^{pjk},\\
   \bar u_{\perp s}^k &= v_{Ts}c_k \sum_{p=0}^\infty \sum_{j=0}^\infty n_a^{pj} I_{\perp s}^{pjk},
\end{align}
 \end{subequations}
\\
where we introduce the velocity integrals

\begin{align}
    \label{eq:I_para_pjk_def} I_{\parallel s}^{pjk} &=  \frac{1}{N_a}\int d \mu  d v_\parallel d \theta\frac{B}{m_a} \frac{H_p(s_{\parallel s})L_j(x_s)}{\sqrt{2^p p!}}F_{Ms} L_k^{3/2}(s_s^2) J_{0s} s_{\parallel s}
\end{align}
\\
and

\begin{align}        
    I_{\perp s}^{pjk} &=  \frac{1}{N_a}\int d \mu  d v_\parallel d \theta\frac{B}{m_a} \frac{H_p(s_{\parallel s})L_j(x_s)}{\sqrt{2^p p!}}F_{Ms} L_k^{3/2}(s_s^2) J_{1s} \sqrt{x_s}.   \label{eq:I_perp_pjk_def} 
\end{align}
\\
To analytically evaluate the velocity integrals in $ I_{\parallel s}^{pjk}$ and $I_{\perp s}^{pjk}$, we expand the Bessel functions $J_{0s}$ and $J_{1s}$ in terms of associated Laguerre polynomials\rev{, $L_n^m(x_s)$ (see \eqref{eq:Laguerre})}, as follows \citep{gradshteyn},
 
\begin{equation}\label{eq:Bessel_relation}
    J_m(b_s\sqrt{x_s}) =\left( \frac{b_s \sqrt{x_s}}{2} \right)^m \sum_{n=0}^\infty \frac{n! \mathcal K_n(b_s)}{(n+m)!} L_n^m(x_s),
\end{equation}
\\
with the $n$th-order kernel function $\mathcal K_n(b_s) = (b_s /2)^{2n} e^{-b_s^2/4} / n!$ describing FLR effects. \rev{Then, using \eqref{eq:Bessel_relation} with

\begin{equation}
    L_n^m(x) L_j(x) x^m = \sum_{f=0}^{n+m+j} d_{njf}^m L_f(x),
\end{equation}
\\
where $L_f$ is the Laguerre polynomial defined in \eqref{eq:Laguerre} with $\alpha =0$ and $ d_{njf}^m$ are numerical coefficients whose closed analytical expressions are given in Ref. \onlinecite{frei2021}}, the velocity integral in $I_{\parallel s}^{pjk}$ can be computed,

\begin{align} \label{eq:Iparallelspjk}
    I_{\parallel s}^{pjk} = \frac{2}{3\sqrt{\pi}}  \sum_{n=0}^\infty \sum_{f=0}^{n+j} (T^{-1})_{pf}^{1k0} \frac{\mathcal{K}_n(b_s)}{\sqrt{2^p p!}} d_{njf}^0   \frac{(k+3/2)!}{k!}\nonumber\\
    \times [(p\geq 1 \cup f\geq 1)\cap (f+\lfloor p/2\rfloor \geq k) ],
\end{align}
\\
with $[\cdot]$ the Iverson bracket ($[A] = 1$ if $A$ is true, and $0$ otherwise). Finally, the velocity integral contained in $ I_{\perp s}^{pjk} $ can be evaluated similarly to the one in \eqref{eq:Iparallelspjk}, and yields

\begin{align} \label{eq:Iperpspjk}
     I_{\perp s}^{pjk} =& \sum_{n=0}^\infty\sum_{f=0}^{n+j+1}\sum_{r=0}^{f+\lfloor p/2 \rfloor}\sum_{q=0}^k \sum_{r_1=0}^{r} \frac{(T^{-1})_{pf}^{0r0}}{\sqrt{\pi}}\frac{b_s\mathcal{K}_n(b_s)}{(n+1)\sqrt{2^pp!}}\nonumber\\
    & \times  d_{njf}^1 L_{kq}^1 L^0_{rr_1} \left( 1/2 + r_1 + q\right)!.
\end{align}
\\
We remark that the expression of the numerical coefficients $(T^{-1})_{pj}^{lkm}$ in Eqs. (\ref{eq:Iparallelspjk}) and (\ref{eq:Iperpspjk}) can be found in Ref. \onlinecite{jorge2019nonlinear}, and arise from the basis transformation from Hermite-Laguerre to associated Legendre-Laguerre polynomials.

We now have all elements necessary to focus on the evaluation of $\Delta \C_{ab}^{Tpj}$ obtained by projecting \eqref{eq:DeltaCabT} onto the Hermite-Laguerre basis. In \eqref{eq:DeltaCabTpjdef}, one recognises the quantities $ I_{\parallel s}^{pjk}$ and $I_{\perp s}^{pjk}$ defined in Eqs. (\ref{eq:Iparallelspjk}) and (\ref{eq:Iperpspjk}). Thus, using their definitions, the gyro-moment expansion of the test component of the correction terms, $\Delta \C_{ab}^{Tpj}$, is deduced

\begin{align} \label{eq:DeltaCabTpj}
    \Delta \C_{ab}^{Tpj} = \sum_{\ell=0}^{L}\sum_{k=0}^K \frac{2c_\ell  }{\bar{\tau}_{ab}} \Delta M_{ab}^{\ell k} \left( \frac{\bar u_{\parallel a}^k}{v_{Ta}}  I_{\parallel a}^{pj \ell} + \frac{\bar u_{\perp a}^k}{v_{Ta}} I_{\perp a}^{pj\ell} \right).
\end{align}
\\
We remark that $\Delta \C_{ab}^{Tpj}$ can then be expressed explicitly in terms of the non-adiabatic gyro-moments $n_a^{lk}$ appearing in the definitions of $u_{\parallel a}^k$ and  $u_{\perp a}^k$ given in \eqref{eq:uparauaperp}. 

Carrying out the same derivation yielding \eqref{eq:DeltaCabTpj} for the gyro-moment expansion of the field component of the correction term, $\Delta \C_{ab}^{Fpj}$, defined in \eqref{eq:DeltaCabFpjdef}, and inverting the species role between $a$ and $b$ in $u_{\parallel s}^k$ and  $u_{\perp s}^k$ yields

\begin{align}  \label{eq:DeltaCabFpj}
   \Delta \C_{ab}^{Fpj} = \sum_{\ell=0}^{L}\sum_{k=0}^K \frac{2 c_\ell}{\bar{\tau}_{ab} } \Delta N_{ab}^{\ell k} \left( \frac{\bar u_{\parallel b}^k}{v_{Ta}}  I_{\parallel a}^{pj\ell} + \frac{\bar u_{\perp b}^k}{v_{Ta}}  I_{\perp a}^{pj\ell} \right),
 \end{align}
 \\
 where $\bar u_{\parallel b}^k$ and $\bar u_{\perp b}^k$ can be expressed in terms of the non-adiabatic gyro-moments of $h_b$ by using \eqref{eq:uparauaperp}. With the gyro-moment expansion of the test and field components of the correction term, given in Eqs. (\ref{eq:DeltaCabTpj}) and (\ref{eq:DeltaCabFpj}), the GK IS collision operator, \eqref{eq:CabISpj}, is expressed in terms of the non-adiabatic gyro-moments $n_a^{pj}$. \rev{We remark that, given the gyro-moment expansion of the GK IS collision operator in Eqs. (\ref{eq:DeltaCabTpj}) and (\ref{eq:DeltaCabFpj}), the test and field components of this operator, Eq. (\ref{eq:DeltaGKCabTF}), can be recovered by applying the inverse transformation of Eqs. (\ref{eq:DeltaCabTpjdef}) and (\ref{eq:DeltaCabFpjdef}) respectively, e.g., for the test component
 
  \begin{align}  \label{eq:inversetransformationCab}
  \Delta \C_{ab}^{T}  = \sum_{p =0}^\infty  \sum_{j=0}^\infty   \Delta \C_{ab}^{Tpj}  \frac{H_p(s_{\parallel a}) L_j(x_a)}{\sqrt{2^p p!}}. 
 \end{align}
}
 
\subsection{Expansion of the DK IS Collision Operator}

We now consider the the gyro-moment expansion of the GK IS collision operator, \eqref{eq:CabISpj}, in the DK limit. The gyro-moment expansion of the DK IS collision operator can be obtained by projecting \eqref{eq:DeltaCabTFDK} or by taking explicitly the zeroth-limit of the gyro-moment expansion of the GK IS collision operator \eqref{eq:CabISpj}. In both cases, it yields the following Hermite-Laguerre projections of the test and field components of the correction term,

\begin{align}
\label{eq:DKdeltaCabTpj}
\Delta \C_{ab}^{Tpj}    & =\sum_{\ell =0}^{\min\left(L,j + \lfloor p/2 \rfloor \right)} \sum_{k=0}^K \frac{\left( T^{-1} \right)^{1\ell}_{pj}}{\sqrt{2^p p!}}  \frac{ (\ell+ 3/2)!}{\ell !} \frac{4 c_\ell  c_k }{3 \sqrt{\pi}\bar{\tau}_{ab}}   \Delta M_{ab}^{\ell k} \nonumber \\
& \times \bm b \cdot  \tens{M}_{a}^{1k} \left[ p \geq 1 \cup j \geq 1\right]
\end{align}

and 
\begin{align}
\label{eq:DKdeltaCabFpj}
\Delta \C_{ab}^{Fpj}    & =\sum_{\ell =0}^{\min\left(L,j + \lfloor p/2 \rfloor \right)} \sum_{k=0}^K \frac{\left( T^{-1} \right)^{1\ell}_{pj}}{\sqrt{2^p p!}}  \frac{ (\ell + 3/2)!}{\ell !} \frac{4 c_\ell  c_k }{3 \sqrt{\pi}\bar{\tau}_{ab}}  \frac{1}{\chi} \Delta N_{ab}^{\ell k} \nonumber \\
&  \bm b \cdot  \tens{M}_{b}^{1k} \left[ p \geq 1 \cup j \geq 1\right].
\end{align}
\\
In Eqs. (\ref{eq:DKdeltaCabTpj}) and (\ref{eq:DKdeltaCabFpj}), the spherical harmonic moments $\tens{M}_{a}^{1k}$, defined by \eqref{eq:momentexpansion} and related to the flow vectors in \eqref{eq:uaktoM1k}, \rev{are expressed as a function of the gyro-moments of the perturbed distribution function $g_a$, namely $N_a^{pj} =  \int d \theta d \mu d v_\parallel B g_a H_p L_j / m_a \sqrt{2^p p!}$ (see \eqref{eq:npjdef})}, by 

 \begin{align} \label{eq:bcdotMs1k}
 \bm b \cdot \tens{M}_s^{1k} & =  \sum_{g=0}^{1+2k}\sum_{h=0} ^{k}  T _{1k}^{gh}  \sqrt{2^g g!} N_{s}^{gh},
\end{align}
\\
where we use the fact that $f_a \simeq g_a$ in the DK limit. The basis transformation coefficients, $T_{1k}^{gh}$, are the DK basis transformation coefficients defined and derived in Ref. \onlinecite{jorge2017drift}. With the gyro-moment expansion of the DK OS collision operator derived in Ref. \onlinecite{frei2021}, the DK IS collision operator follows by adding Eqs. (\ref{eq:DKdeltaCabTpj}) and (\ref{eq:DKdeltaCabFpj}) to the former. \rev{Finally, we remark that the DK IS operator can be recovered by applying the same inverse transformation as the one given in \eqref{eq:inversetransformationCab} by using Eqs. (\ref{eq:DKdeltaCabTpj}) and (\ref{eq:DKdeltaCabFpj}) instead.}

\label{subsec:expansionofDKIS}

\section{Numerical Tests and Comparison between Collision operators}
\label{sec:numericaltest}

Using the gyro-moment expansion of the IS collision operator, we perform the first numerical tests and comparisons between the IS, OS and the Coulomb collision operators. The discussion of the numerical results is organized as follows. First, we discuss the numerical implementation of the closed analytical formulas appearing in the IS operator, in particular of the Braginksii matrices in \secref{sec:numericalimplementation}. We show that the numerical implementation satisfies the conservation laws and associated symmetry properties of Braginskii matrices to machine precision, regardless of the values of $(L,J)$. Then, as a first application of the IS collision operator using the gyro-moment approach, we investigate the collisionality dependence of TEM that develops at steep pressure gradients, \rev{such as those in H-mode pedestal}, and compare the IS with the OS and Coulomb collision operators in \secref{sec:TEM}. \rev{The Coulomb and OS collision operators and their GK limits used in the present comparisons are detailed in Appendix \ref{appendix:Abis}}. In \secref{sec:ZF}, we perform tests to study the collisional ZF damping and compare the numerical results with analytical predictions. \rev{Finally, in \secref{sec:spitzer}, we compare the parallel electrical Spitzer conductivities predicted by the OS, IS and Coulomb operators.} The numerical tests show that the IS collision operator approaches better the Coulomb operator than the OS collision operator in the Pfirsch-Schlüter regime. In all cases investigated, $L = K \simeq 3$ terms in $\Delta C_{ab}$ are required for convergence.

\subsection{Numerical Implementation}
\label{sec:numericalimplementation}

\begin{figure}
    \centering
    \includegraphics[scale = 0.55]{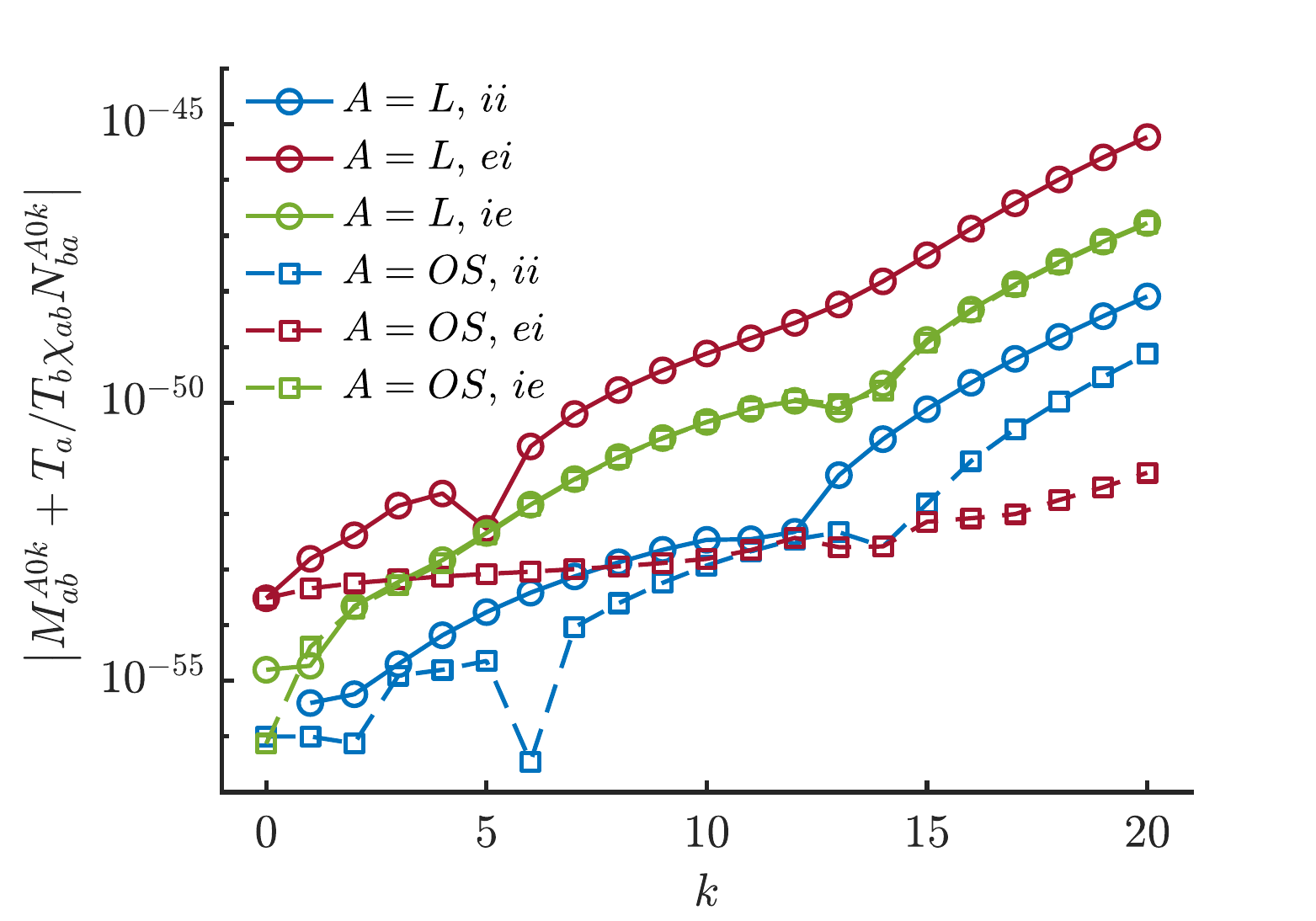}
    \caption{\rev{Verification of the momentum conservation, \eqref{eq:momentumconsbraginskii}, obtained by using the closed expressions of the Braginksii matrices (numerically computed using $50$ significant digits) as a function of $k$ for the Coulomb ($A = L$, solid lines) and OS ($A = S$, dotted lines) collision operators. Collisions between electrons and ions with $T_e = T_i$ and $m_e / m_i = 0.0027$ are considered.} }
    \label{fig:fig_conservation_momentum}
\end{figure}

\begin{figure}
    \centering
    \includegraphics[scale = 0.55]{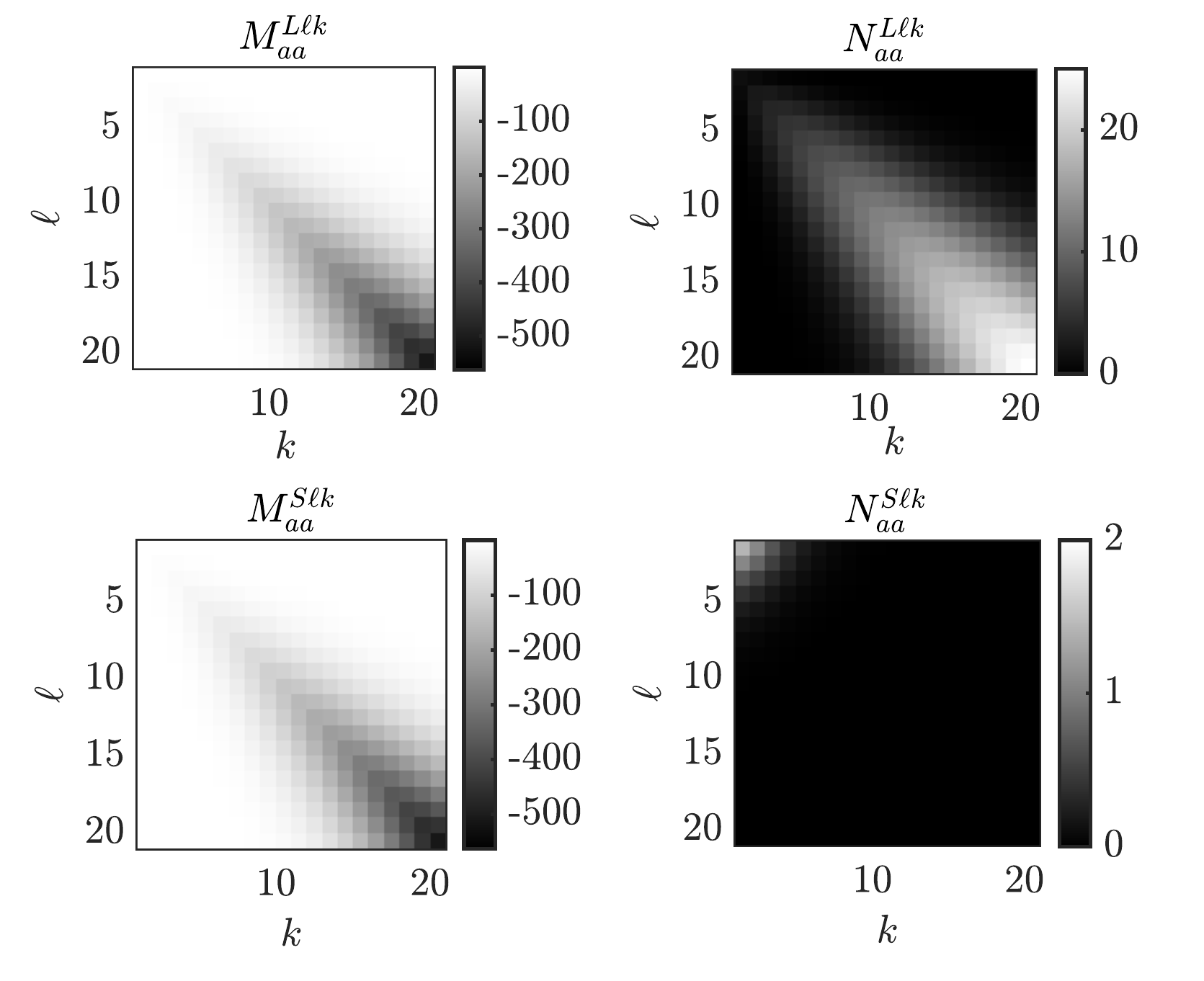}
    \caption{Braginskii matrices of (top) Coulomb and (bottom) OS collision operators for like-species ($a=b$) associated with the test (left) and the field (right) components, respectively defined by $M_{aa}^{A\ell k}$ and $N_{aa}^{A\ell k}$, as a function of $(\ell, k)$.}
    \label{fig:fig_MNbraginskii_self}
\end{figure}

\revv{In the present work, the need of the numerical integration for the evaluation of the velocity integrals appearing in the definitions of $M_{ab}^{A \ell k}$ and $N_{ab}^{A \ell k}$, given in \eqref{eq:matrixelements}, and of the velocity integrals involved in the gyro-moment expansion of the IS operator, Eqs. (\ref{eq:I_para_pjk_def}) and (\ref{eq:I_perp_pjk_def}), is removed by using the closed analytical expressions given, for instance, in Eqs. (\ref{eq:Iparallelspjk}) and (\ref{eq:Iperpspjk}) to evaluate the velocity integrals contained in Eqs. (\ref{eq:I_para_pjk_def}) and (\ref{eq:I_perp_pjk_def}).} We note that the numerical error stemming from the numerical integration of the velocity integrals become arbitrarily large as the order of polynomials increases, and thus alter the accuracy of the Hermite-Laguerre projection for high-order gyro-moments. However, the evaluation of a large number of numerical coefficients is still required in the gyro-moment approach, which is affected by cancellation and round-off errors. Therefore, the analytical expressions in the present work are evaluated using an arbitrary-precision arithmetic software. 

\rev{In \figref{fig:fig_conservation_momentum}, we verify the momentum conservation law, given in \eqref{eq:momentumconsbraginskii}, using the analytical expressions derived in \secref{sec:braginksii} using the arbitrary-precision arithmetic software with $50$ significant digits (numerical experiments show that fewer digits could lead to round off numerical errors in the Braginskii matrices and, more generally, in the gyro-moment expansions) as a function of $k$ for both the Coulomb and OS collision operators. The momentum conservation law is shown to be satisfied within an error of less than $10^{-50}$, illustrating the robustness of the numerical framework used in this work.} Additionally, the Braginskii matrices, $M_{aa}^{A\ell k}$ and $N_{aa}^{A\ell k}$, of the Coulomb and OS collision operators are shown in \figref{fig:fig_MNbraginskii_self} for the case of like-species collisions. Since the test components of both operators (see Eqs. (\ref{eq:CLT}) and (\ref{eq:CabST})) are equal in the case of like-species collisions (and, more generally, in the  $T_a = T_b$ case), the Braginskii matrices $M_{aa}^{L\ell k}$ and $M_{aa}^{S\ell k}$ are equal as shown in the left panel of \figref{fig:fig_MNbraginskii_self}, implying $\Delta M_{ab}^{\ell k} = 0$. On the other hand, the difference in the Braginskii matrices associated with the field components, i.e. $N_{aa}^{A\ell k}$ (right panel of \figref{fig:fig_MNbraginskii_self}), arises due to the difference in the field component of the OS with respect to the Coulomb collision operator.

The gyro-moments approach allows us to investigate the coupling between gyro-moments induced by the IS and OS collision operators. This can be done by truncating the Hermite-Laguerre expansion of $h_a$, see \eqref{eq:hHL} (or $g_a$ in the DK limit) at $(p,j) = (P,J)$ therefore, assuming that higher-order gyro-moments, $p > P$ and $j > J$, vanish. \rev{Given $(P,J)$, the gyro-moment expansion of the collision operator $A$ can be written as 

\begin{align} \label{eq:CabISmatricprod}
     \C_{ab}^{Apj} = \sum_{p' =0}^P  \sum_{j' =0}^J \C_{ab p'j'}^{ATpj} n_a^{p'j'} + \sum_{p' =0}^P  \sum_{j' =0}^J \C_{ab p'j'}^{AF pj} n_b^{p'j'}.
\end{align}
\\
The coefficients $\C_{ab p'j'}^{ATpj}$ and $ \C_{ab p'j'}^{AF pj}$ associated with the Coulomb ($A = L$) and OS ($A = S$) collision operators can be obtained from Ref. \onlinecite{frei2021}. In the case of the IS collision operator ($A = IS$), \eqref{eq:CabISmatricprod} becomes

\begin{align} \label{eq:CabISmatricprodIS}
     \C_{ab}^{ISpj} & = \sum_{p' =0}^P  \sum_{j' =0}^J \left( \C_{ab p'j'}^{STpj} + \Delta  \C_{ab p'j'}^{Tpj} \right) n_a^{p'j'} \nonumber \\
     & + \sum_{p' =0}^P  \sum_{j' =0}^J \left( \C_{ab p'j'}^{SF pj} +\Delta  \C_{ab p'j'}^{Fpj}   \right) n_b^{p'j'}.
\end{align}
\\
\revv{In the case of the GK IS, the explicit expressions of the coefficients, $\Delta \C_{ab p'j'}^{Tpj}$ and $ \Delta \C_{ab p'j'}^{F pj}$, can be obtained by inserting the analytical expressions of $\bar{ u}_{\parallel s}^k$ and $\bar{ u}_{\perp s}^k$, given in \eqref{eq:uparauaperp}, into Eqs. (\ref{eq:DeltaCabTpj}) and (\ref{eq:DeltaCabFpj}), respectively. In the case of the DK IS, the coefficients, $\Delta \C_{ab p'j'}^{Tpj}$ and $ \Delta \C_{ab p'j'}^{F pj}$, are obtained by using \eqref{eq:bcdotMs1k} (given $L = K$) into Eqs. (\ref{eq:DKdeltaCabTpj}) and (\ref{eq:DKdeltaCabFpj}), respectively.} We note that Appendix \ref{appendix:B} reports the closed analytical expressions of the lowest order coefficients, $\C_{ab p'j'}^{ATpj}$ and $ \C_{ab p'j'}^{AF pj}$, associated with the DK Coulomb and OS collision operators as well as $\Delta \C_{ab p'j'}^{Tpj}$ and $ \Delta \C_{ab p'j'}^{F pj}$ in the high-collisional case where the gyro-moments is truncated by neglecting the moments with $p + 2j > 3$. In this case, the analytical expressions of $\Delta  \C_{ab p'j'}^{Tpj} $ and $\Delta  \C_{ab p'j'}^{Fpj} $ become independent of $K$ (and $L$) when $ L = K > 1$ (see \eqref{eq:bcdotMs1k}).}

\begin{figure*}
    \centering
    \includegraphics[scale = 0.67]{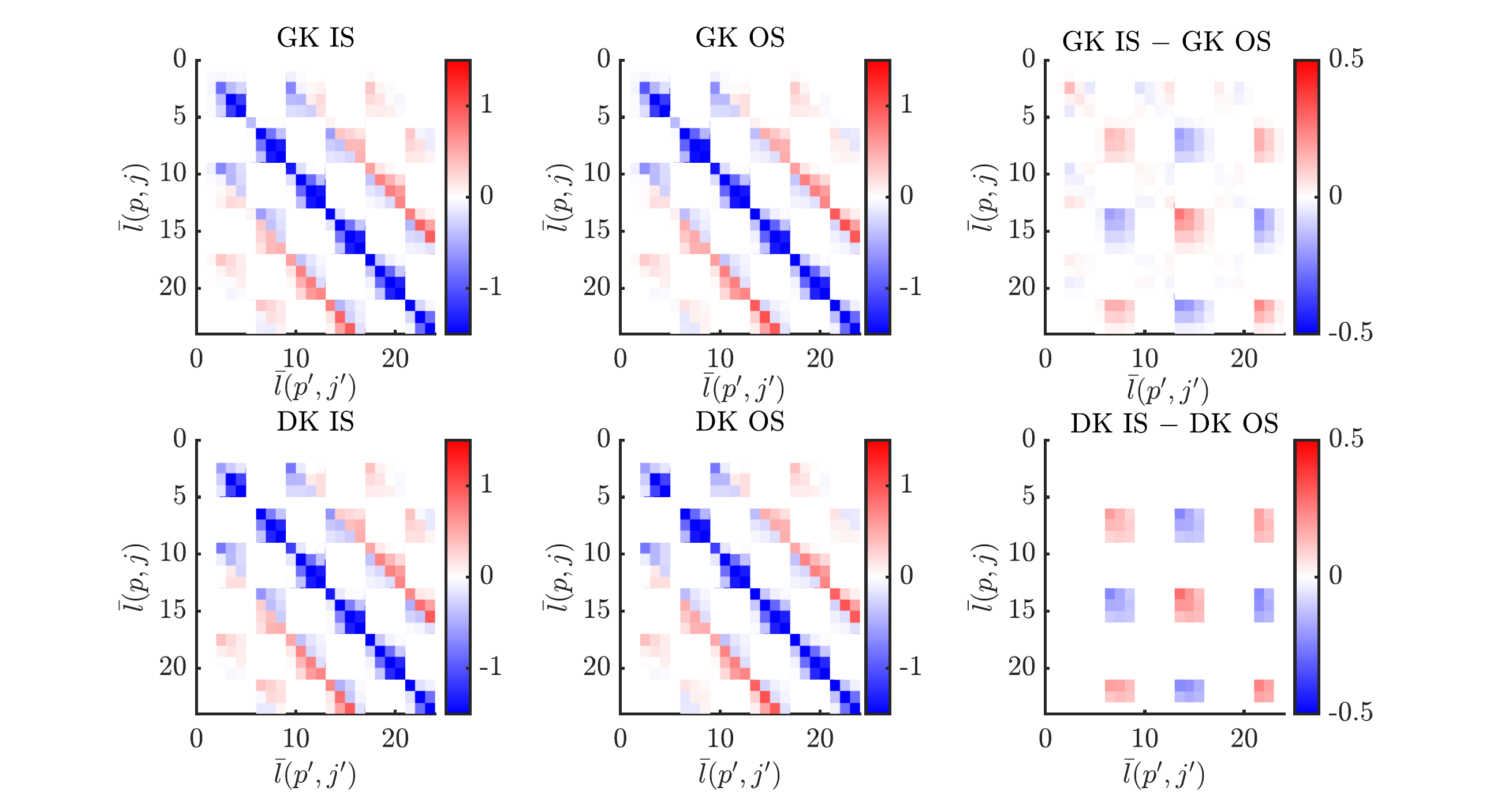}
    \caption{\rev{Matrix representation of the coupling between gyro-moments associated with Eqs. (\ref{eq:CabISmatricprod}) and (\ref{eq:CabISmatricprodIS}). The IS (left), OS (center) and their relative difference associated with the gyro-moment expansion of the correction term $\Delta C_{ab}$ in \eqref{eq:ISugama} (right) are represented using the row index $\bar{l}(p,j) = (J+1) p + j +1$ and column index $\bar{l}(p',j') = (J+1) p' + j' +1$. The matrix elements are obtained by evaluating numerically the analytical expressions derived in Sev. V.} Both the GK (top) and DK (bottom) limits fo the IS and OS operator are shown. The colobars associated with the differences of the OS and IS operator are adjusted for better visualization. Here, we consider $(P,J) = (6,3)$ gyro-moments and $ L = K = 5$ and $k_\perp = 0.5$. }
    \label{fig:fig_Cabpj}
\end{figure*}

\rev{For a numerical implementation, Eq. (\ref{eq:CabISmatricprod}) can be recast in a matrix form by introducing the one-dimensional row index $\bar{l}(p,j)= (J +1)p + j +1$ (where $p$ and $j$ run from $0$ to $P$ and $J$, respectively) and, similarly, the column index $\bar{l}(p',j')$. This formulation allows us also to illustrate the coupling between the gyro-moments associated with the GK IS and GK OS operators, as well as their difference for the case of like-species (see \figref{fig:fig_Cabpj}).} The same analysis is carried out also for the DK operators in \figref{fig:fig_Cabpj}. We observe that, because of the self-adjoint relations of like-species collisions (see \eqref{eq:selfadjoint}), the gyro-moment matrices are symmetric. We also observe the block structure of the GK IS and OS\cite{frei2021}, a consequence of vanishing polynomial basis coefficients  $(T^{-1})_{pj}^{lkm}$.

\subsection{Trapped electron Mode in Steep Pressure Gradient Conditions}
\label{sec:TEM}

 The study of microinstabilities appearing in steep pressure gradient conditions have gained large interest in the past years because of their role in determining the turbulent transport in H-mode pedestals \cite{fulton2014,kotschenreuther2017pedestal,pueschel2019microinstabilities}. The linear properties of mircroinstabilities at steep pressure gradients can significantly differ from the one at weaker gradients, typically found in the core. For example, unconventional ballooning mode structures can be encountered if the pressure gradients are above a certain linear threshold with the location of the largest mode amplitude being shifted from the outboard midplane position, in contrast to the conventional mode structure found at lower gradients \citep{xie2016global,han2017multiple}. Because the pedestal is characterized by a wide range of collisionalities ranging from the low-collisionality banana (at the top of the pedestal) to the high-collisionality Pfisch-Schlüter regime (at the bottom of the pedestal and in the SOL) \cite{thomas2006effect}, an accurate collision operator is necessary for the proper description and interaction of these modes. Thus, we compare the properties of steep pressure gradient TEM, when the IS, the OS and the Coulomb collision operators are used.

 To carry out this numerical investigation, a linear flux-tube code using the gyro-moment approach has been implemented to solve the linearized electromagnetic GK Boltzmann equation, that we obtain from the full-F GK equation in Ref. \onlinecite{frei2020gyrokinetic},

 \begin{align} \label{eq:linGK}
 \frac{\partial}{\partial t} g_{a} &+i \omega_{B a} h_{a}+v_{\|} \bm b \cdot \grad h_{a}-\frac{\mu}{m_{a}}(\boldsymbol{b} \cdot \nabla B) \frac{\partial}{\partial v_{\|}} h_{a} \nonumber \\
 & -i \omega_{T a}^{*} \left<\chi_{a}(\bm r) \right>_{\bm R} F_{M a}= \sum_b \C^A_{ab},
 \end{align}
 \\
 where $ \chi_a(\bm r) = \phi(\bm r) - v_\parallel \psi(\bm r) $, with $\phi$ the perturbed electrostatic potential and $\psi$ the perturbed magnetic vector potential, being evaluated at $\bm r$, and given by the self-consistent GK quasineutrality and GK Ampere's law,

  \begin{align} \label{eq:Poisson}
     \sum_a \frac{q_a^2}{N_a T_a} \left( 1 - \Gamma_0(a_a) \right) \phi(\bm r)= \sum_a q_a  \int d \mu d v_\parallel d \theta \frac{B}{m_a} J_0(b_a \sqrt{x_a}) \g_a,
 \end{align}
 \\
and
 
 \begin{equation} \label{eq:Ampere}
\left(\frac{ k_\perp^2}{4 \pi}+  \sum_a \frac{q_a^2 N_a}{m_a}  \Gamma_0(a_a) \right) \psi(\bm r)  =  \sum_a q_a    \int d \mu d v_\parallel d \theta \frac{B}{m_a} J_0(b_a \sqrt{x_a} ) v_\parallel \g_a,
\end{equation}
 \\
 respectively. In Eqs. (\ref{eq:linGK}), (\ref{eq:Poisson}) and (\ref{eq:Ampere}), we introduce the magnetic drift frequency $\omega_{Ba}    = v_{Ta}^2 ( x_a+ 2 s_{\parallel a}^2)  \omega_B/ (2 \Omega_a)$ with $\omega_B = \left(  \bm b \times \grad \ln B \right) \cdot \bm k$, the diamagnetic frequency $\omega_{Ta}^*  =  \left[\omega_N + \omega_{T_a}\left( x_a + s_{\parallel a}^2 -3/2\right) \right]$  with $\omega_N = \bm b \times \grad \ln N \cdot \bm k/ B$ and $\omega_{T_a} = \bm b \times \grad \ln T_a \cdot \bm k/B$, $a_a= b_a^2 / 2 $, and $\Gamma_0(x) = I_0(x) e^{- x}$ (with $I_0$ the modified Bessel function). On the right hand-side of Eq. (\ref{eq:linGK}) is the linearized GK collision operator composed by the sum of the GK test ($\mathcal{C}^{AT}_{ab}$) and field ($\mathcal{C}^{AF}_{ab}$) components, such that $\mathcal{C}^A_{ab} = \mathcal{C}^{AT}_{ab} + \mathcal{C}^{AF}_{ab} $, which are defined in Appendix \ref{appendix:Abis}.
 
While a detailed description of the resolution of Eq. (\ref{eq:linGK}) using the gyro-moment approach will be the subject to a future publication. Here we mention that we assume concentric, circular and closed magnetic flux surfaces using the $s-\alpha$ model (with $\alpha = 0$) \citep{lapillonne2009}. In the local flux-tube approach, we assume constant radial density and temperature gradients, $L_N^{-1}$ and $L_{Ta}^{-1}$, with values $R_N = R_0 / L_N = R_0 / L_{Ti} =  R_0 / L_{Te} = 20$, where $R_0$ is the tokamak major radius. Electromagnetic effects are introduced with $\beta_e = 8 \pi P_e^2 / B_0^2 = 0.01 \%$. For numerical reasons, we use a larger electron to ion mass ratio $m_e / m_i = 0.0027$. \rev{The local safety factor $q$ and the inverse aspect ratio $\epsilon$ are fixed at $q = 2.7$ and $\epsilon = 0.18$, while a magnetic shear $s = 0.5$ is used. We remark that the value of magnetic shear, smaller than typical edge values ($s \gtrsim 3$), is chosen to
 limit the additional computational cost related to the rapid increase of the radial wavenumber, $k_x$, with $s$ along the parallel direction.} Additionally, we center the $k_x$ spectrum around $k_x = 0$. \rev{Collisional effects are introduced by using the gyro-moment expansion of the IS operator derived in this work and the gyro-moment expansions of the Coulomb and OS Sugama operators reported in Ref. \onlinecite{frei2021}.}

\begin{figure}
    \centering
    \includegraphics[scale = 0.55]{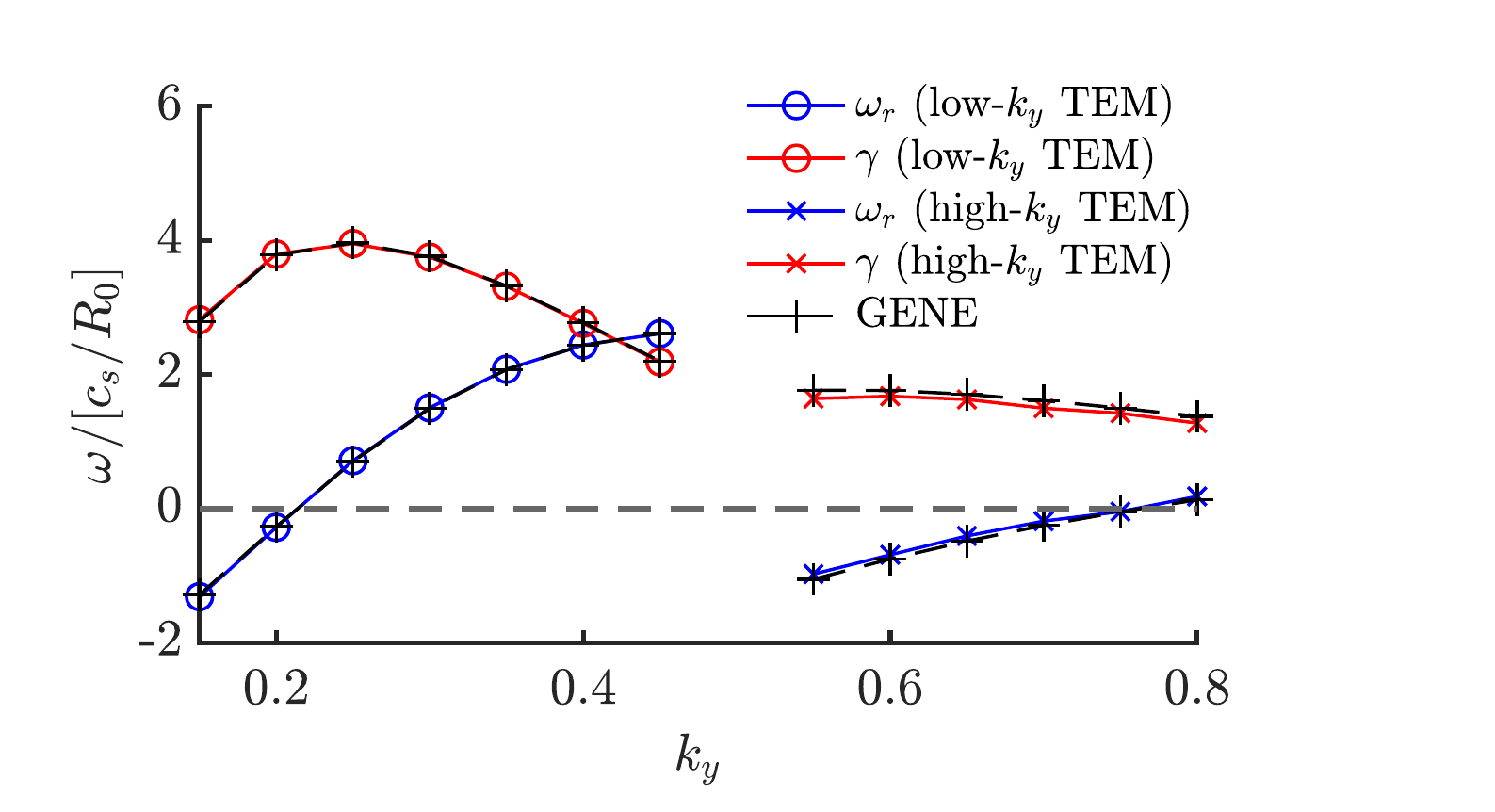}
    \caption{Collisionless growth rate $\gamma$ (red lines) and frequency $\omega_r$ (blue lines) as a function of the binormal wavenumber $k_y$ plotted on the same axis. The colored lines are the results obtained by the gyro-moment approach using $(P,J) = (24,10)$ and the black cross markers are the collisionless results using the GENE eigensolver. A positive mode frequency ($\omega_r>0$) corresponds to the mode propagating in the ion diamagnetic direction and a negative mode frequency ($\omega_r < 0$) to the electron diamagnetic direction. Here, $T_i / T_e = 1$. }
    \label{fig:fig_omegacomplex_omn20_kymin} 
\end{figure}

 \begin{figure}
    \centering
    \includegraphics[scale = 0.55]{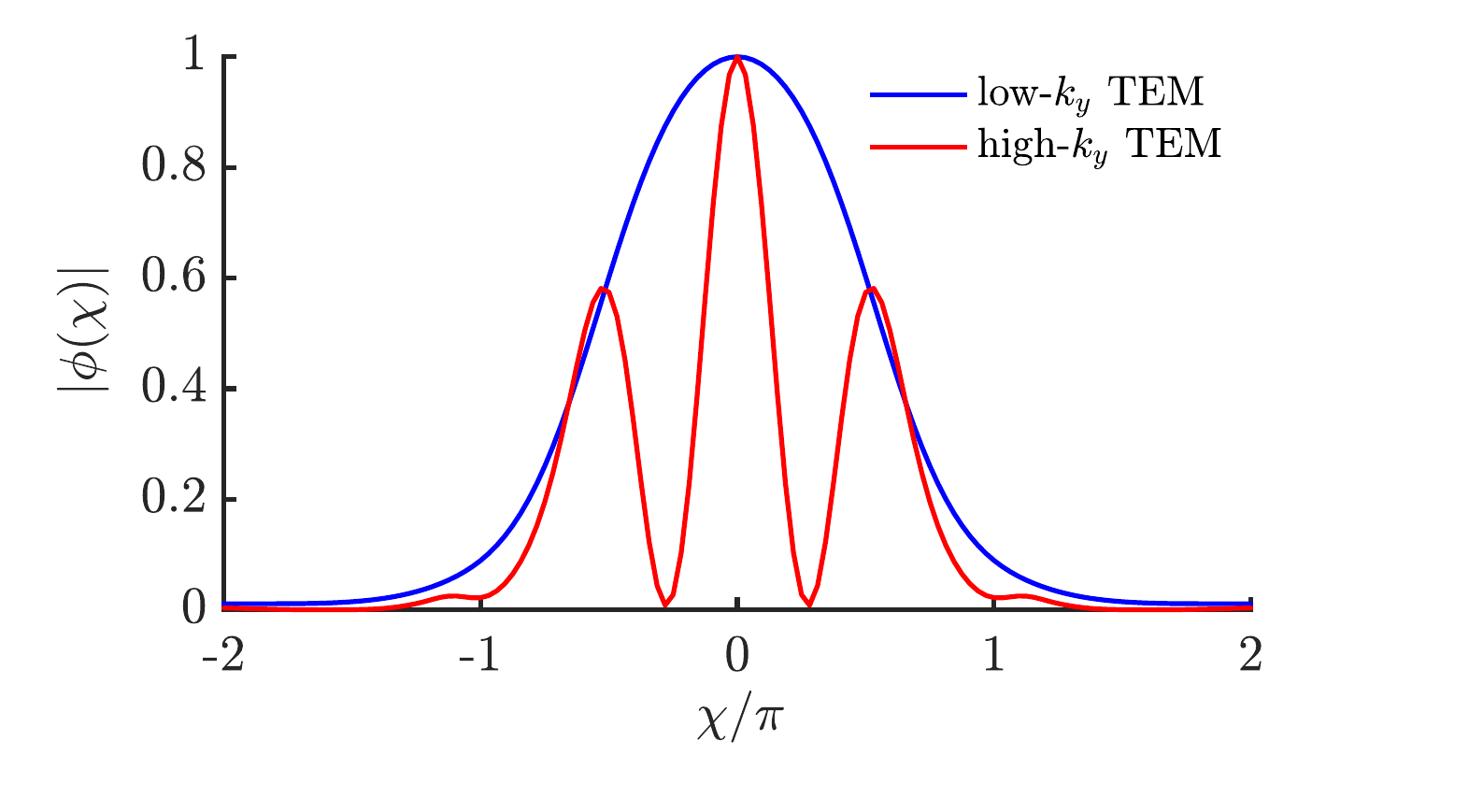}
    \caption{Modulus of the electrostatic ballooning eigenmode function, $\phi(\chi)$ (normalized to $\phi(\chi = 0)$), as a function of the ballooning angle $\chi$ corresponding to the case of the low-$k_y$ TEM at $k_y =0.25$ (blue solid line) and to the high-$k_y$ TEM developing at $k_y =0.6$ (red solid line). }
    \label{fig:fig_phiB}
\end{figure}

Given these parameters, we identify two branches of unstable modes developing at different values of the binormal wavenumber $k_y$ in the collisionless limit. This is shown in \figref{fig:fig_omegacomplex_omn20_kymin} where the growth rate $\gamma$ and the real mode frequency $\omega_r$ are plotted as a function of $k_y$ using $(P,J) = (20,10)$ gyro-moments. We remark that the collisionless GENE simulations are in excellent agreement with the gyro-moment approach, verifying its validity in the collisionless regime. \rev{A discontinuous frequency jump from positive to negative values indicates a mode transition, separating a branch of low-$k_y$ and high-$k_y$ modes. We also notice that the low-$k_y$ and the high-$k_y$ modes change continuously from electron to the ion diamagnetic directions as $k_y$ increases}. While the mode peaking at $k_y = 0.25$ ($k_y$ is normalized to the ion sound Larmor radius) is normally identified as the ITG mode at lower gradients because of its propagation along the ion diamagnetic direction ($\omega_r  >0$), we identity it here as a TEM with a conventional mode structure. Indeed, despite $\omega_r > 0$, the mode persists if the ion and electron temperature gradients are removed from the system, while it is stabilised if the electrons are assumed adiabatic. The discontinuity observed in \figref{fig:fig_omegacomplex_omn20_kymin} is due to a transition to a TEM developing at $k_y \gtrsim 0.5$ with an unconventional ballooning mode structures with secondary peaks located near $ \chi = \pm \pi /2$ ($\chi$ is the ballooning angle) away from the outboard midplane (where the mode at $k_y \sim 0.2$ peaks), as shown in \figref{fig:fig_phiB}. Hence, we refer to the TEM peaking at $k_y = 0.25$, with a conventional mode structure, as the low-$k_y$ TEM and to the TEM peaking near $k_y = 0.6$, with an unconventional mode structure, as the high-$k_y$ TEM. \rev{We remark that the TEM modes identified in \figref{fig:fig_omegacomplex_omn20_kymin} have very similar features to the ones found in, e.g., Refs. \onlinecite{coppi1990candidate,ernst2005new,wang2012linear}.}
 
\begin{figure*}
    \centering
    \includegraphics[scale = 0.51]{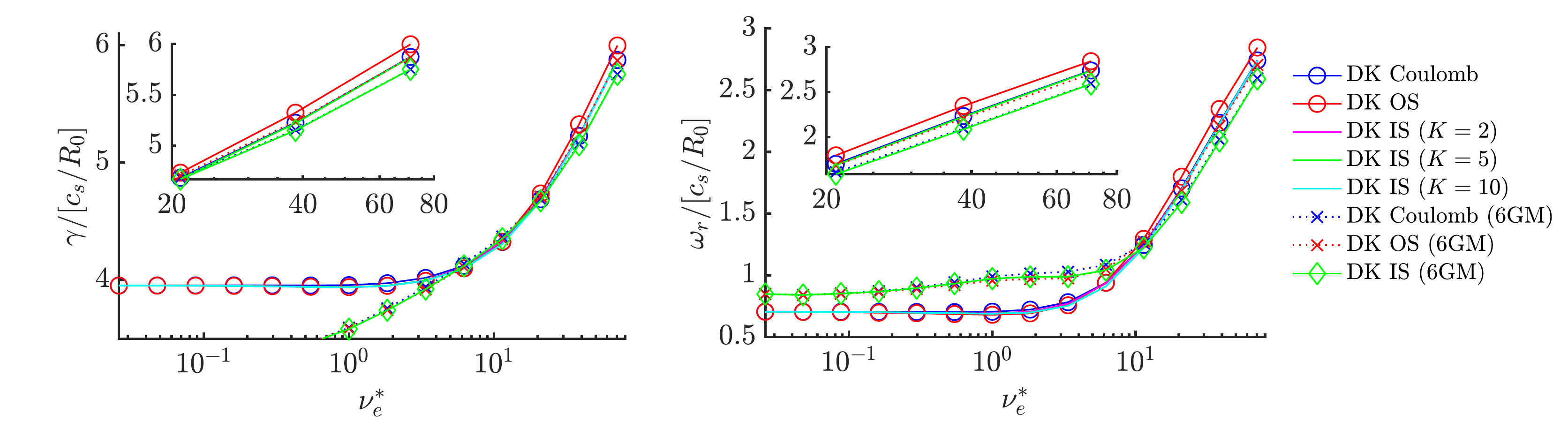}
   \caption{The low-$k_y$ TEM growth rate $\gamma$ (left) and mode frequency $\omega_r$ (right) as a function of $\nu_e^*$ obtained using the DK Coulomb (blue markers), the DK OS  (red markers), the DK IS with $K = 2$ (solid magenta line), the DK IS with $K = 5$ (solid green line) and the DK IS with $K = 10$ collision operators (solid light blue line). We note that the DK IS with $K = 5$ (light green line without marker) overlaps with the $K = 10$ case  (light blue line without marker). The results with the DK Coulomb and DK OS operators using $(P,J) = (16, 8)$ gyro-moments and $6$ gyro-moments ($6$GM) with the analytical expressions reported in Appendix \ref{appendix:B} are shown by the circle and cross markers, respectively. The insets focus on the Pfirsch-Schlüter regime, $20 < \nu_e^* < 80$. Here, the parameters are the same as in \figref{fig:fig_omegacomplex_omn20_kymin} at $k_y = 0.25$ with $T_i / T_e = 1$. }
  \label{fig:fig6}
\end{figure*}

We now investigate the collisionality dependence of both modes. While the electron collisionality is expected to be mainly in the banana regime with $\nu_e^* =  \sqrt{2 } q  \nu / \epsilon^{3/2}   \lesssim 1$ in the middle of the pedestal of present and future tokamak devices (with $\nu$ the ion-ion collision frequency normalized to $c_s / R_0$, see Ref. \onlinecite{frei2021}), the temperature drop yields collisionalities than can be in the Pfirsch-Schlüter regime, $\nu_e^* \sim  1 / \epsilon^{3/2} \gg 1$, at the bottom of the pedestal \cite{thomas2006effect}. Focusing first on the low-$k_y$ TEM, we notice that since it develops at low binormal wavenumber ($k_y = 0.25$), the DK limits of the collision operators are considered (numerical tests show that FLR effects at $k_y = 0.25$ change the value of the growth rate $\gamma$ by less than $1 \%$ at the level of collisionalities explored). \rev{The growth rate and the real mode frequency obtained by using the DK Coulomb, DK OS and DK IS using $K = 2,5$ and $10$ terms in \eqref{eq:DeltaCabTFDK} are shown as a function of $\nu_e^*$ in \figref{fig:fig6} in the case of $T_i / T_e =1$. Here, we consider the case $(P,J) = (16, 8) $ and the case where only $6$ gyro-moments ($6$GM) are retained. For the $6$GM model, the closed analytical expressions of the collision operators reported in Appendix \ref{appendix:B} are used to evaluate the collisional terms. We first observe that the low-$k_y$ TEM is destabilized by collisions in the Pfirsh-Schlüter regime where both the growth rate and frequency increase with collisionality. We remark that the low-$k_y$ TEM is weakly affected by collisions when $\nu_e^* < 1$, with collisions that have a stabilizing effects on the mode when $R_N < 20$. Second, it is remarkable that the deviation (of the order or smaller than $10 \%$) between the DK OS and DK Coulomb increases with $\nu_e^{*}$, while the DK IS is able to correct the DK OS and to approach the DK Coulomb as $L$ and $ K$ increase. This is better shown in the insets of \figref{fig:fig6} where a good agreement between the DK IS and Coulomb operator is observed. Third, the $6$GM model (cross markers) approach the results of the full calculations, i.e. with $(P, J) = (16, 8)$, as the collisionality increases, while they deviate from each other at low collisionality because of kinetic effects that are not resolved in the $6$GM model. In addition, we note the good agreement between the DK IS and Coulomb in the $6$GM model in the Pfirsch-Schlüter regime, demonstrating the robustness of the present numerical results.}

\begin{figure}
  \centering
    \includegraphics[scale = 0.5]{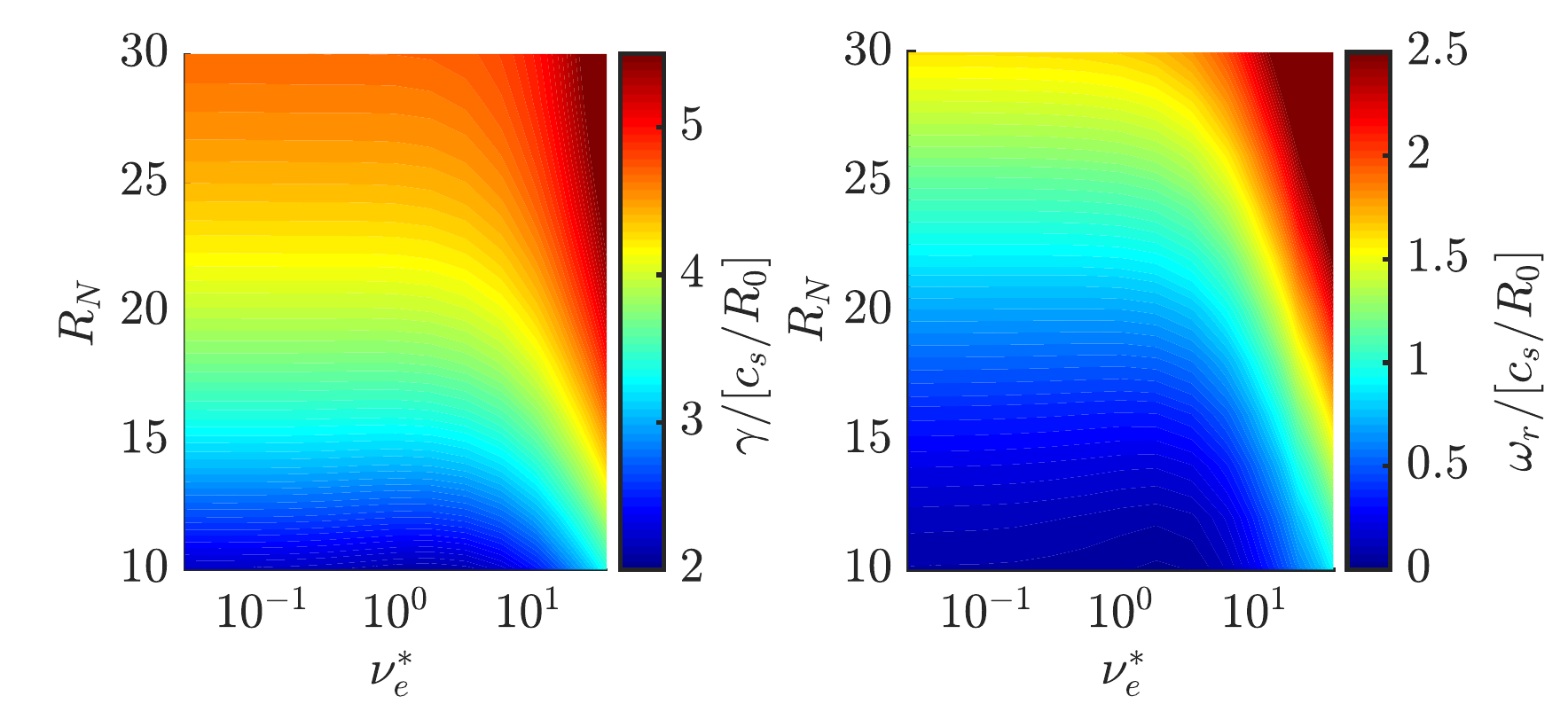}
    \caption{Low-$k_y$ TEM growth rate $\gamma$ (left), and real mode frequency $\omega_r$ (right) as a function of the normalized density gradient, $R_N$, and collisionality, $\nu_e^*$, using the DK Coulomb collision operator.}
    \label{fig:fig_kymin_0.25_dkcoulomb}
\end{figure}
\begin{figure*}
    \centering
    \includegraphics[scale = 0.53]{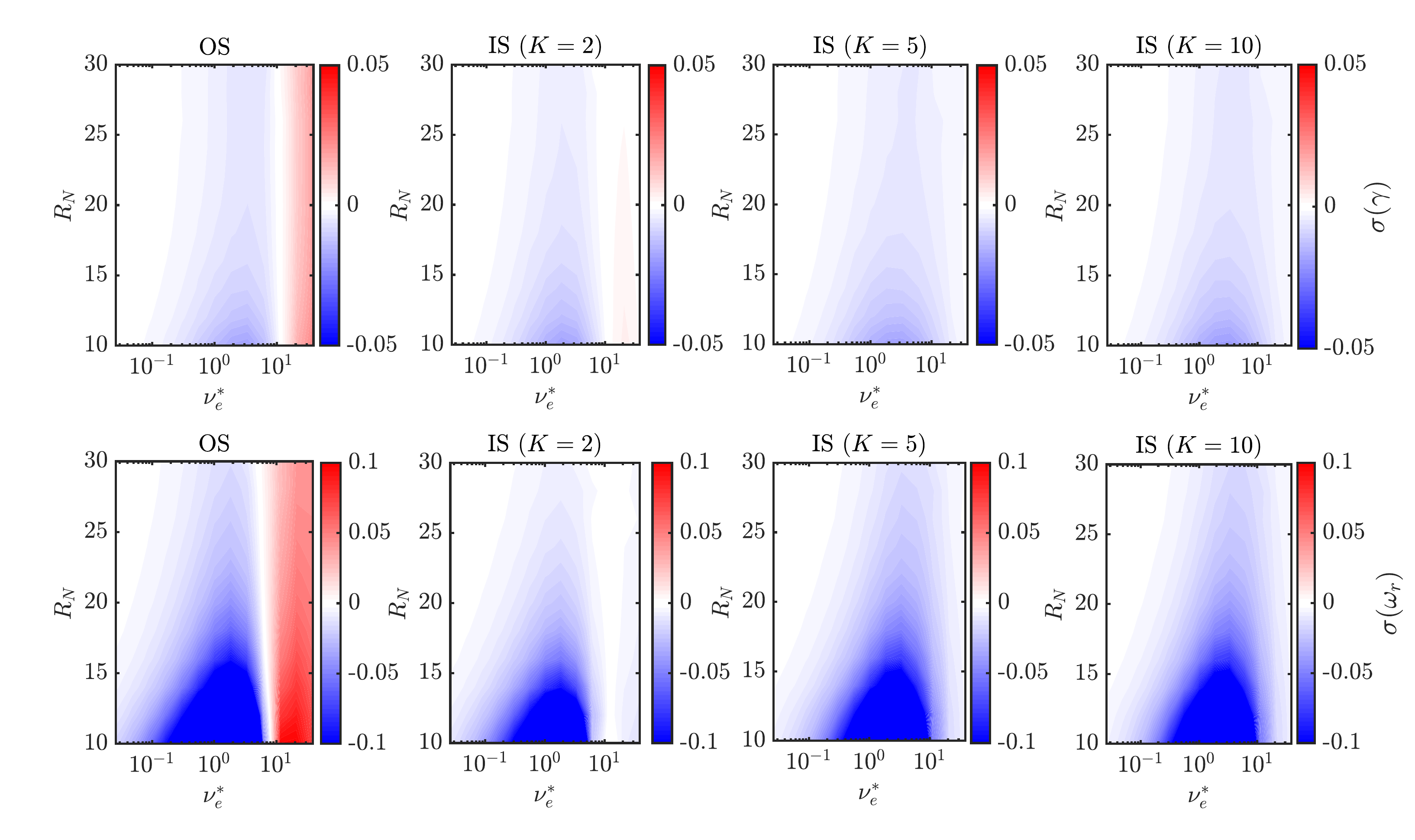}
    \caption{\rev{Signed relative difference of the growth rate, $\sigma( \gamma)$ (top), and frequency, $\sigma( \omega_r)$ (bottom), as a function of the density gradient $R_N$ and electron collisionality $\nu_e^*$ for the OS and IS ($K = 2$, $K=5$ and $K=10$) operators (from left to right, respectively). The results using the DK Coulomb, used as reference, are shown in \figref{fig:fig_kymin_0.25_dkcoulomb}. The other parameters are the same as in \figref{fig:fig_omegacomplex_omn20_kymin}, with $k_y = 0.25$ and $T_i / T_e = 1$.} }
    \label{fig:fig8}
\end{figure*}

 
\rev{To analyse the relative difference of the IS and OS operators with respect to the DK Coulomb, we scan the low-$k_y$ TEM mode over the density gradient $R_N $ (with the temperature gradients, $ R_0 / L_{Ti} =  R_0 / L_{Te} = R_N$) for different collisionalities and compute the signed relative difference of the growth rate $\gamma$, $\sigma( \gamma) = ( \gamma - \gamma_C) / \gamma_C$, where $\gamma$ is obtained using the OS or IS collision operator models and $\gamma_C$ is the one obtained using the DK Coulomb operator (see \figref{fig:fig_kymin_0.25_dkcoulomb}). The same definition $\sigma( \omega_r) = ( \omega_r - \omega_{rC}) / \omega_{rC}$ is used for the real mode frequency. We plot the results in \figref{fig:fig8}. First, we observe that the DK OS operator underestimates both $\gamma$ and $\omega_r$ (compared to the DK Coulomb) at low collisionality (with a peak near $\nu_e^* \sim 1$), and that difference changes sign in the Pfirsch-Schlüter regime, when $\nu_e^* \gtrsim 10$, where the DK OS operator overestimates $\gamma$ and $\omega_r$. A difference of the order of $5 \%$ is found in growth rate and of the order of $10 \%$ in the frequency. While these deviations increase with $\nu_e^*$ (see the red areas in the left panels in \figref{fig:fig8}), the correction terms in the IS operator reduce $\sigma(\gamma)$ (and $\sigma(\omega_r)$) below $2 \%$ for all density gradients at high-collisionalities. \revv{More precisely, the agreement between the DK IS and DK Coulomb improves with $K$ in the Pfirsch-Schlüter regime. In fact, the deviations from the DK Coulomb operator observed by the presence of the red area in $\sigma(\gamma)$ in the case $K = 2$ are reduced in the case of $K =5$ (and $K=10$) when $\nu_e^* \gtrsim 10$. Additionally, the small differences observed between the DK IS with $K=10$ and $K=5$ show that the results of the IS operator are converged when $K \gtrsim 5$.} We notice that, in general, $\sigma (\gamma ) \lesssim \sigma(\omega_r)$ for all operators. Finally, we remark that, in both the DK OS and DK IS operators, the deviations in $\gamma$ and $\omega_r$ peak near $\nu_e^* \sim 1$ and increase at lower gradients with $\sigma(\gamma) \sim 5 \%$ and $ \sigma(\omega_r) \sim 10 \%$. The larger deviations are explained by the fact that the OS and, hence, the IS operators are based on a truncated moment approximation of the Coulomb collision operator (see, e.g., the field component), and that the effects of high-order moments in the collision operator models can no longer be ignored at this intermediate level of collisionality. On the other hand, in the Pfirsch-Schlüter regime, the contribution of high-order moments becomes small as being damped by collisions. The increase of their relative differences with decreasing density gradient, $R_N$, is mainly attributed to the decrease of $\gamma$ and $\omega_r$, as they are of the order of the diamagnetic frequency. }

\begin{figure}
  \centering
    \includegraphics[scale = 0.53]{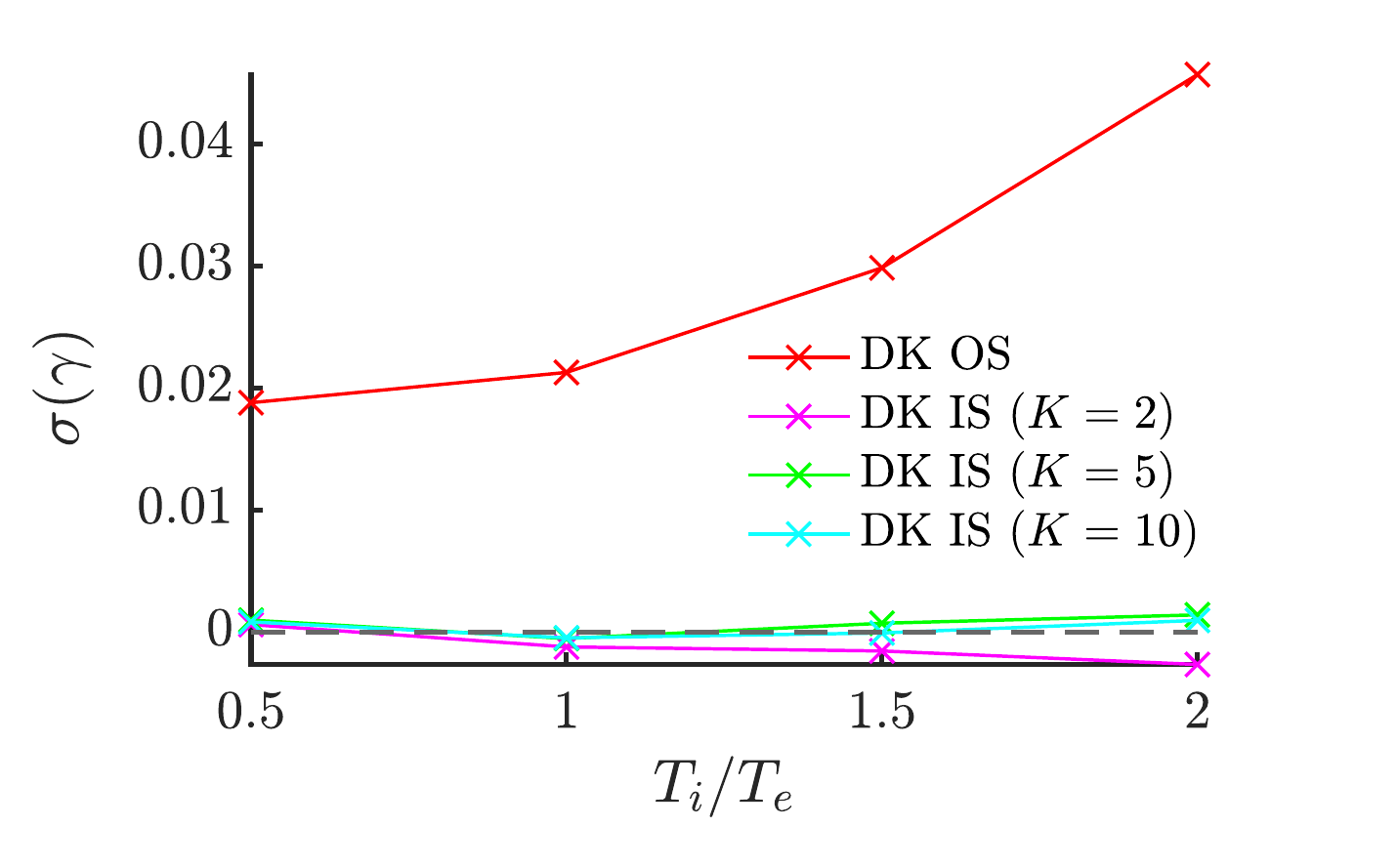}
    \caption{Signed relative difference of the growth rate $\gamma$, $\sigma(\gamma)$, predicted by the OS (red cross) and IS with $K = 2$ (magenta), $K =5$ (green) and $K =10$ (light blue) operators. The dashed back line represents perfect agreement with the DK Coulomb collision operator. The same parameters as in \figref{fig:fig6} are used, except for $\nu_e^* = 50$.}
    \label{fig:fig_gammaratio_vs_tau}
\end{figure}

\rev{The deviation between the DK OS and DK Coulomb operators depends on the temperature ratio of the colliding species (as well as on the mass ratio). In order to study the impact of the ion to electron temperature ratio, we consider $\sigma(\gamma)$ plotted as a function of the temperature ratio $T_i / T_e$ and shown in \figref{fig:fig_gammaratio_vs_tau} in the Pfirsch-Schlüter regime when $\nu_e^* = 50$. It is confirmed that the correction terms (see \eqref{eq:ISugama}) enable the IS operator to approximate the Coulomb collision operator better than the OS operator as $T_i / T_e$ increases, with $\sigma(\gamma) \lesssim 1 \% $. The same observations can be made for the real mode frequency, $\omega_r$.}


\begin{figure*}
  \centering
    \includegraphics[scale = 0.51]{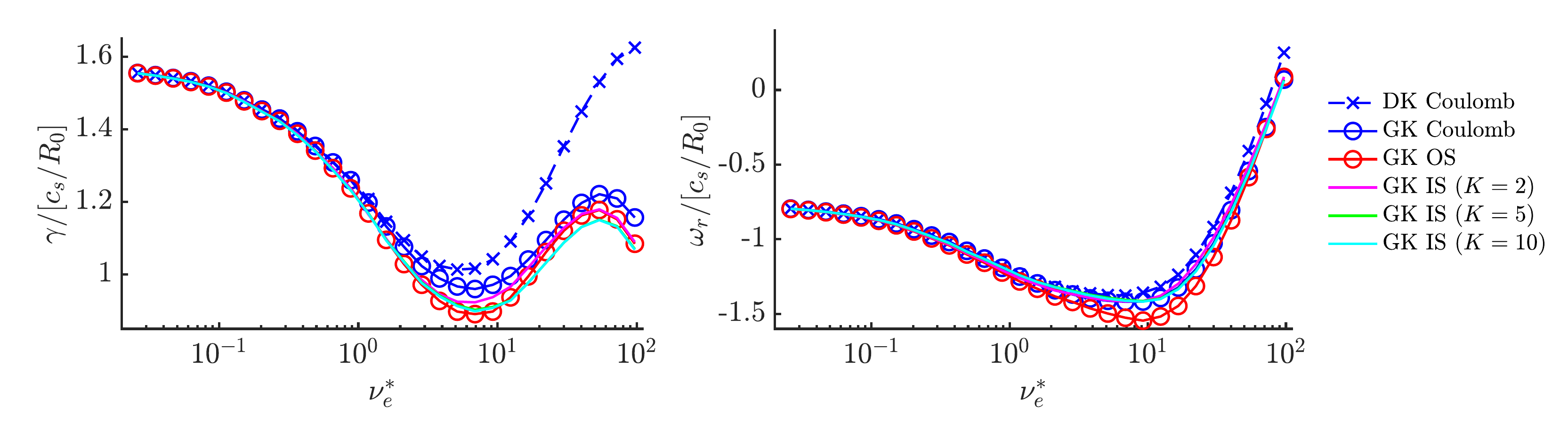}
    \caption{High-$k_y$ TEM growth rate $\gamma$ (left) and the real mode frequency $\omega_r$ (right) as a function of $\nu_e^*$ obtained using the GK Coulomb (blue circle markers), the GK OS (red circle markers), the GK IS with $K = 2$ (solid magenta line), the GK IS with $K =5$ (solid green line) and  the GK IS with $K = 10$ (solid light blue line) collision operators. For comparison, the predictions of the DK Coulomb operator (blue cross markers) are also shown. The parameters are the same as in \figref{fig:fig_omegacomplex_omn20_kymin} at $k_y = 0.6$ with $T_i / T_e = 1$.}
    \label{fig:fig_TEM_kymin.0.6}
\end{figure*}

We now turn to the collisionality dependence of the high-$k_y$ TEM mode developing near $k_y = 0.6$ (see \figref{fig:fig_omegacomplex_omn20_kymin}) using the GK collision operators. In particular, we use the GK Coulomb, GK OS, and GK IS operators, \rev{ using the spherical harmonic expansion detailed in Appendix \ref{appendix:Abis}.} As the perpendicular wavenumber in the argument of the Bessel functions increases, an increasingly large number of terms in the infinite sums arising form the expansion of the Bessel functions, \eqref{eq:Bessel_relation}, is required for convergence \cite{frei2022}. We evaluate numerically these sums in $I_{\parallel s}^{pjk}$ and $I_{\perp s}^{pjk}$, given in Eqs. (\ref{eq:Iparallelspjk}) and (\ref{eq:Iperpspjk}), by truncating them at $n = 6$ (we have verified the convergence of our results). \rev{The collisionality dependence of the high-$k_y$ TEM growth rate, $\gamma$, and mode frequency, $\omega_r$, obtained by using the same parameters in \figref{fig:fig_omegacomplex_omn20_kymin} (for $k_y = 0.6$) are shown in \figref{fig:fig_TEM_kymin.0.6}, as a function of the electron collisionality $\nu_e^*$ using the GK Coulomb, GK OS and GK IS with different values of $L = K$. We also show the results of the DK Coulomb to illustrate the effects of FLR terms in the collision operators. First, we observe that the high-$k_y$ TEM is stabilized by the GK operators compared to the DK Coulomb (and other DK operators) because of the presence of the FLR terms in the former. Second, it is remarkable that, while the GK OS operator is able to capture the trend of the growth rate and of the mode frequency observed with the GK Coulomb, it yields systematically a smaller growth rate at all collisionalities. We remark that, while a direct comparison between the GK operators implemented in the GENE code is outside of the scope of the present work, similar observations are made in TEM simulations at weaker gradients \cite{Pan2020,pan2021importance} based on the GENE code. Third, and finally, we observe that the GK IS yields a growth rate similar (yet smaller) to the GK OS collision operator. In particular, the GK IS with $K = 2$ reproduces the same growth rate as the GK OS, while larger values of $K$, i.e. $K = 5$ and $10$ (convergence is achieved with $L = K \gtrsim 3$), produce a smaller growth rate than the GK OS within $5 \%$ (we have carefully verified that our numerical results are converged by increasing the number of gyro-moments and the number of points in the parallel direction in the simulations). On the other hand, the real mode frequency, $\omega_r$, predicted by the GK Coulomb is well retrieved by the GK IS operators independently for $K$, while it is underestimated by the GK OS (see \figref{fig:fig_TEM_kymin.0.6}). We remark that, in the DK limit, the IS operator yields very similar growth rates and real mode frequencies than the DK Coulomb at high-collisionalities. To understand these observations, we first note that, in the case considered here where $T_i = T_e$, the test components of the GK OS and GK Coulomb are equivalent (see Eqs. (\ref{eq:CLT}) and (\ref{eq:CabST})) since all the terms $\Delta M_{ab}^{\ell k}$ in $\Delta \C_{ab}^T$ (see \eqref{eq:correctionmatrix}) vanish exactly. This implies that the GK OS and GK IS differ only by the GK correction terms in their field components, i.e. by $\Delta \C_{ab}^F$ given in Eq. (\ref{eq:DeltaGKCabF}). To illustrate the contribution from the FLR terms, we repeat the simulations in \figref{fig:fig_TEM_kymin.0.6} considering the DK limits of the field components in all operators, but retain the GK test components being equivalent to the one of the GK Coulomb operator. The results are displayed in \figref{fig:gtdf}, and show that the IS operator yields a growth rate larger and closer to the GK Coulomb compared than the GK OS in the absence of FLR terms in the field components, while (not shown) the real mode frequency agree between the GK IS and GK Coulomb operators. We remark that the high-$k_y$ TEM is strongly damped at high-collisionality if the FLR terms in the field component are neglected. Overall, the GK IS and GK OS yield a similar collisionality dependence of the high-$k_y$ TEM with a good agreement in the mode frequency between the GK OS and GK Coulomb, despite that the growth rate predicted by the GK IS and GK OS differ from the GK Coulomb within $ 10 \%$.

\begin{figure}
  \centering
    \includegraphics[scale = 0.52]{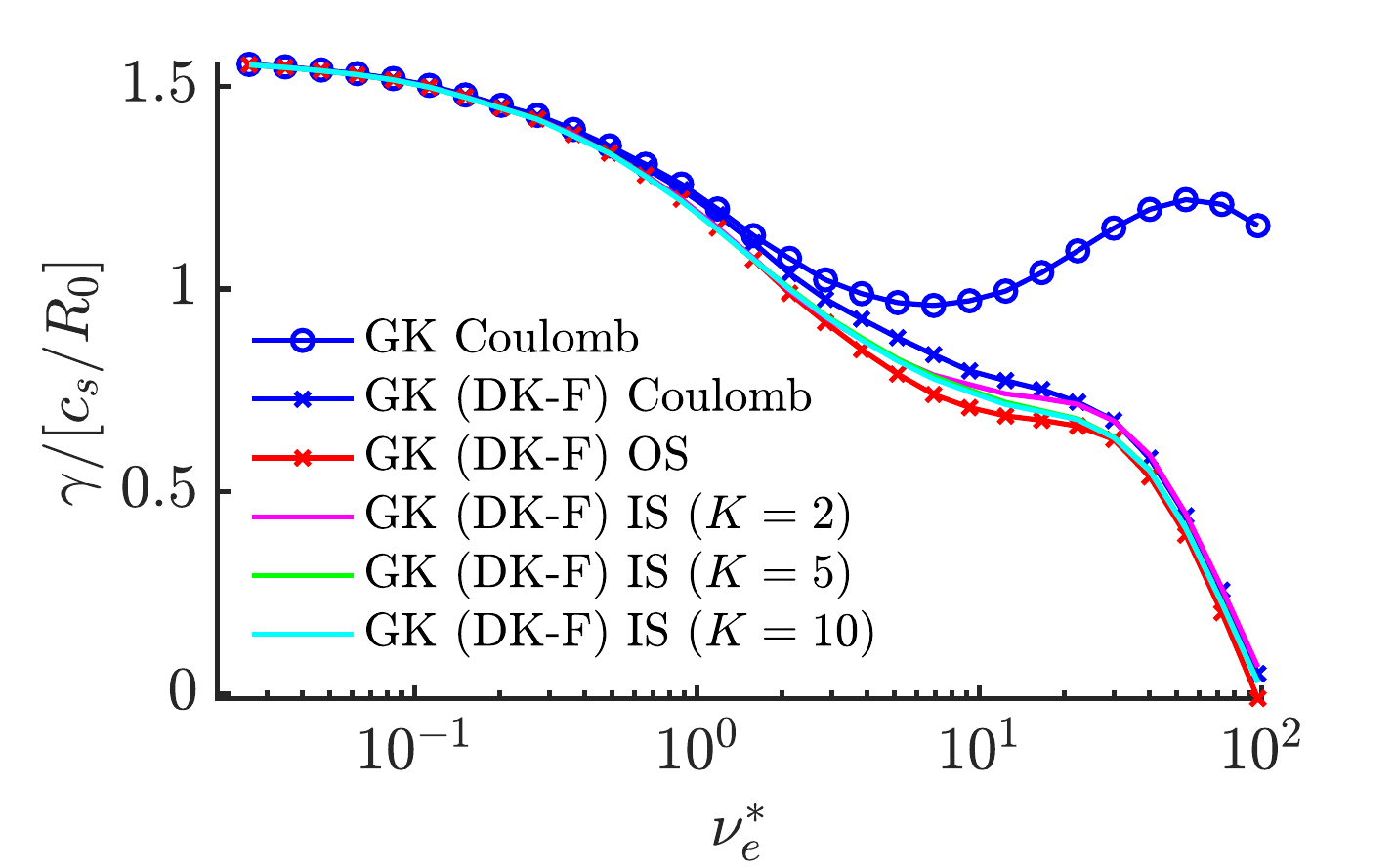}
    \caption{Same as \figref{fig:fig_TEM_kymin.0.6}, but using the field component in the DK limit (DK-F) in all operators.}
    \label{fig:gtdf}
\end{figure}}

\subsection{Collisional Zonal Flow Damping}
\label{sec:ZF}

Axisymmetric, poloidal zonal flows (ZFs) are believed to be among the key physical mechanisms at play in the L-H mode transition by, ultimately, regulating the level of turbulent transport \cite{diamond2005zonal}. It is therefore of primary importance to test and compare the effect of collisions, modelled by using the IS, OS and Coulomb collision operators, on their dynamics.

\begin{figure}
    \centering
    \includegraphics[scale = 0.56]{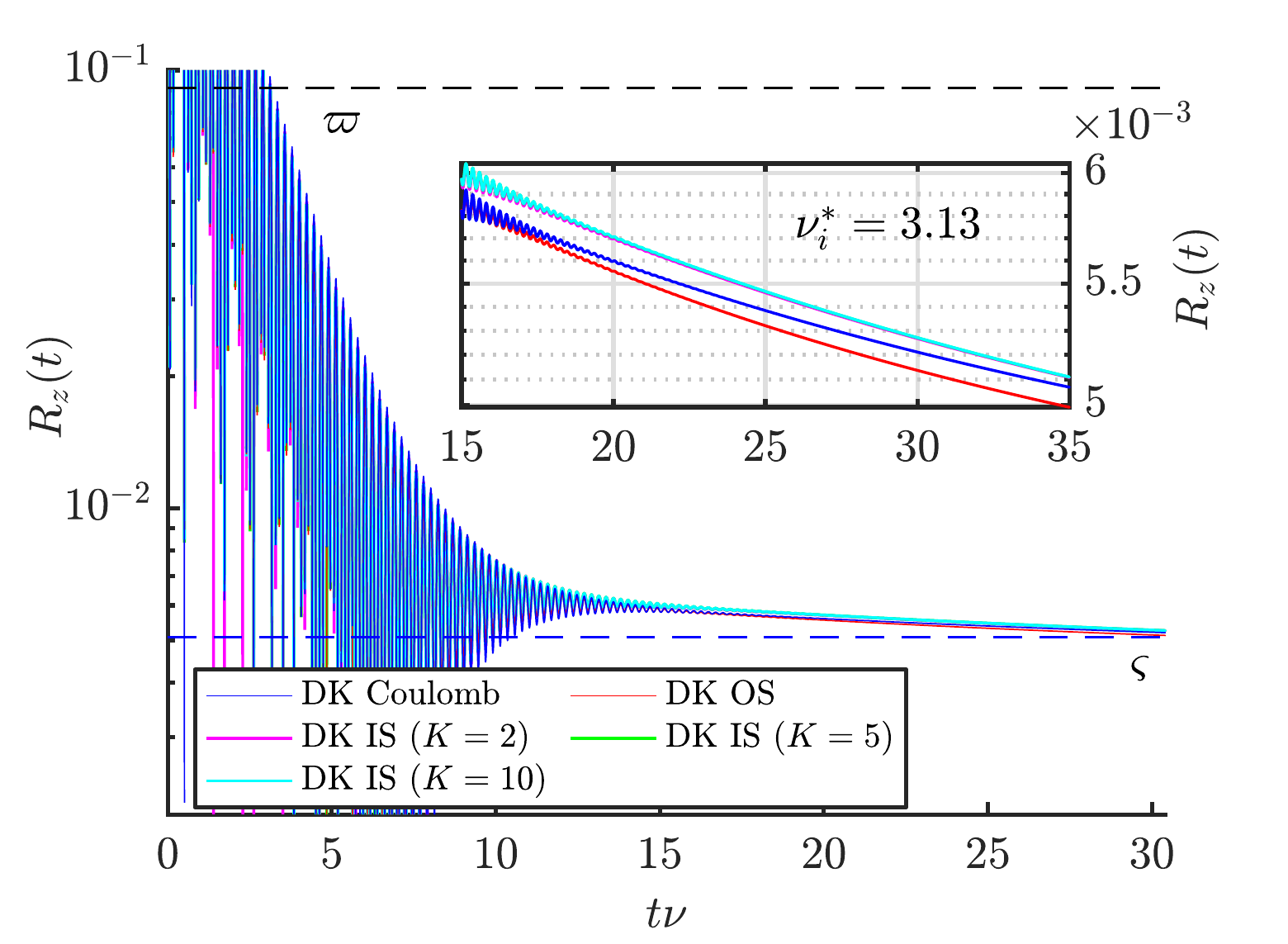}
    \caption{Collisional ZF damping obtained using the DK Coulomb (blue line), DK OS (red line) and DK IS with $L = K = 2, 5, 10$ (magenta, green and light blue lines, respectively) collision operators in the Pfirsch-Shlüter regime $\nu_i^* =  3.13$. The collisionless (see Ref. \onlinecite{Rosenbluth1998a}) and collisional (see \eqref{eq:Xiaoresidual}) long time predictions, $\varpi = 1 / (1 + 1.6 q / \sqrt{\epsilon})$ and $\varsigma$, are plotted by the dashed black and blue lines, respectively. Here, $k_x = 0.05$, $q  = 1.4$ and $\epsilon = 0.1$.  }
    \label{fig:fig_ZF_dk}
\end{figure}

While the damping to a residual level of the ZFs has been originally studied in the collisionless regime \cite{Rosenbluth1998a}, the later study was extended to the assess the collisional ZF damping in the banana regime when $\nu_i^* \lesssim 1$ (with $\nu_i^*$ the ion-ion collisionality) for radial wavelengths much longer than the ion polodial gyroradius, assuming adiabatic electrons, and using a pitch-angle scattering operator mimicking the collisional drag of energetic ions \cite{Hinton1999} . A refinement of the exponential decay in Ref. \onlinecite{Hinton1999} of the ZF residual prediction $R_z(\infty) = \phi_z(\infty) / \phi_z(0)$ (with $\phi_z(t)$ the flux-surface averaged potential) with a momentum restoring pitch-angle scattering operator was later derived in Ref. \onlinecite{Xiao2007} for long wavelength modes,
 
 \begin{align} \label{eq:Xiaoresidual}
    R_z(\infty) \to 
\varsigma =     \frac{ \epsilon^2 }{q^2} \frac{1}{(1 + \epsilon^2/ q^2)}.
 \end{align}
\\
Even if \eqref{eq:Xiaoresidual} does not include energy diffusion as well as the effects of GK terms in the collision operator, it still provides a good estimate asymptotic ZF residual predicted by the DK IS, as shown below.

\begin{figure*}
  \centering
    \includegraphics[scale = 0.55]{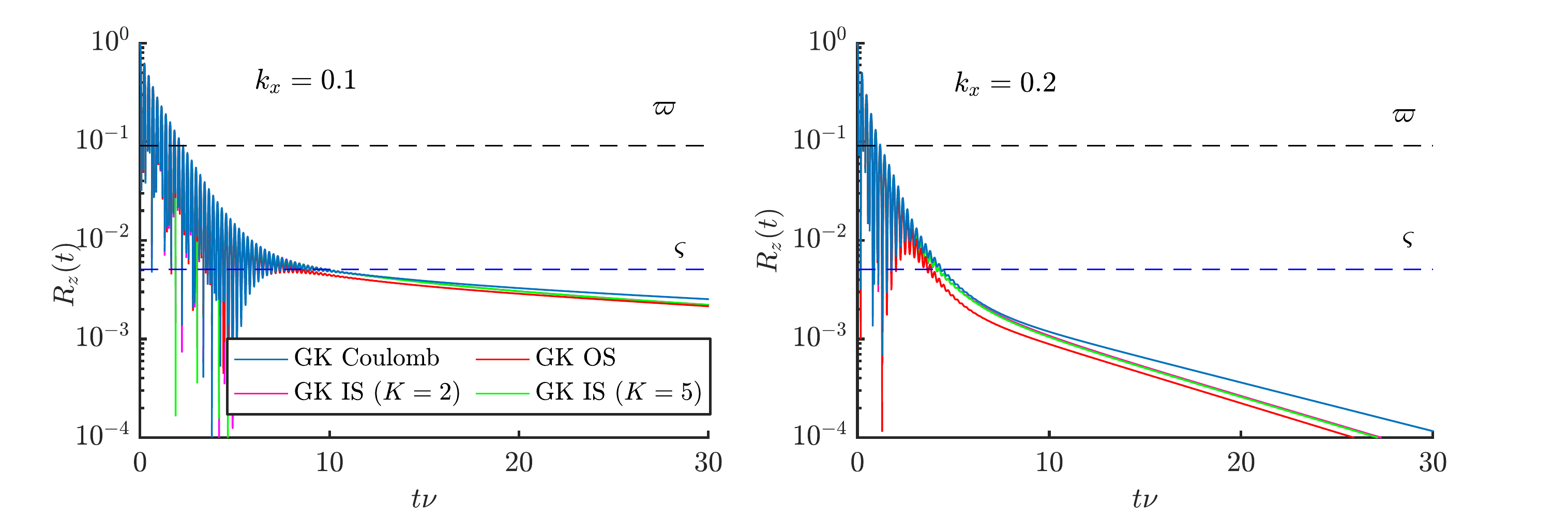}
    \caption{Collisional ZF damping obtained using the GK Coulomb (blue lines), GK OS (red lines) and GK IS collision operators with $K = 2$ (magenta lines) and $K = 5$ (green lines). The damping of an initial density perturbation with a radial wavenumber $k_x = 0.1$ (left) and $k_x = 0.2$ (right) are shown. The analytical collisionless and  collisional predictions, $\varpi$ and $\varsigma$, are plotted for comparison. It is observed that the GK IS collision operator provides better approximation to the GK Coulomb operator. The parameters and the number of gyro-moments are the same as in \figref{fig:fig_ZF_dk}.}
    \label{fig:fig_ZFgk}
\end{figure*}

For our tests, we consider only ion-ion collisions and, by including the IS operator, we extend the study of Ref. \onlinecite{frei2021}, where the differences between the GK OS and GK Coulomb operators in the collisional ZF damping are illustrated, finding a stronger damping by the former. We focus on the Pfirsch-Schlüter regime with $\nu_i^* = 3.13$, with convergence being achieved with $(P,J) =(24,10)$ gyro-moments. The collisional time traces of the ZF residual, $R_z(t) = \phi_z(t) / \phi_z(0)$, obtained for the DK IS (with $ K =L =  2,5, 10$), DK OS and DK Coulomb collision operators are shown in \figref{fig:fig_ZF_dk} for $k_x = 0.05 $ and in \figref{fig:fig_ZFgk} for the $k_x = 0.1$ and $k_x =0.2$ using the GK operators. Starting from the case of small radial wavenumber $k_x = 0.05$, we first observe that all DK operators agree with the analytical long time prediction given in \eqref{eq:Xiaoresidual}, despite the absence of energy diffusion in the latter. Consistently with Ref. \onlinecite{frei2021}, the DK Sugama yields a stronger damping of the ZF than the DK Coulomb. The addition of the correction terms to the OS operator allows the DK IS to better approximate the DK Coulomb operator, yielding a weaker damping of the ZF. We remark that only a small difference between the $L = K = 2,5$ cases is noticeable showing that $L = K \simeq 3$ is necessary for the DK IS to converge also in this case.

\begin{figure*}
  \centering
    \includegraphics[scale = 0.59]{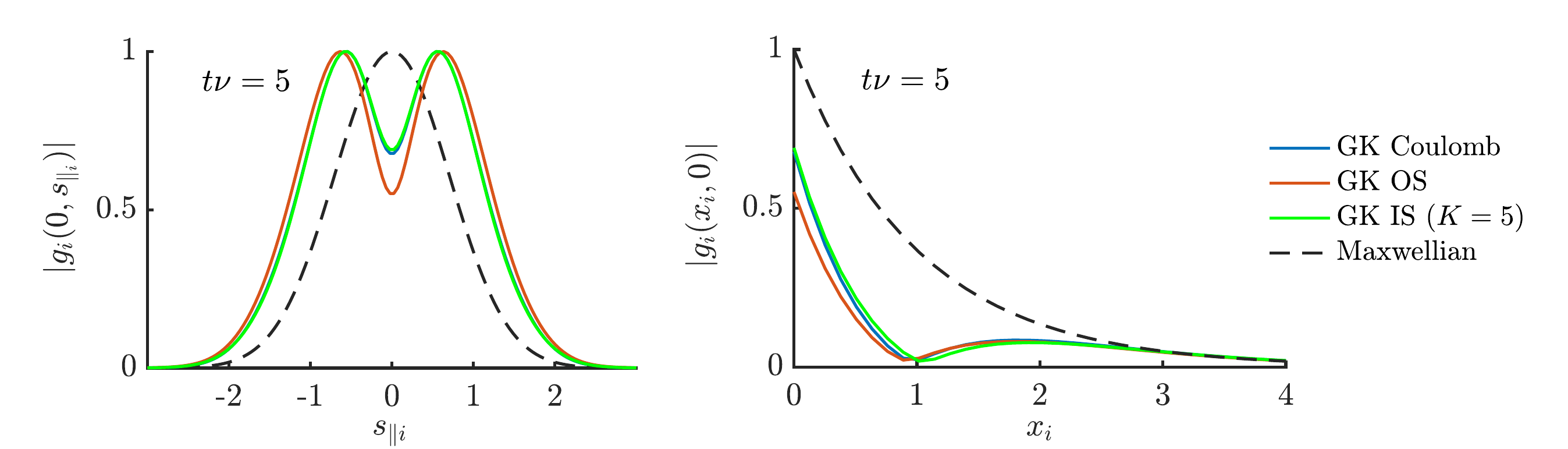}
    \caption{Modulus of the perturbed ion gyrocenter distribution function  $|g_i|$ (normalized to its maximum value), at time $t \nu = 5$ after the damping of the GAM oscillations and at the outboard midplane in \figref{fig:fig_ZFgk}, plotted as a function of $s_{\parallel i}$ at $x_i =0$ (left) and as a function of $x_i$ at $s_{\parallel i} =0$ (right). The case of a Maxwellian distribution function $e^{- s_{\parallel i}^2 - x_i}$ is shown for comparison (dashed black lines).}
    \label{fig:fig_vsp}
\end{figure*}

We now consider the collisional ZF damping at larger $k_x$ values using the GK IS, GK OS and GK Coulomb collision operators, $k_x = 0.1$ and $k_x = 0.2$, and plot the results in \figref{fig:fig_ZFgk}. The same collisionality of the $k_x = 0.05$ case is used. Only the $K = 2, 5$ cases are considered for the GK IS for simplicity, since convergence is achieved with these parameters. First, consistently with Ref. \onlinecite{frei2021}, it is observed that the GK OS produces a stronger ZF damping with respect to the GK Coulomb. \rev{We note that a similar observation based on the results of the GENE code is reported in Refs. \onlinecite{Pan2020,pan2021} (the collisional damping predicted by the gyro-moment method and the GENE code using the GK OS \cite{sugama2009} is successfully benchmarked in Ref. \onlinecite{frei2021}).} Second, the GK IS collision operator provides a better approximation to the GK Coulomb collision operator than the GK OS operator. This is particularly true in the early phase of the damping i.e. $t \nu \lesssim 10$. Although the GK IS still yields a better approximation than the GK OS, larger GK IS departures from the GK Coulomb are observed at later times when the GAM oscillations are completely suppressed. For larger collisionalities and smaller scale lengths, the same observations about the rapid ZF decay made at lower collisionalities hold. 

Finally, we investigate the effects of the GK collision operators on the ion velocity-space distribution function. We consider the modulus of the perturbed ion distribution function, $|g_i|$, obtained by using \eqref{eq:hHL}, at time $t \nu = 5$ after the damping of the GAM oscillations, for the case $k_x = 0.2$. We plot $|g_i|$ as a function of $s_{\parallel_i}$ (at $x_i = 0$) and $x_i$ (at $s_{\parallel i} =0$) in \figref{fig:fig_vsp}, respectively, for the different GK collision operators. It is observed that the GK IS yields similar velocity-space structures along both the parallel and perpendicular directions than the GK OS operator with the distribution function being depleted in the region of the velocity-space $|v_{\parallel }| \lesssim v_{Ti}$ more strongly than the GK Coulomb operator. A similar observation can be made when $v_\perp \lesssim v_{Ti}$, as shown in the right panel of \figref{fig:fig_vsp}.

\rev{\subsection{Spitzer Electrical Conductivity }
\label{sec:spitzer}

\begin{figure}
    \centering
    \includegraphics[scale = 0.65]{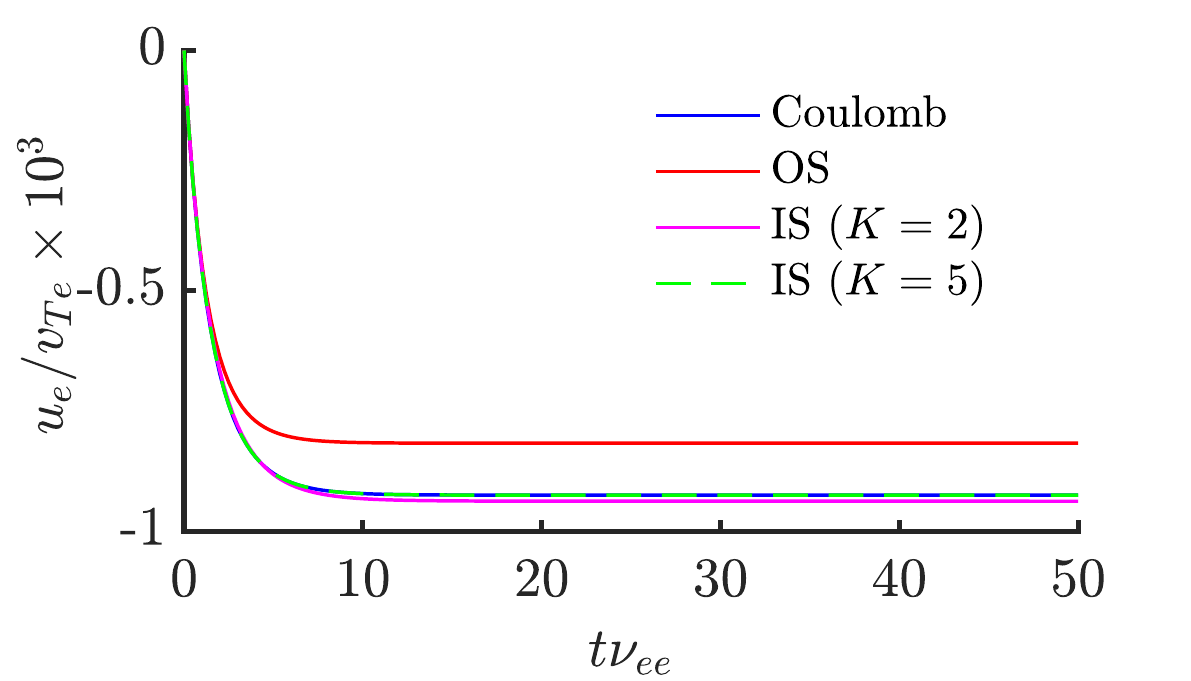}
    \caption{Normalized saturated electron parallel velocity, $u_e / v_{Te}$, as a function of $t \nu_{ee}$ obtained using the Coulomb (blue lines), OS (red line) and IS with $K = 2$ (magenta line) and $K =5$ (green dashed line) operators. Here, we use $(P,J) = (20,5)$ gyro-moments. }
    \label{fig:fig_jsat}
\end{figure}

As a final numerical test, we consider the evaluation of the Spitzer electrical conductivity. This is obtained from the stationary current resulting from the balance between the collisional drag and a constant electric force. We consider an unmagnetized fully-ionized plasma with a fixed background of ions with electrical charge $q_i Z$ (with $Z$ the ion ionization degree), subject to an electric field, $\bm E= E \bm e_z$, with $E$ a constant amplitude. Hence, the kinetic equation describing the evolution of the electron distribution function, $f_e$, is given by

\begin{align} \label{eq:spitzerprob}
    \partial_t f_e  - \frac{e}{m_e} E \partial_{v_\parallel} f_e = C_{ee}^{AT}( f_e )  + C_{ee}^{AF}(f_e) + C_{ei}^{AT}(f_e),
\end{align}
\\
with $v_\parallel = \bm v \cdot \bm e_z$. Collisional effects between electrons ($ C_{ee}^{AT}$ and $C_{ee}^{AF}$) and the stationary background of ions ($ C_{ei}^{AT}$) are modelled using the Coulomb, OS and IS collision operators.  Projecting \eqref{eq:spitzerprob} onto the Hermite-Laguerre basis yields the evolution equation of the gyro-moments of $f_e$ denoted by $N_e^{pj}$, i.e.

\begin{align} \label{eq:spitzerprob}
    \partial_t N_e^{pj}  + \frac{e}{ v_{Te} m_e} \sqrt{2 p} E N_e^{p-1j} = \C_{ee}^{ATpj} + \C_{ee}^{AFpj} + \C_{ei}^{ATpj}.
\end{align}
\\
\revv{We remark that the term associated with electric force, proportional to $\sqrt{p} E N_e^{p-1j}$, vanishes when $p=0$. We evolve the gyro-moment hierarchy, given in Eq. (\ref{eq:spitzerprob}), with $(P,J) = (20,5)$ gyro-moments in time until a stationary electron current, $j_{\parallel e} = -e N_e u_e $ (with $u_e =  \int d \bm v v_\parallel f_e / N_e = N_e^{10} v_{Te}/ \sqrt{2}$ the parallel electron velocity), is established resulting from an applied electric field of normalized amplitude $  e E /[\sqrt{m_e T_e} \nu_{ee}] = 10^{-3}$ (see \figref{fig:fig_jsat}). From the saturated current, the electrical conductivity, $\sigma_{\parallel e } = j_{\parallel e} / E $, can be computed.}

In \figref{fig:fig_spitzer}, we compare our numerical estimates of $\sigma_{\parallel e}$ with the analytical prediction of the Spitzer conductivity \cite{spitzer1953transport,helander2005collisional}, given by $\sigma_{\parallel e} = 16 T_e^{3/2} / [ (2 \pi)^{3/2}\sqrt{m_e} Z^2 e^2 \ln \Lambda ])$ obtained in the large ion ionization $Z$ limit. We note that in the $Z \to \infty$ limit, the electron and ion collisions can be modelled by the pitch-angle scattering collision operator, $C_{ei}(f_e) \simeq - \nu_{ei}^D(v) \mathcal{L}^2 f_e$ (where $  \nu_{ei}^D(v)$ is the velocity-dependent deflection collision frequency, defined below \eqref{eq:Cab0}), and the collisions between electrons can be neglected since $\nu_{ee} / \nu_{ei} \sim 1 / Z^2$. First, we observe that the OS operator produces an electrical conductivity that is approximately $10 \%$ smaller than the one obtained by the Coulomb operator at low $Z$, and that the difference between the predictions of the IS and Coulomb operators is less than $1 \%$ for all values of $Z$. While the deviations in the electrical conductivity decrease with $Z$ because the contribution of the pitch-angle scattering that dominates with $Z$ in all operators, the deviation at low $Z$ observed between OS and Coulomb is associated with the different field components of these operators, i.e. $C_{ee}^{LF}$ and $C_{ee}^{SF}$, given in Eqs. (\ref{eq:CLF}) and (\ref{eq:CabSF}) respectively, since the test components of these operators are equal in the case of like-species collisions. Second, we observe that all operators converge to the analytical $Z \to \infty$ Spitzer conductivity as $Z$ becomes large, within less than $8 \%$ for $Z = 10$, providing a verification of the numerical implementation. Given the importance of the plasma resistivity in setting the level of turbulent transport in the scrape-off-layer \cite{giacomin2022first}, the present test shows that the OS operator underestimates the parallel current (see \figref{fig:fig_jsat}), which might have a significant effect on boundary turbulent simulations.}


\begin{figure}
    \centering
    \includegraphics[scale = 0.5]{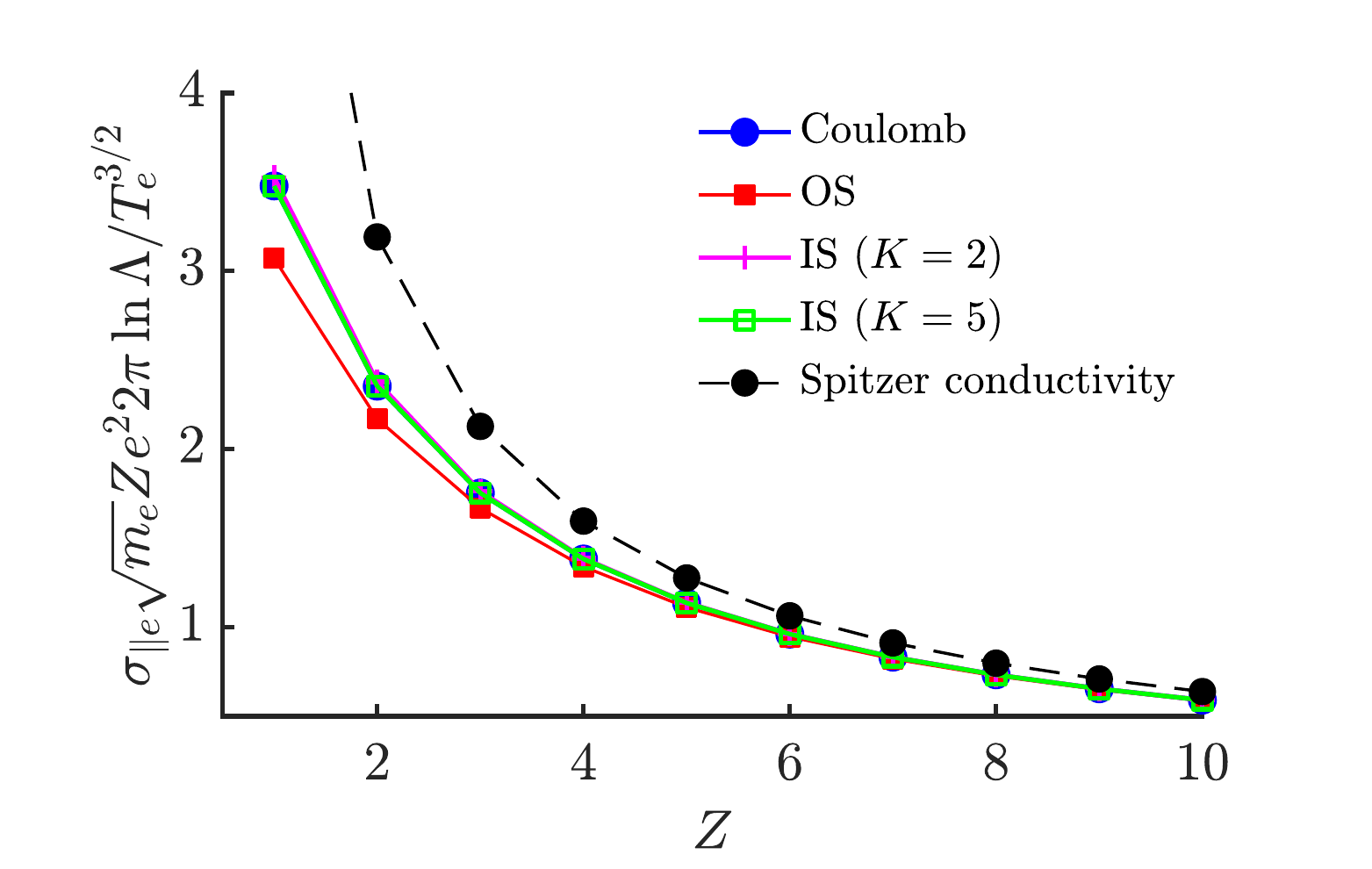}
    \caption{\rev{Normalized electrical conductivity, $\sigma_{\parallel e}$, as a function of the ion ionization degree $Z$ obtained using the same operators \revv{(and number of $(P,J) = (20,5)$ gyro-moments)} as in \figref{fig:fig_jsat}. The numerical results are compared with the Spitzer analytical conductivity \cite{spitzer1953transport} as shown by the black markers.}}.
    \label{fig:fig_spitzer}
\end{figure}


\section{Conclusion}
\label{sec:conclusion}

In this work, the gyro-moment method has been applied to implement the recently developed improved Sugama (IS) collision operator in both the gyrokinetic (GK) and drift-kinetic (DK) regimes \cite{sugama2019improved}. Designed to extend the validity of the original Sugama (OS) collision operator to the Pfirsch-Schlüter regime, the Hermite-Laguerre expansion of the perturbed distribution function allows expressing the IS collision operator as a linear combination of gyro-moments, with coefficients that are analytical functions of the mass and temperature ratios of the colliding species and, in the GK limit, of the perpendicular wavenumber. Analytical expressions of the Braginskii matrices, $M_{ab}^{A \ell k}$ and $N_a^{A \ell k}$, associated with the Coulomb and OS collision operators are obtained for arbitrary $(\ell,k)$ using the spherical harmonic expansion \cite{frei2021}. This allows the evaluations of the correction terms that are added to the OS operator yielding the IS collision operators for arbitrary gyro-moments.

We describe the numerical implementation of the IS collision operator. This is based on an arbitrary precision arithmetic library to avoid the numerical loss of precision and round-off errors when evaluating the Braginskii matrices. We demonstrated that the conservation laws (particle, momentum, and energy) are satisfied at arbitrary order in $(\ell, k)$. The IS collision operator is tested and compared with the OS and Coulomb collision operator in edge conditions at steep pressure gradients and high collisionality. Three test cases are considered, that involve the evaluation of the TEM growth rate and frequency at steep gradients, the collisional damping of ZF and, finally, the evaluation of the Spitzer electrical conductivity. The analysis of the linear properties of a conventional TEM, developing at long wavelengths, reveals that the IS is able to approximate the TEM growth rate and frequency predicted by the DK Coulomb operator better than the OS, particularly in the Pfirsch-Schlüter regime. At small perpendicular wavelengths and high-collisionality, the study an unconventional TEM shows that the GK IS and GK OS essentially yields to the same FLR damping, within a $10 \%$ of difference. The collisional ZF damping is also explored and compared with analytical results. The analysis shows that the IS operator yields an intermediate value of the ZF residual, between the weaker value produced by the OS operator and the larger ZF residual predicted by the Coulomb collision operator. \rev{Finally, the electrical conductivity predicted by the IS operator is found to be in good agreement with the one produced by the Coulomb operator (within less than $1 \%$), while deviations larger than $10 \%$ are obtained when using the OS operator. In conclusion, the IS and OS produce linear results that differ by around $10 \%$ in all cases explored in the present work. Nonlinear turbulent simulations are required to investigate the impact of these differences on the saturated turbulent state, in particular near the nonlinear stability threshold.}

We demonstrate that the computational cost of the gyro-moment approach decreases with collisionality, as illustrated in the case of TEM at steep gradients (see \figref{fig:fig6}). In a future publication, we show that the gyro-moment approach is particularly efficient in steep pressure gradient conditions, where the number of gyro-moments can be significantly reduced compared to weaker gradients (typically found in the core) even at low collisionality. This is particular relevant for H-mode pedestal applications and offers an ideal framework to construct reduced fluid-like models to explore turbulent transport in the boundary of fusion devices. As an example, the analytical expression derived in this work allow us to evaluate the lowest-order gyro-moment terms of the IS, OS and Coulomb collision operators in Appendix \ref{appendix:B}. Finally, we remark that the expressions presented in this work can be easily extended to the study of multicomponent plasmas with species of different mass and temperatures.

\section*{Acknowledgement}

This work has been carried out within the framework of the EUROfusion Consortium, funded by the European Union via the Euratom Research and Training Programme (Grant Agreement No 101052200 — EUROfusion). Views and opinions expressed are however those of the author(s) only and do not necessarily reflect those of the European Union or the European Commission. Neither the European Union nor the European Commission can be held responsible for them. The simulations presented herein were carried out in part on the CINECA Marconi supercomputer under the TSVVT421 project and in part at CSCS (Swiss National Supercomputing Center). This work was supported in part by the Swiss National Science Foundation.

\appendix

\section{Linearized Coulomb and Original Sugama Collision Operators}
\label{appendix:A}

The linearized particle Coulomb collision operator, denoted by $C_{ab}^L$, is defined as the sum of its test and field components, $C_{ab}^{LT}$ and  $C_{ab}^{LF}$ respectively, given in pitch-angle coordinates $(\r, v,\xi, \theta)$ (with $v$ the modulus of $\bm v$ and $\xi = v_\parallel/v$ the pitch angle) by \cite{Rosenbluth1957}

\begin{align} \label{eq:CLT}
 C^{LT}_{ab}(f_a) & =  \frac{m_a \nu_{ab} v_{Ta}}{n_b} \left[ \left( \frac{2  \partial_v G(f_{Mb})}{ v^2} + \left( 1 -  \frac{m_a}{m_b} \right) \partial_v H(f_{Mb}) \right)  \partial_v f_{a} \right. \nonumber \\
& \left.  - \frac{1}{v^3} \partial_v G(f_{Mb}) \mathcal{L}^2 f_a +  \partial_{v}^2 G(f_{Mb})  \partial^2_v f_{a}  + \frac{m_a}{m_b} 8 \pi f_{Mb}  f_a\right],
\end{align}
\\
and 

\begin{align} \label{eq:CLF}
C^{LF}_{ab}(f_b) &= \frac{2 \nu_{ab} v_{Ta} f_{Ma}}{n_b} \left[ 2 s_a^2 \partial_v^2 G(f_b) - H(f_b) \right. \nonumber \\ 
&\left. - \left( 1 -  \frac{m_a}{m_b} \right) v \partial_v H(f_b) + \frac{m_a}{m_b} 4 \pi v_{Ta}^2 f_b \right],
\end{align}
\\
respectively. In \eqref{eq:CLT}, we introduce the Rosenbluth potentials $H(f) = 2 \int d \bm v' f(\bm v') /u$ and $G(f)  = \int d \bm v' f(\bm v')u$ (with $u = |\vi - \vi'|$) and the operator $\mathcal{L}^2 f  = \partial_{v}(v^2 \partial_v f) - v^2 \grad_{\bm v}^2 f$. When evaluated with a Maxwellian distribution function, analytical expressions of the Rosenbluth potentials can be obtained, and are given by $G(f_{Mb}) = n_b v_{Tb} \left[ \left( 1 + 2 s_b^2 \right)\erf(s_b)/s_b + \erf'(s_b)\right]/2$ and $H(f_{Mb}) = 2 n_b \erf(s_b)/v$ (with the error function, $\erf(x) = 2 \int_0^x d s e^{-s^2}/\sqrt{\pi}$, and its derivative, $\erf'(x) = 2 e^{-x^2}/\sqrt{\pi}$). We remark that the velocity-space derivatives and integrals, appearing in Eqs. (\ref{eq:CLT}) and (\ref{eq:CLF}), act on the perturbed particle distribution function $f_a$ and are evaluated holding $\bm r$ constant, such that $ C^{LT}_{ab}(f_a) =  C^{LT}_{ab}(f_a (\bm r, \bm v)) = C^{LT}_{ab}(\bm r, \bm v)$ (and the same for $C^{LF}_{ab}(f_b)$).

The original particle Sugama (OS) collision operator, denoted by $C_{ab}^{S}$, is expressed by the sum of its test and field components, $C_{ab}^{ST}$ and  $C_{ab}^{SF}$, given by \cite{sugama2009}

\begin{align} \label{eq:CabST}
C_{ab}^{ST}(f_a) = C_{ab}^0(f_a)+  \sum_{n=1}^3  X^n_{ab}
\end{align}
\\
where

\begin{align} \label{eq:Cab0}
C_{ab}^0(f_a ) = - \nu_{ab}^D(v)\mathcal{L}^2 f_a  +  \frac{1}{v^2} \partial_v\left[  \nu_{ab}^{\parallel}(v) v^4 F_{aM} \partial_v \left( \frac{f_a}{f_{aM}}\right)\right],
\end{align}
\\
and

\begin{align}
X_{ab}^1 &= 2 (\theta_{ab} -1 )   f_{Ma} \left[ \frac{m_a}{T_a} \frac{1}{n_a} \int d \bm v' \bm v'  C_{ab}^0(f_a) \cdot \bm v  \right. \nonumber \\
& \left. +  \left( s_a^2 - \frac{3}{2}\right)   \frac{1}{n_a} \int  d\vi  \frac{2}{3} s_a^2 C_{ab}^0(f_a) \right], \\
X_{ab}^2 &= 2 (\theta_{ab} -1 )  \left[  \frac{m_a}{T_a} \bm u_a(f_a) \cdot C_{ab}^0( f_{Ma} \bm v) + \frac{\delta T_a (f_a)}{T_a} C_{ab}^0(f_{Ma}s_a^2) \right], \label{eq:X2ab} \\
X_{ab}^3 &= - \frac{8 \nu_{ab}}{3 \sqrt{\pi}}f_{Ma} \frac{\chi(\theta_{ab} -1 )^2}{ (1 + \chi_{ab}^2)^{1/2}} \left[ \frac{m_a}{T_a} \bm v \cdot \bm u_a(f_a)   \right. \nonumber \\
& \left.  + \frac{\delta T_a(f_a)}{T_a}\left( s_a^2 -\frac{3}{2}\right)  \frac{2 }{ (1 + \chi_{ab}^2)} \right],
\end{align}
\\
and by

\begin{align} \label{eq:CabSF}
C_{ab}^{SF}(f_b) = - \frac{m_a \bm V_{ab}(f_b)}{T_a} \cdot C_{ab}^{ST}\left( f_{Ma} \bm v\right)   - W_{ab}(f_b) C_{ab}^{ST}( f_{Ma} s_a^2).
\end{align}
\\
respectively. In \eqref{eq:CabST}, we introduce the velocity-dependent collision frequencies, $\nu_{ab}^D(v ) =  \nu_{ab}  \left[ \erf(s_b) - \Phi(s_b)\right]/s_a^3$ and $\nu_{ab}^{\parallel}(v)  = 2 \nu_{ab} \Phi(s_b) / s_a^3$ (with  $\Phi(x) =  \left[ \erf(x) - x \erf'(x)\right]/(2 x^2)$ being the Chandrasekhar function), the fluid quantities, $\bm u_a(f_a) = \int d \bm v \bm v  f_a / n_a$, $\delta T_a(f_a) = T_a \int d \bm v f_a \left( 2s_a^2/3 - 1 \right) /n_a$, the coefficients $\theta_{ab} = \sqrt{ (T_a /T_b + \chi_{ab}^2)/(1 + \chi_{ab}^2)}$, $\tau = T_a / T_b $, $\chi_{ab} = v_{Ta} / v_{Tb} = \sqrt{\tau / \sigma}$ and $\sigma = m_a / m_b$. Finally, in \eqref{eq:CabSF}, we have 

\begin{subequations}\label{eq:VabWab}
\begin{align} 
\bm V_{ab}(f_b)& = \frac{m_b}{\gamma_{ab}} \int d \vi \frac{f_b}{f_{Mb}} C_{ba}^{ST}\left( f_{Mb} \vi \right), \\ 
W_{ab}(f_b) & = \frac{T_b}{\eta_{ab}} \int d \vi \frac{f_b}{f_{Mb}} C_{ba}^{ST}(f_{Mb} s_b^2),
\end{align}
\end{subequations}
\\
with \rev{ $ \gamma_{ab} = -  (n_a m_a \chi_{ab}) \left( T_a /T_b + \chi_{ab}^2\right) /[  \bar{\tau}_{ab} (1 + \chi_{ab}^2)^{3/2}]$} and $\eta_{ab} = - 3 \chi_{ab} T_a (T_a / T_b + \chi_{ab}^2) / [ \bar{\tau}_{ab}(1 + \chi_{ab}^2)^{5/2}]$. The expressions of $\bm V_{ab}(f_b)$ and $W_{ab}(f_b)$ given in Eqs. (\ref{eq:VabWab}) are chosen such that particle, momentum and energy conservations are satisfied (see \eqref{eq:conservationlaws}). The closed analytical expressions of $C_{ab}^0(f_{Ma} \bm v)$, $C_{ab}^0(f_{Ma} s_a^2)$ and $C_{ab}^{ST}\left( f_{Ma} \vi \right)$ with $C_{ab}^{ST}(f_{Ma} s_a^2)$ can be found in Ref. \onlinecite{sugama2009}. We remark that $\theta_{ab} =1$ when $T_a = T_b$ such that the test component of the OS operator reduces to $C_{ab}^{ST}(f_a) = C_{ab}^{0}$, with $C_{ab}^{0}$ being the test component with the linearized Coulomb collision operator given in \eqref{eq:CLT}. Hence, the OS operator differs from the Coulomb collision operator only by its field component, given in \eqref{eq:CabSF}, when $T_a = T_b$. Similarly to the linearized Coulomb collision operator, we remark that the velocity-space derivatives and integrals contained in $C_{ab}^0$ and $C_{ab}^{SF}$ given in Eqs. (\ref{eq:Cab0}) and (\ref{eq:CabSF}) respectively, act on the perturbed particle distribution function $f_a$.

\section{GK Formulation of Collision Operators}
\label{appendix:Abis}

We now discuss the GK formulations of the GK Coulomb and GK OS operators based on the spherical harmonic technique used in this work. The GK linearized collision operator associated with the operator $C_{ab}$ (see \eqref{eq:linCab}), $\C_{ab}$, is given by the sum of the GK test and field components, i.e.

\begin{align} \label{eq:gkoperatorpushforeward}
    \mathcal{C}_{ab}^T  & = \left< (\mathcal{T}^{-1} C_{ab}^{T})(\bm Z) \right>_{\bm R} = \int_{0}^{2 \pi} \bigg \rvert_{\bm R} \frac{d \theta}{2 \pi} C_{ab}^{T}(\bm z(\bm Z)),
    \end{align}
    \\
    and
        
 \begin{align}\label{eq:gkoperatorpushforewardfield}
     \mathcal{C}_{ab}^F & = \left< \left( \mathcal{T}^{-1} C_{ab}^{F} \right)( \bm Z)  \right>_{\bm R} = \int_{0}^{2 \pi} \bigg \rvert_{\bm R} \frac{d \theta}{2 \pi} C_{ab}^{F}(\bm z(\bm Z)),
\end{align}
\\
where we note that $C_{ab}^{T} = C_{ab}^{T}(\bm z) = C_{ab}^{T} \left(f_a(\bm z),  f_{Mb} \right)$ and, similarly, $C_{ab}^{F} = C_{ab}^{F}(\bm z) = C_{ab}^{F} \left( f_{Ma},f_b(\bm z) \right)$. In Eqs. (\ref{eq:gkoperatorpushforeward}) and (\ref{eq:gkoperatorpushforewardfield}), we define the push-forward operator, $\mathcal{T}^{-1}$ (inverse of $\mathcal{T}$) that allows us to express a scalar function, such as $f(\bm z)$, defined on the particle phase-space $\bm z$, in terms of the gyrocenter phase-space coordinates $\bm Z$, i.e. $ f(\bm z(\bm Z)) = (\mathcal{T}^{-1} f)(\bm Z)   = f( \mathcal{T}^{-1} \bm Z )$, being $ \bm z (\bm Z) = \mathcal{T}^{-1} \bm Z$.

 In the spherical harmonic expansion, $C_{ab}^{T}( \bm r, \bm v )$ and $C_{ab}^{F}( \bm r, \bm v )$ are expressed as linear combinations of particle fluid moments, $\tens{M}_a^{pj}(\bm r)$, that are written in terms of $h_a$, i.e.\citep{jorge2017drift,jorge2019nonlinear,frei2021}

\begin{align} \label{eq:Mapjwithha}
 \tens{M}_a^{pj}(\bm r)   & = \frac{1}{N_a} \int d \bm R d v_\parallel d \mu d \theta \frac{B}{m_a} \delta ( \bm R + \bm \rho_a - \bm r) \nonumber \\
   & \times h_a(\bm R(\r, \bm v), \mu, v_\parallel ) \tens{Y}^p(\bm s) L_j^{p + 1/2}(s_a^2),
\end{align}
\\
with $\bm R(\r, \bm v) \simeq \bm r - \bm{\rho}_a$. We notice that Eq. (\ref{eq:Mapjwithha}) reduces to \eqref{eq:Mawithha} with $p = 1$. The analytical  expression of the $\tens{M}_a^{pj}$ moments in terms of gyro-moments, $n_a^{pj}$, is detailed in Ref. (\onlinecite{frei2021}). Using \eqref{eq:Mapjwithha} into the spherical harmonic expansion of the collision operator, \eqref{eq:cmomentexpansion}, allows us to obtain the linearized collision operator (Coulomb or OS operators) in the particle phase-space $\bm z = ( \bm r, \bm v)$ in terms of the gyrocenter distribution function $h_a$.

In order to evaluate the gyro-average in Eqs. (\ref{eq:gkoperatorpushforeward}) and (\ref{eq:gkoperatorpushforewardfield}), the collision operator is first transformed to gyrocenter coordinates using the push-forward operator $\mathcal{T}^{-1}$. Because the collision operator is a scalar phase-space function defined on the particle phase-space, i.e. $C_{ab}^T = C_{ab}^T(\bm z)$, it transforms as 

\begin{align} \label{eq:CabTgyrocenter}
C_{ab}^T(\bm z(\bm Z)) =\left( \mathcal{T}^{-1} C_{ab}^T \right) (\bm Z)& = C_{ab}^T \left(\mathcal{T}^{-1} \bm Z \right) \nonumber \\
 & \simeq C_{ab}^T \left( \bm{R} + \bm{\rho}_a(\mu, \theta) , \mu, v_\parallel, \theta \right)
\end{align}
\\
where we use the lowest order gyrocenter coordinate transformations, in particular $\bm r ( \bm Z) \simeq \bm R + \bm{\rho}_a(\mu, \theta) $. Eq. (\ref{eq:CabTgyrocenter}) allows the expression of the linearized collision operator in gyrocenter phase-space coordinates $\bm Z$, as needed to perform analytically the gyro-average. For instance, evaluating the gyro-average of the test component of the Coulomb operator for a single Fourier component using \eqref{eq:cmomentexpansion}, where the particle position $\bm r$ appearing in $\tens{M}_a^{pj}$ is transformed to $\bm R + \bm {\rho}_a(\mu, \theta)$ according to \eqref{eq:CabTgyrocenter}, yields 

\begin{align} \label{eq:averCLT}
    \mathcal{C}_{ab}^{LT} & = \sum_{p = 0}^\infty \sum_{j= 0}^\infty \sum_{m  =-p}^p \sum_{n = -\infty}^\infty \frac{f_{Ma}}{\sigma_j^p} \nu_{ab}^{Tpj}(v)  e^{i \bm k \cdot \bm R} \tens{M}_a^{pj}(\bm k) \cdot \bm e^{pm} \nonumber  \\ 
    & \times i^n J_n( b_a \sqrt{x_a})  \sqrt{\frac{2 \pi^{3/2} p!}{2^p (p+1/2)!}} \left< e^{i n \theta } Y_p^m(\xi, \theta) \right>_{\bm R} ,
\end{align}
\\
where \eqref{eq:Yp} and the Jacobi-Anger identity, $e^{i \bm k \cdot \bm{\rho}_a } = \sum_n i^n J_n( b_a \sqrt{x_a}) e^{i n \theta}$, are used. A similar derivation for the GK field component can be performed. Finally, the gyro-moment expansion of the GK Coulomb collision operator that we use in this work is derived by projecting \eqref{eq:averCLT} onto the Hermite-Laguerre basis, as detailed in Ref. \onlinecite{frei2021}. The gyro-averaging produces FLR terms, associated with the spatial shift $\bm \rho_a(\mu, \theta)$ and the $\theta$-dependence of $Y_p^m(\xi, \theta)$, in both the test and field components of the linearized GK collision operators, that appear through Bessel functions of all orders, i.e. $J_n(b_a \sqrt{x_a}) = \sum_{j=0}^\infty c_j ( b_a \sqrt{x_a}/2)^{2j+n }$, with $c_j = (-1)^j / [j! ( n + j)!]$. These FLR terms ultimately damp small scale fluctuations \cite{frei2022}, and vanish when the DK limit is considered.
 
We remark that the spherical harmonic expansion allows us to evaluate the velocity derivatives (and integrals contained in the Rosenbluth potentials \citep{frei2021}) exactly and express them as linear combination of $\tens{M}_a^{pj}(\bm r)$ given in \eqref{eq:Mapjwithha}. \revvv{For instance, the velocity derivative contained in the test component is computed using the spherical harmonic expansion, \eqref{eq:momentexpansion}, and evaluated in gyrocenter coordinates, $\bm Z = (\bm R, \mu, v_\parallel, \theta)$, as follows \cite{jorge2019nonlinear}

    \begin{align}\label{eq:dvfa} 
    & \frac{\partial }{\partial \bm  v} \bigg \rvert_{\bm r }  f_a (\bm r, \bm v) =\sum_{p=0}^\infty\sum_{j=0}^\infty\sum_{\ell=0}^j \frac{ L_{j\ell}^p}{\sigma_j^p} \cdot\tens{M}_a^{pj}(\bm r(\bm Z))   \frac{\partial }{\partial \bm  v} \bigg \rvert_{\bm r }  \left( s_a^{2 \ell} \tens{Y}^{p}(\bm s_a) f_{Ma} \right),
\end{align}
\\
with $\bm r(\bm Z) \simeq \bm R + \bm \rho_a$. \revv{A similar expression as \eqref{eq:dvfa} can be derived for the second order velocity derivatives contained in \eqref{eq:CLT}.}} 

On the other hand, in previous formulations \cite{sugama2009,LiB2011} and numerical implementations \cite{Crandall2020,Pan2020} of GK collision operators, the velocity derivatives and integrals contained in the particle collision operators are evaluated in terms of the lowest order gyrocenter coordinates, $\bm Z = (\bm R, \mu, v_\parallel, \theta)$, and in terms of $h_a$ by using the chain rule while holding $\r$ constant, i.e. 
 
 \begin{align}
 \label{eq:partialZpartialv}
     \frac{\partial }{\partial \bm  v} \bigg \rvert_{\bm r }  f_a (\bm r, \bm v) &  = \frac{\partial \bm Z}{\partial \bm v} \bigg \rvert_{\bm r }   \frac{\partial}{\partial \bm Z}   f_a (\bm r(\bm Z), \bm v(\bm Z)) \nonumber \\
     & \simeq  \frac{\partial}{\partial \bm v} \bigg \rvert_{\bm R }   h_a   -    \frac{\partial \bm \rho_a}{\partial \bm  v} \bigg \rvert_{\bm r } \cdot \grad  h_a,
\end{align}
\\
with $f_a \simeq h_a$ (see \eqref{eq:fatoha}), and by approximating

\begin{align} \label{eq:Ghb}
    G(f_b(\bm r, \bm v)) & \simeq G(h_b( \bm R(\bm r, \bm v), \mu, v_\parallel)) \nonumber \\
    & = \int d \bm v' \left| \bm v' - \bm v \right| h_b( \bm r - \bm \rho_b (\theta', \mu'), \mu', v_\parallel') ,
\end{align}
\\
in the Rosenbluth potential $G$ (and similarly in $H(f_b)$). Using the transformation in Eq. (\ref{eq:partialZpartialv}) to express the second order velocity derivatives in \eqref{eq:CLT} yields FLR terms of the order of $ \sim \nu_{ab } k_\perp^2 \rho_a^2 h_a$ in the test component \revvv{when gyro-averaged}, while the transformation in Eq (\ref{eq:Ghb}) produces FLR terms proportional to Bessel functions $J_{n}$ in the field components \cite{sugama2009,LiB2011}. However, despite the difference in the analytical treatment of the velocity derivatives using the spherical harmonic expansion (see \eqref{eq:dvfa}) and the lowest order gyrocenter coordinates (see \eqref{eq:partialZpartialv}), both approaches contain the same lowest order FLR terms, proportional to the spatial gradients of $h_a$ if proper approximations are applied to the spherical harmonic approach. \revvv{In fact, \eqref{eq:partialZpartialv} can be recovered from \eqref{eq:dvfa} by Taylor expanding the spatial dependence of particle spherical moments, such that 

\begin{align} \label{eq:momentTaylor}
\tens{M}_a^{pj}(\bm r(\bm Z)) \simeq \tens{M}_a^{pj}(\bm R) + \bm \rho_a \cdot \grad \tens{M}_a^{pj}(\bm R) .
\end{align}
\\
Using \eqref{eq:momentTaylor} into \eqref{eq:dvfa}, the fact that that $\bm \rho_a$ is a velocity-dependent function and the spherical harmonic expansion of $f_a(\bm R, \bm v)$ (i.e., \eqref{eq:momentexpansion} with $\bm r $ replaced by $\bm R$), we derive that 

\begin{align} \label{eq:partialvfaLWL}
 \frac{\partial }{\partial \bm  v} \bigg \rvert_{\bm r }  f_a (\bm r, \bm v) & \simeq  \frac{\partial }{\partial \bm  v} \bigg \rvert_{\bm r }  f_a(\bm R  + \bm \rho_a, \bm v ) - \frac{\partial \bm \rho_a}{\partial \bm  v} \bigg \rvert_{\bm r } \cdot \grad f_a(\bm R, \bm v) \nonumber \\
 & \simeq  \frac{\partial}{\partial \bm v} \bigg \rvert_{\bm R }   h_a   -    \frac{\partial \bm \rho_a}{\partial \bm  v} \bigg \rvert_{\bm r } \cdot \grad  h_a.
\end{align}
\\
In \eqref{eq:partialvfaLWL}, we also use the scalar invariance of the velocity-space derivative, i.e. $\partial_{\bm v}\rvert_{\bm r} f_a(\bm R + \rho_a, \bm v ) \simeq \partial_{\bm v}\rvert_{\bm R}   h_a(\bm R, \mu , v_\parallel )$, and $f_a \simeq h_a$ in the second term. Eq. (\ref{eq:partialvfaLWL}) shows that the velocity-space derivative evaluated using the spherical moments in \eqref{eq:dvfa} agrees with the transformation in \eqref{eq:partialZpartialv} at the lowest order in $k_\perp \rho_a$. Using the first order derivative, given in \eqref{eq:partialvfaLWL}, to express the second order derivatives in the GK test component yields a FLR term of the order of $\sim \nu_{ab} k_\perp^2 \rho_a^2 n_a^{pj}$ when gyro-averaged, similar to the transformation in \eqref{eq:partialZpartialv}.

We finally remark the the last term in \eqref{eq:partialvfaLWL}, proportional to $\rho_a ^2 k_\perp^2$, depends on the perpendicular energy coordinate $v_\perp^2$. Hence, at large $k_\perp^2 v_\perp^2$, a large number of Laguerre gyro-moments is required to capture accurately these FLR effects. In these cases, a careful truncation of the sum over $j$ in \eqref{eq:cmomentexpansion} and in the Bessel function expansions (e.g., the sum over $n$ in \eqref{eq:Bessel_relation}) contained in the GK collision operators  must be performed to obtain a well-behaved numerical results at larger $v_\perp^2$ and $k_\perp^2$, respectively.  Numerical tests reveal that truncating these sums at $j \sim 12$ and $n \sim 8$ is sufficient in the cases discussed herein where $k_\perp \rho_i \lesssim 1$.}

\section{Lowest Order Gyro-Moment Analytical Expressions}
\label{appendix:B}

At high-collisionality, the number of gyro-moments necessary to describe the perturbed distribution function $g_a$ is reduced since higher-order gyro-moments are strongly damped by collisions. In this regime, the perturbed distribution function is well approximated by a perturbed Maxwellian, with its perturbation that has a relative amplitude of the order of the ratio of the particle mean-free path to the typical parallel scale length, i.e. $\lambda_{mfp} / L_\parallel \ll 1$. The Maxwellian gyro-moments $(p,j)$, i.e. the gyrocenter density $(0,0)$, the parallel gyrocenter velocity $(1,0)$, the parallel and perpendicular temperatures, $(2,0)$ and $(0,1)$ respectively are leading order in $\lambda_{mfp} / L_\parallel$. On the other hand, the gyro-moments, associated with the non-Maxwellian component of $\g_a$, i.e. the parallel and perpendicular heat fluxes  $(3,0)$ and $(1,1)$, respectively are first order in $\lambda_{mfp} / L_\parallel$.

 By projecting the GK Boltzmann equation on the Hermite-Laguerre basis \cite{frei2020gyrokinetic} (or the linearized GK Boltzmann equation given \eqref{eq:linGK}), the gyro-moment expansion of the IS collision operator, presented in this work, allows us to evaluate explicitly the collisional terms that enter in the evolution equations of the lowest order gyro-moment enumerated above. We consider the DK IS collision operator (FLR effects yield complicated coefficients that rely on sums with the number of significant terms that depends on $k_\perp$). Using the closed analytical formulas of the DK IS, given in Eqs. (\ref{eq:DKdeltaCabTpj}) and (\ref{eq:DKdeltaCabFpj}), the non-vanishing terms $\Delta \C_{abp'j'}^{Tpj}$ in the gyro-moment expansion of the test component $\Delta \C_{ab}^{Tpj}$, are given by

\begin{widetext}
\begin{subequations} \label{eq:ISTexplicit}
\begin{align}
    \Delta \C_{ab30}^{T10} & = \frac{4 \tau^{3/2}  \nu_{ab} }{5 (\sigma +\tau )^{5/2}}  \sqrt{\frac{6}{\pi }}  \left[\tau  \left(-\sqrt{\frac{(\sigma+1) \tau }{\sigma +\tau }}+\sigma +1\right)  -\sigma  \sqrt{\frac{(\sigma +1) \tau
   }{\sigma +\tau }}\right], \\
    \Delta \C_{ab11}^{T10} & = \frac{ 8 \tau ^{3/2} \nu_{ab}}{5 \sqrt{\pi } (\sigma +\tau )^{5/2}} \left[\sigma  \sqrt{\frac{(\sigma +1) \tau }{\sigma +\tau }} +\tau  \left(\sqrt{\frac{(\sigma
   +1) \tau }{\sigma +\tau }}-\sigma -1\right)\right], \\
    \Delta \C_{ab10}^{T30} & =- \frac{4 \sqrt{\tau } \nu_{ab}}{5 (\sigma +\tau )^{5/2}}  \sqrt{\frac{2}{3 \pi }} \left[10 \sigma ^2 (\tau -1)+3 \sqrt{\frac{(\sigma +1) \tau ^5}{\sigma +\tau }}+\sigma  \tau 
   \left(3 \sqrt{\frac{(\sigma +1) \tau }{\sigma +\tau }}+\tau -4\right)-3 \tau ^2\right], \\
      \Delta \C_{ab30}^{T30} & = -\frac{12\nu_{ab} \sigma  (\tau -1) \sqrt{\tau } \left(10 \sigma ^2-2 \sigma  \tau +3 \tau ^2\right)}{25 \sqrt{\pi } (\sigma +\tau )^{7/2}}, \\
            \Delta \C_{ab11}^{T30} & = \frac{4\nu_{ab} \sqrt{\frac{6}{\pi }} \sigma  (\tau -1) \sqrt{\tau } \left(10 \sigma ^2-2 \sigma  \tau +3 \tau ^2\right)}{25 (\sigma +\tau )^{7/2}}, \\
                \Delta \C_{ab10}^{T11} & = \frac{8 \nu_{ab}\sqrt{\tau }}{15 \sqrt{\pi } (\sigma +\tau )^{5/2}  } \left[10 \sigma ^2 (\tau -1)+3 \sqrt{\frac{(\sigma +1) \tau ^5}{\sigma +\tau }}+\sigma  \tau  \left(3 \sqrt{\frac{(\sigma +1)
   \tau }{\sigma +\tau }}+\tau -4\right)-3 \tau ^2\right], \\
      \Delta \C_{ab30}^{T11} & =  \Delta \C_{ab11}^{T30}, \\
            \Delta \C_{ab11}^{T11} & = -\frac{8 \nu_{ab} \sigma  (\tau -1) \sqrt{\tau } \left(10 \sigma ^2-2 \sigma  \tau +3 \tau ^2\right)}{25 \sqrt{\pi } (\sigma +\tau )^{7/2}},
\end{align}
\end{subequations}
\end{widetext}
\noindent
where $\tau = T_a / T_b$, $\sigma = m_a / m_b$ are the temperature and mass ratios of the colliding species, respectively. Similarly, the non-vanishing terms \rev{$\Delta \C_{ab,p'j'}^{Fpj}$} in the gyro-moment expansion of the field component $\Delta \C_{ab}^{Fpj}$ (see \eqref{eq:CabISmatricprod}) are

\begin{widetext}
\begin{subequations} \label{eq:ISFexplicit}
\begin{align}
    \Delta \C_{ab30}^{F10} & = -\frac{4 \sqrt{\frac{6}{\pi }} \sigma ^{3/2} \tau  \left(-\sqrt{(\sigma +1)
   (\sigma +\tau )}+\sigma +1\right)}{5 (\sigma +\tau )^{5/2}} , \\ 
    \Delta \C_{ab11}^{F10} & =  \frac{8 \sigma  \sqrt{\sigma  \tau } \left(-\sqrt{(\sigma +1) \tau ^3}+\sigma
    \left(\sqrt{\tau  (\sigma +\tau )}-\sqrt{(\sigma +1) \tau
   }\right)+\sqrt{\tau  (\sigma +\tau )}\right)}{5 \sqrt{\pi } (\sigma +\tau
   )^3}, \\
      \Delta \C_{ab10}^{F30} & = \frac{4 \sqrt{\frac{6}{\pi }} \sqrt{\sigma } \tau  \left(\sqrt{(\sigma +1)
   \tau ^3}+\tau  \left(-\sqrt{\sigma +\tau }\right)+\sigma  \left(-3 \tau 
   \sqrt{\sigma +\tau }+\sqrt{(\sigma +1) \tau }+2 \sqrt{\sigma +\tau
   }\right)\right)}{5 (\sigma +\tau )^3}, \\
     \Delta \C_{ab30}^{F30} & =\frac{36 \sigma ^{3/2} \tau  \left(\sigma  \left(5 \tau -\sqrt{\tau
   }-2\right)-\left(\sqrt{\tau }-3\right) \tau \right)}{25 \sqrt{\pi }
   (\sigma +\tau )^{7/2}}, \\
      \Delta \C_{ab11}^{F30} & = \frac{12 \sqrt{\frac{6}{\pi }} \sigma ^{3/2} \tau  \left(\sigma  \left(-5
   \tau +\sqrt{\tau }+2\right)+\left(\sqrt{\tau }-3\right) \tau \right)}{25
   (\sigma +\tau )^{7/2}}, \\
   \Delta \C_{ab10}^{F11} & = -\frac{8 \sqrt{\sigma  \tau } \left(\sigma  \left(-3 \sqrt{\tau ^3 (\sigma
   +\tau )}+\sqrt{\sigma +1} \tau +2 \sqrt{\tau  (\sigma +\tau
   )}\right)-\sqrt{\tau ^3 (\sigma +\tau )}+\sqrt{\sigma +1} \tau
   ^2\right)}{5 \sqrt{\pi } (\sigma +\tau )^3}, \\
      \Delta \C_{ab30}^{F11} & = \Delta \C_{ab11}^{F30}, \\
     \Delta \C_{ab11}^{F11} & = \frac{24 \sigma ^{3/2} \tau  \left(\sigma  \left(5 \tau -\sqrt{\tau
   }-2\right)-\left(\sqrt{\tau }-3\right) \tau \right)}{25 \sqrt{\pi }
   (\sigma +\tau )^{7/2}}.
\end{align}
\end{subequations}
\end{widetext}
\noindent
We notice that, from the conservation of particle in \eqref{eq:particlecons}, it follows $\Delta \C_{ab,p'j'}^{T00} = \Delta \C_{ab,p'j'}^{F00} $. \rev{Finally, we give the non-vanishing terms in the case of like-species collision. Because the test component of the OS operator reduces to the Coulomb collision operator when $a = b$, $\Delta C_{aa p'j'}^{Tpj} =0$, while the non-vanishing terms of the field component $\Delta C_{aa p'j'}^{Fpj}$ are given by

\begin{widetext}
\begin{subequations}
\begin{align}
\Delta C_{aa30}^{F30} & = \frac{9}{25} \sqrt{\frac{2}{\pi }} , \\
\Delta C_{aa11}^{F30} & =  -\frac{6}{25} \sqrt{\frac{3}{\pi }}, \\
\Delta C_{aa11}^{F11} & = \frac{6}{{25}} \sqrt{\frac{2}{\pi }}.
\end{align}
\end{subequations}
\end{widetext}

}

We also provide the non-vanishing lowest order terms in the gyro-moment expansion of the OS collision operator, that are given by 

\begin{subequations} \label{eq:OSTexplicit}
\begin{align}
    \C^{ST10}_{ab10} &= -\frac{8 \nu_{ab} (\sigma +1) }{3\sqrt{\pi }} \left(\frac{\tau }{\sigma +\tau }\right)^{3/2},\\
    \C^{ST10}_{ab30} &=  \frac{4 \nu_{ab}}{5}\sqrt{\frac{6}{\pi }} \frac{ \sqrt{\sigma +1} \tau ^2}{ (\sigma +\tau)^2},\\
    \C^{ST10}_{ab11} &= -\frac{8 \nu_{ab} \sqrt{\sigma +1} \tau ^2}{5 \sqrt{\pi } (\sigma +\tau )^2},\\
    \C^{ST20}_{ab20} &= -\frac{16 \nu_{ab} \sqrt{\tau } \left(5 \sigma ^2 (\tau +2)+21 \sigma  \tau +6\tau ^2\right)}{45 \sqrt{\pi } (\sigma +\tau )^{5/2}},\\
    \C^{ST20}_{ab01} &= -\frac{16 \nu_{ab} \sqrt{\frac{2}{\pi }} \sqrt{\tau } \left(-5 \sigma ^2 (\tau-1)+3 \sigma  \tau +3 \tau ^2\right)}{45 (\sigma +\tau )^{5/2}},\\
    \C^{ST01}_{ab01} &= -\frac{16 \nu_{ab} \sqrt{\tau } \left(5 \sigma ^2+2 (5 \sigma +9) \sigma  \tau +3\tau ^2\right)}{45 \sqrt{\pi } (\sigma +\tau )^{5/2}},\\
    \C^{ST30}_{ab30} &= -\frac{4 \nu_{ab} \sqrt{\tau } \left(70 \sigma ^2+56 \sigma  \tau +31 \tau^2\right)}{35 \sqrt{\pi } (\sigma +\tau )^{5/2}},\\
    \C^{ST30}_{ab11} &= -\frac{4 \nu_{ab} \sqrt{\frac{2}{3 \pi }} \tau ^{3/2} (28 \sigma +\tau )}{35(\sigma +\tau )^{5/2}},\\
    \C^{ST11}_{ab11} &= -\frac{8 \nu_{ab} \sqrt{\tau } \left(105 \sigma ^2+98 \sigma  \tau +47 \tau^2\right)}{105 \sqrt{\pi } (\sigma +\tau )^{5/2}},
\end{align}
\end{subequations}

with $\C^{STpj}_{ablk} = \C^{STlk}_{abpj}$, and

\begin{subequations} \label{eq:OSTexplicit}
\begin{align}
    \C^{SF10}_{ab10} &= \frac{8 \nu_{ab} \sigma  (\sigma +1) \tau }{3 \sqrt{\pi } \sqrt{\sigma  (\sigma+\tau )^3}},\\
    \C^{SF10}_{ab30} &= -\frac{4 \nu_{ab} \sqrt{\frac{6}{\pi }} \sqrt{\sigma ^3 (\sigma +1)} \tau }{5(\sigma +\tau )^2},\\
    \C^{SF10}_{ab11} &= \frac{8 \nu_{ab} \sqrt{\sigma ^3 (\sigma +1)} \tau }{5 \sqrt{\pi } (\sigma +\tau)^2},\\
    \C^{SF20}_{ab20} &= \frac{16 \nu_{ab} \sigma  (\sigma +1) \tau ^{3/2}}{9 \sqrt{\pi } (\sigma +\tau)^{5/2}},\\
    \C^{SF20}_{ab01} &= -\frac{16 \nu_{ab} \sqrt{\frac{2}{\pi }} \sigma  (\sigma +1) \tau ^{3/2}}{9(\sigma +\tau )^{5/2}},\\
    \C^{SF01}_{ab20} &= -\frac{16 \nu_{ab} \sqrt{\frac{2}{\pi }} \sigma  (\sigma +1) \tau ^{3/2}}{9(\sigma +\tau )^{5/2}},\\
    \C^{SF01}_{ab01} &= \frac{32 \nu_{ab} \sigma  (\sigma +1) \tau ^{3/2}}{9 \sqrt{\pi } (\sigma +\tau)^{5/2}},\\
    \C^{SF30}_{ab10} &= -\frac{4 \nu_{ab} \sqrt{\frac{6}{\pi }} \sqrt{\sigma  (\sigma +1) \tau ^3}}{5(\sigma +\tau )^2},\\
    \C^{SF30}_{ab30} &= \frac{36 \nu_{ab} (\sigma  \tau )^{3/2}}{25 \sqrt{\pi } (\sigma +\tau )^{5/2}},\\
    \C^{SF30}_{ab11} &= -\frac{12 \nu_{ab} \sqrt{\frac{6}{\pi }} (\sigma  \tau )^{3/2}}{25 (\sigma +\tau)^{5/2}},\\
    \C^{SF11}_{ab10} &= \frac{8 \nu_{ab} \sqrt{\sigma  (\sigma +1) \tau ^3}}{5 \sqrt{\pi } (\sigma +\tau)^2},\\
    \C^{SF11}_{ab30} &= -\frac{12 \nu_{ab} \sqrt{\frac{6}{\pi }} (\sigma  \tau )^{3/2}}{25 (\sigma +\tau)^{5/2}},\\
    \C^{SF11}_{ab11} &= \frac{24 \nu_{ab} (\sigma  \tau )^{3/2}}{25 \sqrt{\pi } (\sigma +\tau )^{5/2}}.
\end{align}
\end{subequations}
\\
for the test and field components, respectively. \rev{In the case of like-species collision, the non-vanishing terms are, given that 

\begin{widetext}
\begin{subequations}
\begin{align}
    \C^{S20}_{aa20}    & = - \frac{64}{45} \sqrt{\frac{2}{ \pi}} , \\    
      \C^{S20}_{aa01}  & = - \frac{64}{45} \sqrt{\frac{1}{ \pi}} , \\  
                      \C^{S01}_{aa01}  & = - \frac{32}{45} \sqrt{\frac{2}{ \pi}} , \\ 
                            \C^{S30}_{aa30}  & =  - \frac{361}{175} \sqrt{\frac{2}{\pi}}, \\ 
                                          \C^{S30}_{aa11}   & =  - \frac{208}{175} \sqrt{\frac{1}{3\pi}}, \\ 
                                           \C^{S11}_{aa11}   & =  - \frac{1187}{525} \sqrt{\frac{2}{\pi}}.
\end{align}
\end{subequations}
\end{widetext}
}

Finally and for completeness, we report the non-vanishing coefficients of the DK Coulomb collision operator of the test $\C_{ab}^{LT}$ and the field $\C_{ab}^{LF}$ components respectively, i.e.

\begin{subequations} \label{eq:ISTexplicit}
\begin{align}
   \C_{ab10}^{LT10} & = -\frac{8 \nu_{ab} (\sigma +1)}{3 \sqrt{\pi }} \left(\frac{\tau }{\sigma +\tau }\right)^{3/2}, \\ 
 \C_{ab30}^{LT10} & =    \frac{4}{5} \sqrt{\frac{6}{\pi }} (\sigma +1) \left(\frac{\tau }{\sigma +\tau
   }\right)^{5/2}, \\
  \C_{ab11}^{LT10} & =  -\frac{8 \nu_{ab}  (\sigma +1) }{{5 \sqrt{\pi } }\left(\frac{\tau }{\sigma +\tau }\right)^{5/2}}, \\
    \C_{ab00}^{LT20} & = -\frac{8  \nu_{ab} \sqrt{\frac{2}{\pi }} (\tau -1) \sqrt{\frac{\tau }{\sigma }}}{3
   \left(\frac{\sigma +\tau }{\sigma }\right)^{3/2}}, \\
    \C_{ab20}^{LT20} & = -\frac{ \nu_{ab}8 \left(\frac{\tau }{\sigma }\right)^{3/2} (\sigma  (10 \sigma +\tau +13)+4 \tau
   )}{15 \sqrt{\pi } \sigma  \left(\frac{\sigma +\tau }{\sigma }\right)^{5/2}}, \\
      \C_{ab01}^{LT20} & = \frac{8 \nu_{ab} \sqrt{\frac{2}{\pi }} \left(\frac{\tau }{\sigma }\right)^{3/2} (-3 \sigma  \tau
   +\sigma -2 \tau )}{15 \sigma  \left(\frac{\sigma +\tau }{\sigma }\right)^{5/2}}, \\ \C_{ab00}^{LT01} & =  \frac{16  \nu_{ab}(\tau -1) \sqrt{\frac{\tau }{\sigma }}}{3 \sqrt{\pi } \left(\frac{\sigma +\tau
   }{\sigma }\right)^{3/2}} , \\
   \C_{ab20}^{LT01} & = \frac{8 \sqrt{\frac{2}{\pi }} \left(\frac{\tau }{\sigma }\right)^{3/2} (-3 \sigma  \tau
   +\sigma -2 \tau )}{15 \sigma  \left(\frac{\sigma +\tau }{\sigma }\right)^{5/2}} , \\
    \C_{ab01}^{LT01} & =-\frac{16 \left(\frac{\tau }{\sigma }\right)^{3/2} (\sigma  (5 \sigma -\tau +7)+\tau
   )}{15 \sqrt{\pi } \sigma  \left(\frac{\sigma +\tau }{\sigma }\right)^{5/2}}, \\
   \C_{ab10}^{LT30} & =-\frac{4 \nu_{ab}}{5 \sigma ^2}  \sqrt{\frac{2}{3 \pi }} \sqrt{\frac{\tau }{\sigma }} \left(\frac{\sigma
   }{\sigma +\tau }\right)^{5/2} \left(10 \sigma ^2 (\tau -1)+\sigma  (\tau -4) \tau -3
   \tau ^2\right), \\
      \C_{ab30}^{LT30} & = -\frac{4 \nu_{ab} \left(\frac{\tau }{\sigma +\tau }\right)^{3/2} \left(14 \sigma ^2 (\tau +8)+70
   \sigma ^3+\sigma  \tau  (19 \tau +68)+31 \tau ^2\right)}{35 \sqrt{\pi } (\sigma
   +\tau )^2}, \\
         \C_{ab11}^{LT30} & = \frac{4 \nu_{ab} \sqrt{\frac{2}{3 \pi }} \left(\frac{\tau }{\sigma +\tau }\right)^{3/2}
   \left(\sigma ^2 (14-42 \tau )+\sigma  \tau  (3 \tau -32)-\tau ^2\right)}{35 (\sigma
   +\tau )^2}, \\
   \C_{ab10}^{LT11} & = \frac{8\nu_{ab} \sqrt{\frac{\tau }{\sigma }} \left(10 \sigma ^2 (\tau -1)+\sigma  (\tau -4)
   \tau -3 \tau ^2\right)}{15 \sqrt{\pi } \sigma ^2 \left(\frac{\sigma +\tau }{\sigma
   }\right)^{5/2}}, \\
      \C_{ab30}^{LT11} & = \frac{4 \nu_{ab} \sqrt{\frac{2}{3 \pi }} \left(\frac{\tau }{\sigma +\tau }\right)^{3/2}
   \left(\sigma ^2 (14-42 \tau )+\sigma  \tau  (3 \tau -32)-\tau ^2\right)}{35 (\sigma
   +\tau )^2}, \\
         \C_{ab11}^{LT11} & =-\frac{8 \nu_{ab} \left(\frac{\tau }{\sigma +\tau }\right)^{3/2} \left(7 (15 \sigma +23) \sigma
   ^2+(27 \sigma +47) \tau ^2+2 (21 \sigma +59) \sigma  \tau \right)}{105 \sqrt{\pi }
   (\sigma +\tau )^2}, 
\end{align}
\end{subequations}

and

\begin{subequations} \label{eq:ISTexplicit}
\begin{align}
   \C_{ab10}^{LF10} & =  \frac{8 \nu_{ab} (\sigma +1) \tau }{3 \sqrt{\pi
   } \sigma } \left(\frac{\sigma }{\sigma +\tau }\right)^{3/2} , \\
   \C_{ab30}^{LF10} & = -\frac{4  \nu_{ab}\sqrt{\frac{6}{\pi }} (\sigma +1) \tau }{5 \sigma } \left(\frac{\sigma }{\sigma +\tau }\right)^{5/2}  , \\
   \C_{ab11}^{LF10} & = \frac{8 \nu_{ab} (\sigma +1) \tau }{5 \sqrt{\pi } \sigma  } \left(\frac{\sigma }{\sigma +\tau }\right)^{5/2} , \\
     \C_{ab00}^{LF20} & = -\frac{8 \nu_{ab}\sqrt{\frac{2}{\pi }} (\tau -1) \sqrt{\frac{\tau }{\sigma }}}{3
   \left(\frac{\sigma +\tau }{\sigma }\right)^{3/2}}, \\
     \C_{ab20}^{LF20} & = \frac{8\nu_{ab} \sqrt{\frac{\tau }{\sigma }} (\sigma  (3 \tau -1)+2 \tau )}{5 \sqrt{\pi }
   \sigma  \left(\frac{\sigma +\tau }{\sigma }\right)^{5/2}} , \\
       \C_{ab01}^{LF20} & = \frac{8\nu_{ab} \sqrt{\frac{2}{\pi }} \sqrt{\frac{\tau }{\sigma }} (-3 \sigma  \tau +\sigma -2
   \tau )}{15 \sigma  \left(\frac{\sigma +\tau }{\sigma }\right)^{5/2}}, \\ 
     \C_{ab00}^{LF01} & = \frac{16 (\tau -1) \sqrt{\frac{\tau }{\sigma }}}{3 \sqrt{\pi } \left(\frac{\sigma +\tau
   }{\sigma }\right)^{3/2}}, \\
      \C_{ab20}^{LF01} & = \frac{8 \sqrt{\frac{2}{\pi }} \sqrt{\frac{\tau }{\sigma }} (-3 \sigma  \tau +\sigma -2
   \tau )}{15 \sigma  \left(\frac{\sigma +\tau }{\sigma }\right)^{5/2}}, \\
       \C_{ab01}^{LF01} & = \frac{32 \sqrt{\frac{\tau }{\sigma }} (\sigma  (3 \tau -1)+2 \tau )}{15 \sqrt{\pi }
   \sigma  \left(\frac{\sigma +\tau }{\sigma }\right)^{5/2}}, \\ 
   \C_{ab10}^{LF30} & = -\frac{ 4 \nu_{ab} \sqrt{\frac{6}{\pi }} \tau  (\sigma  (3 \tau -2)+\tau )}{5 \sigma ^2
   \left(\frac{\sigma +\tau }{\sigma }\right)^{5/2}}, \\
      \C_{ab30}^{LF30} & = \frac{12 \nu_{ab} \tau  (\sigma  (5 \tau -2)+3 \tau )}{7 \sqrt{\pi } (\sigma +\tau )^2
   \left(\frac{\sigma +\tau }{\sigma }\right)^{3/2}}, \\
         \C_{ab11}^{LF30} & = -\frac{12 \nu_{ab} \sqrt{\frac{6}{\pi }} \tau  (\sigma  (5 \tau -2)+3 \tau )}{35 (\sigma +\tau
   )^2 \left(\frac{\sigma +\tau }{\sigma }\right)^{3/2}}, \\
     \C_{ab10}^{LF11} & = \frac{8  \nu_{ab} \tau  (\sigma  (3 \tau -2)+\tau )}{5 \sqrt{\pi } \sigma ^2 \left(\frac{\sigma
   +\tau }{\sigma }\right)^{5/2}},  \\
        \C_{ab30}^{LF11} & = -\frac{12 \nu_{ab}  \sqrt{\frac{6}{\pi }} \tau  (\sigma  (5 \tau -2)+3 \tau )}{35 (\sigma +\tau
   )^2 \left(\frac{\sigma +\tau }{\sigma }\right)^{3/2}}, \\        
         \C_{ab11}^{LF11} & = \frac{48 \nu_{ab} \tau  (\sigma  (5 \tau -2)+3 \tau )}{35 \sqrt{\pi } (\sigma +\tau )^2
   \left(\frac{\sigma +\tau }{\sigma }\right)^{3/2}} ,
\end{align}
\end{subequations}
\\
respectively. \rev{Finally, the non-vanishing terms for like-species collisions are 

\begin{widetext}
\begin{subequations}
\begin{align}
   \C_{ab20}^{L20}& = - \frac{16}{15} \sqrt{\frac{2}{\pi}}, \\
  \C_{ab01}^{L20}& = - \frac{16}{15} \sqrt{\frac{1}{\pi}}, \\
             \C_{ab01}^{L01}& =  -\frac{8}{15} \sqrt{\frac{2}{\pi}}, \\
 \C_{ab30}^{L30}& = - \frac{8}{5} \sqrt{\frac{2}{\pi}}, \\
              \C_{ab11}^{L30}& = - \frac{8}{5} \sqrt{\frac{1}{3\pi}}, \\
                           \C_{ab11}^{L11}& = - \frac{28}{15} \sqrt{\frac{2}{\pi}}, 
\end{align}
\end{subequations}
\end{widetext}

}

The lowest order gyro-moments of the IS, OS and Coulomb reported above can be used to obtain reduced-fluid models to study the plasma dynamics in the Pfirsch-Schlüter regime. 

\bibliography{aipsamp}

\end{document}